%% file: BL.tex
\def\DIRvalue{BeniniLeFloch}
\def\IDvalue{BL}
\def\titlevalue{Supersymmetric localization in two dimensions}

\def\authorvalue{Francesco Benini$^{1,2}$ and Bruno Le Floch$^3$}
\def\shortauthorvalue{Francesco Benini and Bruno Le Floch}

\def\addressvalue{%
$^{1}$International School for Advanced Studies (SISSA),
via Bonomea 265, 34136 Trieste, Italy \\
$^{2}$Blackett Laboratory, Imperial College London,
London SW7 2AZ, United Kingdom \\
$^{3}$Princeton Center for Theoretical Science, Princeton University,
Princeton, NJ 08544, USA \\
{\tt fbenini@sissa.it}, {\tt blefloch@princeton.edu}
}

\def\abstractvalue{%
This is an introductory review to localization techniques in supersymmetric two-dimensional gauge theories.
In particular we describe how to construct Lagrangians of $\cN{=}(2,2)$ theories on curved spaces, and how to compute their partition functions and certain correlators on the sphere, the hemisphere and other curved backgrounds. We also describe how to evaluate the partition function of $\cN{=}(0,2)$ theories on the torus, known as the elliptic genus. Finally we summarize some of the applications, in particular to probe mirror symmetry and other non-perturbative dualities.
}

\def\preprintvalue{%
SISSA 27/2016/FISI \\
Imperial/TP/2016/FB/01}

\ifx\ifLONG\undefined
\documentclass[12pt]{article}
\input{header.tex}

\input{BLdef.tex}
\input{BLdefcommon.tex}

\hypersetup{
  pdfauthor = {Francesco Benini and Bruno Le Floch},
  pdftitle = {Supersymmetric localization in two dimensions},
}

\hypersetup{
  bookmarksnumbered,
  linktoc = all,
  pdfborderstyle = {/S/U/W 0.5},
  allcolors = darkgray,
}

\begin{document}
\thispagestyle{empty}
\documentheader
\else

\chapterheader

\fi


\newcommand{\textsuperscripttwo}{\texorpdfstring{\textsuperscript{2}}{\000\262}}


\usetikzlibrary{automata}
\usetikzlibrary{arrows}
\usetikzlibrary{calc}
\usetikzlibrary{decorations.markings}
\usetikzlibrary{decorations.pathreplacing}
\usetikzlibrary{intersections}
\usetikzlibrary{positioning}
\usetikzlibrary{topaths}
\usetikzlibrary{shapes.geometric}
\usetikzlibrary{shapes.misc}
\usetikzlibrary{snakes}
\tikzset{->-/.style = {
    decoration = {markings, mark = at position #1 with {\arrow{>}}},
    postaction = {decorate}}}
\tikzset{color-group/.style = {
    shape = circle,
    minimum size = 2.5ex,
    inner sep = .5ex,
    draw}}
\tikzset{flavor-group/.style = {
    shape = rectangle,
    minimum size = 2.5ex,
    inner sep = .5ex,
    draw}}

\section{Introduction}
\label{BLsec:intro}

Two-dimensional theories, despite the low dimensionality, are interesting for a number of reasons: they often appear in statistical physics and condensed matter physics; they share many properties with four-dimensional theories but are more tractable and yet quite non-trivial; they play a central role in string theory; they endow intricate mathematical problems and structures. Particularly tractable are supersymmetric theories, and we will be mostly concerned with 2d $\cN{=}(2,2)$ and $\cN{=}(0,2)$ supersymmetry. Those theories appear on the worldsheet of strings compactified down to four dimensions with $\cN{=}2$ or $\cN{=}1$ supersymmetry. They also exhibit dualities, which identify low-energy limits of pairs of theories, similar to 4d Seiberg duality \cite{BLSeiberg:1994pq}. Two-dimensional non-linear sigma models (NLSMs) with K\"ahler or Calabi-Yau target space, or bundles on such spaces, are related to mathematical problems such as mirror symmetry, Gromov-Witten invariants and quantum sheaf cohomology. Gauged linear sigma models (GLSMs), namely two-dimensional supersymmetric gauge theories, can provide convenient ultraviolet (UV) descriptions of NLSMs \cite{BLWitten:1993yc} thus proving to be extraordinary computational tools. Finally, GLSMs are also used as microscopic descriptions of surface operators in higher dimensions and as worldsheet theories for brane intersections (such as M-strings).

In this review we summarize recent results in supersymmetric localization techniques for two-dimensional theories, and applications to Calabi-Yau manifolds and dualities.
The first half of the review concerns $\cN{=}(2,2)$ theories on curved (compact) spaces: their construction and the computation of the corresponding Euclidean path-integral---that we will generically call a ``partition function''. Curved-space Lagrangians can be obtained by coupling the supersymmetric theory to supergravity, and then switching on background values for the metric and the other bosonic fields in the graviton multiplet \cite{BLFestuccia:2011ws} (also \volcite{DU}). While the flat space $\cN{=}(2,2)$ supersymmetry algebra admits both a vector and an axial $U(1)$ R-symmetry, a mixed anomaly prevents them from being simultaneously gauged: the curved space background must break one, giving rise to A-type and B-type backgrounds, respectively. Supersymmetric supergravity backgrounds on compact orientable Riemann surfaces and preserving the vector R-symmetry were classified in \cite{BLClosset:2014pda}.%
\footnote{A more general classification was provided in \cite{BLBae:2015eoa}.}
With the topology of the sphere, one finds the well-known A- (and $\overline{\text{A}}$-) twist \cite{BLWitten:1988xj} with $\pm 1$ units of R-symmetry flux, as well as an ``$\Omega$-deformation'' thereof \cite{BLBenini:2015noa, BLClosset:2015rna} (see \autoref{BLssec:curved-omega})---but also ``untwisted'' backgrounds \cite{BLBenini:2012ui, BLDoroud:2012xw, BLGomis:2012wy} (see \autoref{BLsec:sphere} and \autoref{BLssec:curved-twisted}) with zero net R-symmetry flux, analogous to the seminal setup of Pestun on $S^4$ \cite{BLPestun:2007rz}. The genus $g=1$ case includes flat tori (see \autoref{BLsec:torus}). For all these cases, we show how partition functions can be computed. For $g>1$, the only solution is the A-twist and we will not discuss it further.

To begin with, in \autoref{BLsec:sphere} we follow \cite{BLBenini:2012ui, BLDoroud:2012xw, BLGomis:2012wy} and perform supersymmetric localization for $\cN{=}(2,2)$ chiral and vector multiplets on squashed-sphere untwisted backgrounds preserving the vector R-symmetry $R$. Since continuous deformations of the coefficients in kinetic and superpotential terms in the action do not affect the path-integral (as those terms are $\cQ$-exact with respect to a supercharge $\cQ$), the partition function is independent of gauge couplings and wave-function renormalization and it is thus a renormalization group (RG) invariant. It is a non-trivial non-perturbative function of R-charges and twisted chiral parameters: twisted masses and flavor fluxes (background field strengths coupled to flavor symmetries) as well as Fayet-Iliopoulos (FI) parameters and theta angles (appearing in the twisted superpotential).

To showcase the supersymmetric localization method, we go through this relatively tractable case in detail.  The path integral localizes to fixed-points of $\supercharge$ (see \autoref{BLssec:sphere-bps}) and quadratic fluctuations around these. Their contribution (see \autoref{BLssec:sphere-det}) is found by adding to the action a $\supercharge$-exact and $\supercharge$-closed deformation term $t\delta_{\supercharge}V$: the limit $t\to\infty$ localizes the path integral further to saddle-points of $\delta_{\supercharge}V$. It turns out that using two different deformation terms one can get different-looking expressions \eqref{BLZCoulomb} and \eqref{BLZHiggs}. The first one (see \autoref{BLssec:sphere-coulomb}), called \emph{Coulomb branch formula}, is a sum over gauge fluxes and an integral over a Coulomb branch parameter of the theory, which converges for generic FI/theta parameters. We show in \autoref{BLssec:sphere-pde} that it obeys a system of differential equations called the A-system. The second one (see \autoref{BLssec:sphere-higgs}), called \emph{Higgs branch formula}, is an expansion in some corner of the FI/theta moduli space: it involves a sum over solutions (dubbed Higgs branches) of the D-term equations, with non-perturbative contributions from point-like \mbox{(anti-)}vortices at the (South) North pole.  $\supercharge$-invariant operators can also be included in both expressions (see \autoref{BLssec:operators}).

The two forms are useful in different settings. Higgs branch expressions are used to confirm Seiberg-like dualities, as discussed in \autoref{BLssec:dual-seiberg}. For instance, $U(K)$ and $U(N-K)$ gauge theories with $N$ fundamental chiral multiplets are expected to have the same low-energy limit.  Their sphere partition functions are shown to be equal by mapping the $( \begin{smallmatrix} N \\ K \end{smallmatrix} )$ solutions of D-term equations of one theory to the $( \begin{smallmatrix} N \\ N-K \end{smallmatrix} )$ solutions for the other, and equating \mbox{(anti-)}vortex contributions order by order in the number of vortices. More complicated variants of this duality can also be checked using Higgs branch expressions (see \autoref{BLssec:dual-variants}).

On the other hand, Coulomb branch expressions are useful to characterize ``phases'' of GLSMs.
Let us consider briefly a famous example: a $U(1)$ vector multiplet and chiral multiplets $P,X_1,\ldots,X_5$ of charges $(-5,+1,\ldots,+1)$ with a superpotential $W=P\,G_5(X)$ for some homogeneous degree~$5$ polynomial~$G_5$. For FI parameter $\zeta\gg 0$ this GLSM flows to an NLSM on the quintic hypersurface $\{G_5(X)=0\}\subset\mathbb{CP}^4$, while for $\zeta\ll 0$ the GLSM flows to an orbifolded Landau-Ginzburg model with a single classical vacuum.  The distinction between these two phases can be seen in the sphere partition function: the Coulomb branch integral can be expanded as a sum of residues of poles to one side or the other of the integration contour depending on whether $2\pi\zeta \lessgtr 5\log 5$ (see \cite{BLHalverson:2013eua, BLBonelli:2013mma} for more general discussions).

We also use the Coulomb branch integral as our starting point when investigating mirror symmetry in \autoref{BLssec:dual-mirror}.  As explained in \volcite{MO}, metric deformations of the NLSM's target Calabi-Yau decompose into complex structure deformations and K\"ahler structure deformations, which correspond respectively to superpotential and twisted superpotential terms in the GLSM action. The partition function $Z_A$ preserving $R$ gives the K\"ahler potential $K_K=-\log Z_A$ on the moduli space of K\"ahler structure deformations \cite{BLJockers:2012dk, BLGomis:2012wy, BLGerchkovitz:2014gta, BLGomis:2015yaa}. Important enumerative geometry data of the Calabi-Yau manifold, namely its genus-zero Gromov-Witten invariants, can then be extracted from the $\zeta\gg 0$ expansion of $Z_A$.  The K\"ahler potential on the moduli space of complex structure deformations is similarly $K_C=-\log Z_B$ in terms of the partition function $Z_B$ of the GLSM on a supergravity background that preserves the axial R-symmetry \cite{BLDoroud:2013pka}.  We compute $Z_B$ in \autoref{BLssec:curved-twisted}. Mirror symmetry states that pairs of Calabi-Yau manifolds have identical moduli spaces, with complex structure and K\"ahler structure deformations interchanged.  Accordingly, we describe in \autoref{BLssec:dual-mirror} how $Z_A$ of a GLSM is equal to $Z_B$ of a GLSM flowing to an NLSM on the mirror Calabi-Yau.

Another important case where localization was performed is the hemisphere \cite{BLSugishita:2013jca, BLHonda:2013uca, BLHori:2013ika} (see \autoref{BLssec:curved-hemi}), which is the simplest case of a manifold with boundaries. GLSMs on the hemisphere can be used to describe open strings with Calabi-Yau target space: boundary conditions for fields on the hemisphere are branes in the target. The hemisphere partition function has an integral and a series representations, like the sphere partition function $Z_A$ (although the contour is difficult to work out in general). We do not discuss the real projective plane calculation of \cite{BLKim:2013ola}, which gives information about orientifold planes in the Calabi-Yau target.

As is well-known, K\"ahler potentials are only defined up to K\"ahler transformations $K\to K+f(z)+\overline{f(z)}$ where~$z$ is a holomorphic coordinate on the given K\"ahler manifold (here a moduli space of metric deformations of a Calabi-Yau).  This translates to a multiplicative ambiguity of sphere partition functions, which can be traced to a freedom in choosing how FI parameters (promoted to twisted chiral multiplets) couple to background supergravity \cite{BLGomis:2012wy, BLClosset:2014pda, BLGerchkovitz:2014gta}. Derivatives $\partial\overline{\partial}\log Z$ of the K\"ahler potential remain unambiguous.  An analysis of supergravity counterterms \cite{BLGomis:2015yaa} shows that there is no such universal content for general $\cN{=}(0,2)$ theories on the sphere. In order to find physical observables, one needs to compute correlators. For some $\cN{=}(0,2)$ deformations of $\cN{=}(2,2)$ GLSMs, placed on the sphere using the $A/2$-twist, correlators were computed using supersymmetric localization in \cite{BLClosset:2015ohf}.

The review is organized as follows.
In \autoref{BLsec:sphere} we calculate the (squashed) sphere partition function of $\cN{=}(2,2)$ GLSMs using the two different localization approaches.
In \autoref{BLsec:curved} we extend the results in various ways: we discuss the inclusion of operators, twisted chiral and twisted vector multiplets, the hemisphere, the $\Omega$-deformed A-twist, ending with a general discussion of $\cN{=}(2,2)$ supersymmetry on curved spaces.
In \autoref{BLsec:torus} we turn to a second major localization result in two dimensions: the (equivariant) elliptic genus of $\cN{=}(2,2)$ and $\cN{=}(0,2)$ theories, namely their partition function on a flat torus.  Contrarily to the sphere, the torus has non-trivial cycles and we include flat background connections. The elliptic genus is an important probe of supersymmetry breaking and is one of the rare RG invariant quantities available to test dualities between $\cN{=}(0,2)$ theories.  After defining the elliptic genus and its modularity properties, we describe $\cN{=}(0,2)$ multiplets and Lagrangians in \autoref{BLssec:torus-theory}, then give the localization formula \eqref{BLeg main formula} in \autoref{BLssec:torus-loc} followed by an outline of the derivation in \autoref{BLssec:torus-calc}, and we end with several extensions and applications in \autoref{BLssec:torus-more}.
In \autoref{BLsec:dual} we highlight some applications of the sphere and torus partition functions. We begin with a check of Abelian mirror symmetry in \autoref{BLssec:dual-mirror}.  Then we check that Seiberg-dual $\cN{=}(2,2)$ theories have equal sphere partition functions and equal elliptic genera in \autoref{BLssec:dual-seiberg} before turning to generalizations in \autoref{BLssec:dual-variants}.  We compare elliptic genera for the $\cN{=}(0,2)$ triality in \autoref{BLssec:dual-02}.
We conclude in \autoref{BLsec:conclusion} with a brief discussion of topics that were not included in the review.


\section{\texorpdfstring{\ifvolume{$\cN{=}(2,2)$}{$\boldsymbol{\cN{=}(2,2)}$}}{N=(2,2)} gauge theories on spheres}
\label{BLsec:sphere}

This section is devoted to partition functions of $\cN{=}(2,2)$ Euclidean gauge theories on the round \cite{BLBenini:2012ui, BLDoroud:2012xw} and squashed \cite{BLGomis:2012wy} sphere. The aims are to show localization at work and to obtain two exact expressions, \eqref{BLZCoulomb} and \eqref{BLZHiggs}, for the $S^2$ partition function.

In terms of the standard flat superspace \cite{BLWitten:1993yc}, the basic $\cN{=}(2,2)$ multiplets are: chiral superfields defined by $\overline{D}_\pm \Phi=0$; vector superfields with gauge transformation $V \cong V+\Lambda+\overline{\Lambda}$ for $\Lambda$ chiral; twisted chiral superfields defined by $\overline{D}_+\widetilde{\Phi}= D_-\widetilde{\Phi}=0$; twisted vector superfields with gauge transformation $\widetilde{V} \cong \widetilde{V} + \Lambda_t + \overline{\Lambda}_t$ for $\Lambda_t$ twisted chiral. The field-strength multiplets $\Sigma=\overline{D}_+D_-V$ and $\widetilde{\Sigma}=\overline{D}_+\overline{D}_-\widetilde{V}$ are twisted chiral and chiral, respectively.   Twisted and untwisted multiplets are interchanged by a $\mathbb{Z}_2$ automorphism of the $\cN{=}(2,2)$ superalgebra, that also exchanges the vector $U(1)_R$ and axial $U(1)_A$ R-symmetries. One can write down kinetic terms for the basic multiplets,
$$
\int \dd[^2]{\theta}\dd[^2]{\bar{\theta}} \bigl( \overline{\Phi}\Phi + \overline{\widetilde{\Phi}} \widetilde{\Phi} \bigr) \;,
$$
including (twisted) chirals and field strengths. Besides, one can write superpotential interactions terms (top component of a composite chiral field $\supo$) and twisted superpotential terms (top component of a composite twisted chiral field $\twsupo$).

We focus in this section on gauged linear sigma models built from vector and chiral multiplets. The components of chiral multiplets (a complex scalar $\phi$, a complex Dirac spinor $\psi$ and a complex auxiliary scalar $F$) transform in some representation $\repr$ of a gauge and flavor symmetry group $G \times G_f$, their conjugates $(\bar{\phi}, \bar{\psi}, \bar{F})$ transform in $\overline{\repr}$, while the vector multiplet components (a real gauge field $A_i$, real scalars $\eta$, $\sigma$, complex Dirac fermions $\lambda$, $\bar{\lambda}$ and a real auxiliary scalar $D$) transform in the adjoint representation of $G$.

In \autoref{BLssec:sphere-theory} we place these GLSMs on the sphere in a way that preserves $U(1)_R$, describe how supersymmetries act and write supersymmetric Lagrangians.  Other $\cN{=}(2,2)$ theories and backgrounds are considered in \autoref{BLsec:curved}.  We choose a localization supercharge $\supercharge$ whose square rotates the sphere around its poles.
In \autoref{BLssec:sphere-bps} we find that $\supercharge$-invariant field configurations are generically parametrized by a discrete gauge flux $\flux$ through $S^2$ and a vector multiplet scalar~$\sigma$.  When chiral multiplets are not charged under $U(1)_R$, we note the existence of additional vortex and antivortex configurations near the poles for particular values of $\sigma$ named Higgs-branch roots.
In \autoref{BLssec:sphere-coulomb} we localize using a deformation term that eliminates (anti)vortices and expresses the partition function as a sum over fluxes and an integral over the Coulomb branch parameter $\sigma$ of one-loop determinants computed later.
The partition function obeys a system of differential equations \cite{BLHalverson:2013eua}, shown in \autoref{BLssec:sphere-pde}, that are (anti)holomorphic in certain combinations of Fayet-Iliopoulos (FI) parameters and theta angles.  It must thus be a sum of products of a holomorphic solution and an antiholomorphic solution.
We reproduce the factorization for a large class of GLSMs in \autoref{BLssec:sphere-higgs} (in the absence of R-charges) by a different choice of deformation term which interpolates between the Coulomb branch integral and a sum over Higgs branches.  Each term in this sum factorizes into (anti)holomorphic vortex partition functions due to (anti)vortices at the (South) North pole.  These can be obtained explicitly by expressing the Coulomb branch integral as a sum of residues.
We end in \autoref{BLssec:sphere-det} by outlining how one-loop determinants for fluctuations around saddle points are computed, correcting a sign in the process.


\subsection{Multiplets, Lagrangians and supersymmetry}
\label{BLssec:sphere-theory}

We now place vector and chiral multiplets and their Lagrangians on squashed spheres which preserve a $U(1)\subset SU(2)$ isometry of $S^2$.  The metric, vielbein, and spin connection are%
\footnote{This metric assumes that lengths of meridian circles are monotonic from the equator to each pole.  Final results will only involve the equatorial radius $r$.}
\begin{equation}
\label{BLcoordinatesS2}
  \dd{s}^2 = \delta_{ab} e^a e^b \,, \qquad
  e^{1} = f(\theta)\dd{\theta} \,, \qquad
	e^{2} = r\sin(\theta)\dd{\varphi} \,, \qquad
	\omega = \frac{r\cos\theta}{f(\theta)} \dd{\varphi} \,,
\end{equation}
where $\varphi$ is $2\pi$-periodic, $0\leq\theta\leq\pi$, and $f(0)=f(\pi)=r$ to avoid conical singularities at the North ($\theta=0$) and South ($\theta=\pi$) poles.  The full covariant derivative is $D_i=\nabla_i-\I A_i$ in terms of the metric-covariant derivative~$\nabla_i$ and (dynamical and background) gauge fields~$A_i$.  Using the vielbein, $D_1=f(\theta)^{-1}D_\theta$ and $D_2=(r\sin(\theta))^{-1}D_{\varphi}$.

The metric is conformally flat, hence the generators of superconformal transformations are those of flat space.  Among those, supercharges which square to isometries of the sphere generate the Poincar\'e superalgebra $\mathfrak{su}(2|1)$ for the round sphere $f(\theta)=r$, and $\mathfrak{su}(1|1)$ in general.  Explicitly, we will use a supercharge $\supercharge \in \mathfrak{su}(1|1)$ whose square is $\supercharge^2=J+R/2$, where $J$~is the $U(1)$ rotation and $R$~is a $U(1)$ vector R-symmetry.

Thanks to conformal flatness, the action of superconformal transformations (and $\supercharge$ in particular) on vector and chiral multiplets is known.  It is conveniently written in terms of conformal Killing spinors~$\epsilon$, namely solutions of $\nabla_i \epsilon = \gamma_i \tilde{\epsilon}$ for some~$\widetilde{\epsilon}$.  Unfortunately the conformal map between the squashed sphere and the plane is quite unwieldy, thus the resulting conformal Killing spinors are complicated.  Another approach, which works for non-conformally-flat spaces in higher dimensions \cite{BLHama:2011ea, BLHama:2012bg}, is to keep spinors simple by introducing an R-symmetry background gauge field~$V_i$.  For definiteness we choose%
\footnote{Our conventions for spinor components of chiral and vector multiplets and our choice of Killing spinors follow \cite{BLBenini:2012ui} for consistency with the rest of the review. They differ from \cite{BLDoroud:2012xw, BLGomis:2012wy} by factors of $e^{i\pi (1-\gamma^3)/4}$.  Note that $e^{i\pi(1-\gamma^3)/4}\gamma^i e^{-i \pi (1-\gamma^3)/4 }=\varepsilon^{ij} \gamma_j$. Moreover the contraction of spinor indices is $\psi \chi = \psi_\alpha \varepsilon^ {\alpha\beta} \chi_\beta$, namely the symbol $\psi\chi$ stands for $\psi^\sT \smat{0 & 1 \\ -1 & 0} \chi$ in standard matrix notation.}
\begin{equation}
  \epsilon = e^{\I\theta\gamma_1/2} e^{\I\varphi/2} \epsilon_0 \,,\quad
  \bar{\epsilon} = e^{\I\theta\gamma_1/2} e^{-\I\varphi/2} \bar{\epsilon}_0 \,,\quad
	\text{with $\gamma_3 \epsilon_0 = \epsilon_0$ and $\gamma_3 \bar{\epsilon}_0 = - \bar{\epsilon}_0$}
\end{equation}
and the normalization $\bar{\epsilon}\epsilon=\bar{\epsilon}_0\epsilon_0=1$. These spinors span the space of solutions to $\nabla_i\epsilon=\I\gamma_i\epsilon/2r$ on the round sphere.  On squashed spheres they are solutions to the R-covariant conformal Killing spinor equation
\begin{equation}\label{BLcKs}
  D_i \epsilon = (\nabla_i - \I V_i) \epsilon = \frac{\I\gamma_i\epsilon}{2f(\theta)}
	\qquad
  D_i \bar{\epsilon} = (\nabla_i + \I V_i) \bar{\epsilon} = \frac{\I\gamma_i\bar{\epsilon}}{2f(\theta)}
\end{equation}
with a connection $V = \frac{1}{2} \bigl(1 - \frac r{f(\theta)} \bigr)\dd{\varphi}$ smooth everywhere.  Note that supersymmetry transformations of vector and chiral multiplets must likewise be made covariant by including~$V$ in every covariant derivative, with the R-charge of each field as its coefficient.

Let us now write the supersymmetry variations $\delta_{\supercharge}$ of vector and chiral multiplet components under the supercharge~$\supercharge$ built from $\epsilon$, $\bar{\epsilon})$.  We only list the supersymmetry transformations of fermions (we have also shifted the auxiliary field $D$ by $\sigma/r$), and refer to \cite{BLBenini:2012ui, BLDoroud:2012xw, BLDoroud:2013pka, BLClosset:2014pda} for the complete expressions:
\begin{align}\label{BLdeltalambda}
  &
  \begin{aligned}
	  \delta_{\supercharge}\lambda &= (\I V_m^+ \gamma^m - D)\epsilon
	  \\
	  \delta_{\supercharge}\bar{\lambda} &= (\I V_m^- \gamma^m + D) \bar{\epsilon}
  \end{aligned}
	\qquad \text{where} \qquad
	\begin{aligned}
	  V_i^\pm &= \mp D_i\sigma + \varepsilon_{ij} D^j\eta
		\\
		V_3^\pm &= F_{12} \pm \I [\sigma,\eta] - \eta / f(\theta)
  \end{aligned}
	\\\label{BLdeltapsi}
	&
	\begin{aligned}
		\delta_{\supercharge}\psi & = \bigl( \I\gamma^i D_i\phi + \I\sigma\phi + \gamma_3 \eta \phi - \Rcharge\phi / (2f(\theta))\bigr)\epsilon+ \bar{\epsilon} F
		\\
		\delta_{\supercharge}\bar{\psi} & = \bigl( \I\gamma^i D_i\bar{\phi} + \I\bar{\phi}\sigma - \gamma_3 \bar{\phi} \eta - \Rcharge \bar{\phi} / (2f(\theta))\bigr) \bar{\epsilon} + \epsilon\bar{F} \;.
  \end{aligned}
\end{align}
The implicit summations on the first and second line are over $m=1,2,3$.

The most general renormalizable action with $\cN{=}(2,2)$ supersymmetry involving only vector and chiral multiplets takes the form
\begin{equation}\label{BL22action}
  S = S_{\text{v.m.}} + S_{\twsupo} + S_{\text{c.m.}} + S_{\supo} \;.
\end{equation}
The vector multiplet action~$S_{\text{v.m.}}$, the chiral multiplet action~$S_{\text{c.m.}}$ and the superpotential term~$S_{\supo}$ are dimensional reductions of their 4d $\cN{=}1$ counterparts, with corrections of order $1/r$ and $1/r^2$ to preserve supersymmetry on the squashed sphere.

The twisted superpotential term~$S_{\twsupo}$ is analogous to the superpotential term~$S_{\supo}$: a (twisted) superpotential is the top component of a polynomial in (twisted) chiral multiplets.  In theories of vector and chiral multiplets the only twisted chiral multiplet available is the field strength~$\Sigma$ of the vector multiplet, and the most commonly used twisted superpotential in a gauge theory is linear in~$\Sigma$.  The twisted superpotential term~$S_{\twsupo}$ is then---for each $U(1)$ gauge group---the familiar FI D-term and a topological term measuring the gauge field flux~$\flux$ through~$S^2$.  The coefficients $\zeta$ (FI parameter) and $\vartheta$ (theta angle) combine into a complexified FI parameter $z=e^{-2\pi\zeta+\I\vartheta}$.

Finally, one can endow chiral multiplets with twisted masses by coupling the flavor symmetry group to an external (non-dynamical) vector multiplet and giving it a supersymmetric background value. We solve the BPS equations in \eqref{BLbpscoul} and find that the background is parametrized by a real scalar $\twmass$ and a discrete flux $\flavorflux$. The action $S_{\text{c.m.}}$ and the supersymmetry transformations \eqref{BLdeltapsi} of a chiral multiplet then depend on its R-charge $\Rcharge$, its twisted mass $\twmass$ and the flux~$\flavorflux$.

Except for the twisted superpotential term $S_{\twsupo}$, all terms in the action \eqref{BL22action} are $\supercharge$-exact and $\supercharge$-invariant. Explicitly, the corresponding Lagrangian densities are
\begin{equation}\begin{aligned}\label{BLQexact}
	\mathcal{L}_{\text{v.m.}}
	&= \delta_{\supercharge} \delta_{\bar{\epsilon}} \Tr\Bigl(\bar{\lambda}\lambda/2 - 2\sigma D+\sigma^2/f(\theta)\Bigr)
	\\
	\mathcal{L}_{\text{c.m.}}
	&= \delta_{\supercharge} \delta_{\bar{\epsilon}} \Bigl(\bar{\psi}\psi-2\I\bar{\phi}\sigma\phi + (\Rcharge-1) \bar{\phi}\phi / f(\theta) \Bigr)
	\\
	\mathcal{L}_{\supo}
	&= \delta_{\supercharge} \Bigl(\psi_{(\supo)}^{} \epsilon + \bar{\epsilon} \bar{\psi}_{(\bar{\supo})}^{}\Bigr) \;.
\end{aligned}\end{equation}
Therefore, any $\supercharge$-invariant observable is independent of the coefficients in $S_{\text{v.m.}}$, $S_{\text{c.m.}}$ and $S_{\supo}$, and can only depend on parameters in the twisted superpotential (FI parameters, theta angles) and in supersymmetry transformations (R-charges, twisted masses, background fluxes).  In particular, these observables are independent of the gauge couplings $g_{\text{YM}}$ hence are invariant under the RG flow, which makes them very powerful probes of the low-energy limit of GLSMs.


\subsection{BPS equations}
\label{BLssec:sphere-bps}

The localization argument guarantees that only $\supercharge$-invariant field configurations (and quadratic fluctuations nearby) contribute to $\supercharge$-invariant path integrals.
The variations $\delta_{\supercharge}$ of bosons involve fermionic fields hence vanish automatically and we are left with solving $\delta_{\supercharge}\lambda=\delta_{\supercharge}\bar{\lambda}=0$ and $\delta_{\supercharge}\psi=\delta_{\supercharge}\bar{\psi}=0$ for the spinors $\epsilon$, $\bar{\epsilon}$ defining~$\supercharge$.

The vanishing of gluino variations \eqref{BLdeltalambda} implies
\begin{equation}\label{BLiv3pmetc}
  \I V_3^{\pm} \mp V_1^{\pm} \sin\theta = D \cos\theta
	\qquad\text{and}\qquad
  \I V_2^{\pm} \pm V_1^{\pm} \cos\theta = D \sin\theta \,.
\end{equation}
The integration contour is fixed by convergence of the path integral: the bosons $A_i$, $\eta$, $\sigma$, and $D$ in vector multiplets are real, thus $V_m^{\pm}$ are real as well. In one of the localization calculations we will replace $D$ by its complex on-shell value, thus we now keep $D$ general when solving the BPS equations.
Extracting the real and imaginary parts of~\eqref{BLiv3pmetc} yields $V_1^{\pm} = \re D = 0$, $V_2^{\pm} = \sin\theta \, \im D$ and $V_3^{\pm} = \cos\theta \, \im D$.  Therefore, the BPS equations read
\begin{align}
  \label{BLbpsvec1}
  0 & = D_2 \eta = D_1 \sigma = D_2 \sigma = [\eta,\sigma] \\
  \label{BLbpsvec2}
  D_1 \eta & = -\sin\theta\,\im D
	\qquad\text{and}\qquad
	F_{12} - \frac{\eta}{f(\theta)} = \cos\theta\,\im D \;.
\end{align}
Fixing the gauge $A_\theta=0$, equations~\eqref{BLbpsvec2} imply $\partial_\theta(A_{\varphi}+r\eta\cos\theta)=0$.  Solving in either region $0\leq\theta<\pi$ or $0<\theta\leq\pi$ one gets $A=(k-r\cos\theta\,\eta)\dd{\varphi}$ where $k$~is fixed by continuity at the pole to be $k^+=r\eta(0)$ and $k^-=-r\eta(\pi)$, respectively.  The two~$A$ are gauge equivalent away from the poles provided the flux $\flux=\frac{1}{2\pi}\int F=r\eta(0)+r\eta(\pi)=k^+-k^-$ is GNO quantized \cite{BLGoddard:1976qe} namely has integer eigenvalues on any representation of $G$.  The remaining equations imply that the constant~$\sigma$ commutes with all $\eta(\theta,\varphi)$ and that
\begin{equation}\label{BLetarot}
  \partial_{\varphi}\eta=\I[k^{\pm},\eta]
\end{equation}
with a constant $k^\pm$ depending on the gauge.
Periodicity in~$\varphi$ requires~$\eta$ to lie in integer eigenspaces of~$k^{\pm}$ (in the adjoint representation), which coincide due to GNO quantization of $k^+-k^-$.

We now turn to the BPS equations of the chiral multiplet.  Linear combinations of  $\delta_{\supercharge}\psi=0$ and the complex conjugate of $\delta_{\supercharge}\bar{\psi} = 0$ yield $0 = F = \sigma \phi$ and
\begin{equation}\label{BLbpsd1d2}
	0 = \cos\frac{\theta}{2} (D_1+\I D_2) \phi-\sin\frac{\theta}{2}\biggl(\eta+\frac{\Rcharge}{2f(\theta)}\biggr) \phi
	= \sin\frac{\theta}{2} (D_1-\I D_2) \phi - \cos\frac{\theta}{2}\biggl(\eta-\frac{\Rcharge}{2f(\theta)}\biggr) \phi \;.
\end{equation}
Taking into account $A=(k^{\pm}-r\cos\theta\,\eta)\dd{\varphi}$ from above, the equations imply
\begin{equation}\label{BLbpschir}
  0 = F = \sigma \phi
	= \biggl( \sin\theta \partial_\theta - f(\theta)\eta + \frac{\Rcharge}{2}\cos\theta \biggr) \phi
	=	\biggl( \partial_\varphi - \I k^{\pm} + \frac{\I\Rcharge}{2f(\theta)/r} \biggr) \phi \;.
\end{equation}
Periodicity in~$\varphi$ requires $\phi$~to lie in the integer eigenspaces of $k^{\pm}-\I\Rcharge r/2f(\theta)$ for all~$\theta$, but this is only possible if $\Rcharge=0$ (or if the sphere is round).  At the poles, \eqref{BLbpsd1d2} imply additionally to first order in $\theta$ that $\phi$ is (anti)holomorphic at the (South) North pole.

In \autoref{BLssec:sphere-coulomb} we will keep $D$ real and assume that all R-charges are positive (and the sphere is squashed), so that $\phi=0$.  Since $\im D=0$ we now have $\partial_\theta\eta=0$ hence $\eta$ is equal to its value at the poles and is constant.  Altogether,
\begin{equation}\label{BLbpscoul}
  f(\theta) \, F_{12} = \eta = \frac{\flux}{2r} \;, \qquad
	\sigma = \text{constant} \;, \qquad
	0 = [\eta, \sigma] = D = \phi = F \;.
\end{equation}
The path integral localizes to these ``Coulomb branch'' configurations, so named in analogy to the Coulomb branch of the flat space theory.  Since $\eta$ and $\sigma$ commute, a constant gauge transformation reduces them to the Cartan algebra $\ft$ of $G$.  Another outcome of this computation concerns non-dynamical vector multiplets: one can turn on a flux $\flux^{\text{ext}}=\flavorflux$ and a real twisted mass $\sigma^{\text{ext}}=\twmass$ for each chiral multiplet, as announced in \autoref{BLssec:sphere-theory}.

In \autoref{BLssec:sphere-higgs} we will assume that all R-charges vanish and alter the contour of integration of $D$ (equivalently we evaluate its Gaussian path integral) to localize onto complex saddle points of the deformation term chosen there.  Namely,
\begin{equation}
  D = -\I(\phi\bar{\phi}-\chi)
\end{equation}
where we will choose a ``deformation'' FI parameter~$\chi$ for each $U(1)$ gauge factor.  Besides Coulomb branch configurations similar to \eqref{BLbpscoul} with $\phi=0$, there are now Higgs branch (and mixed branch) configurations with $\phi\neq 0$.  Writing a twisted mass $\twmass_I$~for each chiral multiplet $\phi_I$ explicitly, the constraint $(\sigma+\twmass_I)\phi_I=0$ only allows non-zero~$\phi$ at particular points on the Coulomb branch.  For generic twisted masses at most $\rank G$ different chiral multiplets can be non-zero.  Due to \eqref{BLbpschir}, $\phi(\theta,\varphi)\sim (e^{\I\varphi}\sin\theta)^{k^\pm}\phi_0$ near the poles, hence regular non-zero solutions $\phi$ must additionally lie in the non-negative integer eigenspaces of $k^\pm$.  The remaining BPS equations
\begin{equation}
  \partial_\theta \eta = f(\theta) \sin\theta \, (\phi\bar{\phi}-\chi) \;, \qquad\qquad\qquad
	\sin\theta \, \partial_\theta \phi = f(\theta) \, \eta \phi
\end{equation}
have not been analysed in full generality.  For $G=U(N)$ with (anti)fundamental matter and generic twisted masses we will find that all contributions to the localized path integral are suppressed as $\chi\to\infty$ except those in which the group is fully Higgsed.  The condition is that the non-vanishing chiral multiplets span $\mathbb{C}^N$: then $(\sigma+\twmass_I)\phi_I=0$ fixes~$\sigma$, and more importantly all eigenvalues of~$k^{\pm}$ must be non-negative integers.  While in the Coulomb branch localization scheme only the difference $k^+-k^-$ was GNO quantized, in the Higgs branch localization scheme $\chi\to\infty$ both $k^+$ and~$k^-$ are quantized (and non-negative).  We will interpret $k^\pm$~as counting vortices at the North pole and antivortices at the South pole.


\subsection{Coulomb branch localization}
\label{BLssec:sphere-coulomb}

In this section we assume for simplicity that all R-charges are in the range $0<\Rcharge<2$. In all models of interest this condition can be made to hold by mixing the R-charge with $U(1)$ gauge charges if needed.  Other values for the R-charges can be reached by analytic continuation.

Recall the localization argument: we add a $\supercharge$-invariant deformation term $t\delta_{\supercharge} \cV$ to the action and take $t\to\infty$ thus making the saddle-point approximation exact.  Any $\supercharge$-invariant observable then reduces to an integral over saddle points of its classical value at these saddles, with a measure given by a Gaussian integral (one-loop determinant) of quadratic fluctuations around the saddles. Additionally, saddle points that are not $\supercharge$-invariant cannot contribute since the Grassmann integral of a constant vanishes.  This second argument would not be necessary if we used the canonical deformation term $\delta_{\supercharge} \bigl(\lambda \, \overline{\delta_{\supercharge}\lambda} + \psi \, \overline{\delta_{\supercharge}\psi}\bigr)$ since the saddle points of its bosonic part are precisely $\supercharge$-invariant configurations.  However, we use the $\supercharge$-closed and $\supercharge$-exact deformation term $S_{\text{v.m.}} + S_{\text{c.m.}}$ in \eqref{BLQexact}.  It is straightforward to check that all $\supercharge$-invariant configurations~\eqref{BLbpscoul} are saddle points.

The (squashed) sphere partition function of a 2d $\cN{=}(2,2)$ GLSM with gauge group~$G$ and chiral multiplets in the representation $\repr=\bigoplus_I \repr_I$ (each with an R-charge $\Rcharge_I$, a background flux $\flavorflux_I$, and a twisted mass~$\twmass_I$) is then a sum over GNO-quantized fluxes ($\flux$ has integer eigenvalues on any representation of $G$) and an integral over the Cartan algebra $\lie{t}$ of $G$ of classical and one-loop factors
\begin{equation}
  \label{BLZCoulomb}
  Z_{S^2}
  = \frac{r^{c/3}}{\abs{\Weyl(G)}} \sum_{\flux} \int_{\lie{t}} \frac{\dd{(r\sigma)}}{(2\pi)^{\rank G}} \,
	Z_{\text{cl}}(z,\bar{z};r\sigma,\flux) \,
	Z_{\text{gauge}}(r\sigma,\flux) \,
	Z_{\text{matter}}(\Rcharge,r\sigma+r\twmass,\flux+\flavorflux) \;.
\end{equation}
The order $\abs{\Weyl(G)}$ of the Weyl group appears due to residual discrete gauge redundancy in~$\lie{t}$.
The result only depends on the squashed sphere through its equatorial radius~$r$.
We explain in \autoref{BLssec:sphere-det} how to compute vector multiplet and chiral multiplet one-loop determinants: they are products over positive roots $\alpha$ of $G$ and over weights $\rho$ of each representation $\repr_I$,
\begin{equation}
	\label{BLZ1l}
  \begin{aligned}
    Z_{\text{gauge}} & = \prod_{\alpha>0} \; (-1)^{\alpha(\flux)} \biggl[ r^2 \alpha(\sigma)^2 +\frac{ \alpha(\flux)^2}{4}\biggr] 
		\\
		Z_{\text{matter}} & =
    \prod_{I,\rho} \;
	  \frac{\Gamma\bigl(\frac{\Rcharge_I}{2}-\I r\twmass_I-\frac{\flavorflux_I}{2}-\I r \rho(\sigma) -\frac{\rho(\flux)}{2} \bigr)}
	  {\Gamma\bigl(1-\frac{\Rcharge_I}{2}+\I r\twmass_I-\frac{\flavorflux_I}{2}+\I r \rho(\sigma) - \frac{\rho(\flux)}{2}\bigr)} \;.
	\end{aligned}
\end{equation}
For the common case of a linear twisted superpotential~$\twsupo$ with an FI parameter~$\zeta_{\ell}$ and a theta term~$\vartheta_{\ell}$ for each $U(1)$ gauge factor, the classical contribution is
\begin{equation}\label{BLZcl}
  Z_{\text{cl}}
  = \prod_{\ell} z_{\ell}^{\Tr_{\ell}(\I r\sigma+\frac{\flux}{2})} \, \bar{z}_{\ell}^{\Tr_{\ell}(\I r\sigma-\frac{\flux}{2})}
\end{equation}
where $z_{\ell}=e^{-2\pi\zeta_{\ell}+\I\vartheta_{\ell}}$ and we denote $\Tr_{\ell}$ the projection onto the $\ell$-th $U(1)$ factor: for $G=\prod_\ell U(N_\ell)$ these really are traces.  To be more precise, \eqref{BLZcl} involves renormalized FI parameters at the scale~$1/r$,
\begin{equation}\label{BLzrenorm}
  z_{\ell} = (rM_{\text{UV}})^{\sum_I Q_I^\ell} \, z_{\ell}^{\text{UV}} \;,
\end{equation}
where $z^{\text{UV}}$ are bare parameters at some UV scale~$M_{\text{UV}}$ and $Q_I^\ell$ are charges of chiral multiplets under the $\ell$-th $U(1)$ factor.  We obtain this dependence on~$r$ from zeta function regularization when computing one-loop determinant in \autoref{BLssec:sphere-det}, and also obtain the overall power $r^{c/3}$,
\begin{equation}\label{BLcover3}
  \frac{c}{3} = \sum_I (1-\Rcharge_I)\dim \repr_I - \dim G \;.
\end{equation}
For theories that flow to a superconformal field theory (SCFT), $c$ is the central charge. The equatorial radius~$r$ is also used as a scale for twisted masses and~$\sigma$.

Several comments are in order.
The partition function only depends on parameters in the twisted superpotential (here $z$, $\bar z$), on R-charges, twisted masses and background fluxes.  This is expected since the coupling constants $g_{\text{YM}}$ and superpotential couplings multiply $\supercharge$-exact terms.  In particular any superpotential simply specializes $Z_{S^2}$ by fixing some linear combinations of the R-charges to $2$.  The partition function depends holomorphically on the combinations $\frac{\Rcharge_I}{2}-\I r\twmass_I$ hence it can be extended to R-charges beyond $0<\Rcharge<2$.

Another extension is to include $\supercharge$-invariant operators in the path integral, such as $\Tr_\ell(\I\sigma+\eta)$ at the North pole or $\Tr_\ell(\I\sigma-\eta)$ at the South pole: this is achieved by including their on-shell values $\Tr_\ell\bigl(\I\sigma\pm\frac{\flux}{2r}\bigr)$ in the integrand \eqref{BLZCoulomb}, and will be further discussed in \autoref{BLssec:operators}. These insertions can be realized by taking derivatives with respect to $\log z_\ell$ and $\log \bar{z}_\ell$, respectively. This is a manifestation of the fact that the integrand in~\eqref{BLZCoulomb} factorizes as a function of $z$ and $\I\sigma+\frac{\flux}{2r}$ times a function of $\bar{z}$ and $\I\sigma-\frac{\flux}{2r}$.  This factorization will play an important role later.

\begin{table}
  \centering
	\begin{tabular}[t]{l}\toprule
	Groups with $(-1)^{2\delta(\flux)}=1$\\\midrule
	$SU(MN)/\mathbb{Z}_M$ with $M$ odd or $N$ even\\
	$SO(2N)$\\
	$SO(8N)/\mathbb{Z}_2$, $SO(8N+2)/\mathbb{Z}_2$\\
	$Sp(4N)/\mathbb{Z}_2$, $Sp(4N+3)/\mathbb{Z}_2$, $E_6/\mathbb{Z}_3$\\
	Simply-conneted groups: $SU(N)$, \\
	\hfill $Spin(N)$, $Sp(N)$, $E_6$, $E_7$, $E_8$, $F_4$, $G_2$\\
	\bottomrule
	\end{tabular}
	\begin{tabular}[t]{l}\toprule
	Groups with $(-1)^{2\delta(\flux)}$ a discrete theta angle\\\midrule
	$SU(MN)/\mathbb{Z}_M$ with $M$ even and $N$ odd\\
	$SO(2N+1)$\\
	$SO(8N+4)/\mathbb{Z}_2$, $SO(8N+6)/\mathbb{Z}_2$\\
	$Sp(4N+1)/\mathbb{Z}_2$, $Sp(4N+2)/\mathbb{Z}_2$, $E_7/\mathbb{Z}_2$\\
	Quotients $Spin(4N)/\mathbb{Z}_2$ other than $SO(4N)$\\
	\bottomrule
	\end{tabular}
	\caption{\label{BLtab:sign}Effect of the vector multiplet one-loop determinant sign $(-1)^{2\delta(\flux)}$ for connected compact simple groups.  It is trivial when the Weyl vector $\delta$ is a weight, and otherwise corresponds to a $\bZ_2$ discrete theta angle.  For $U(N)$ the sign shifts the (continuous) theta angle by~$\pi$ if $N$~is even.}
\end{table}

Note that the sign
\begin{equation}
  \prod_{\alpha>0} (-1)^{\alpha(\flux)} = e^{2\pi\I \delta (\flux)} \;,
\end{equation}
which we will derive later, was originally missed. It was correctly predicted in \cite{BLHori:2013ika, BLHori:2013gga, BLHori:2013ewa}.
For many groups this sign is $+1$ because the Weyl vector $\delta$ (half sum of positive roots) is a weight of $G$.  An important exception is $U(N)$ with $N$ even: then the sign is equivalent to a shift of the theta angle by $\pi$. \autoref{BLtab:sign} gives a list for simple groups.

We expect the one-loop determinants of two chiral multiplets $X$ and $Y$ with opposite gauge and flavor charges and with R-charges $\Rcharge$ and $2-\Rcharge$ to cancel.  Indeed, such chiral multiplets can be integrated out by including a superpotential mass term $\supo=\mu XY$ with $\mu\to\infty$, and $\supo$ does not affect the partition function.  Omitting external sources here for brevity,
\begin{equation}
  \prod_\rho \,
	\frac{\Gamma\bigl(\frac{\Rcharge}{2}-\I r \rho(\sigma) -\frac{\rho(\flux)}{2}\bigr)}
	{\Gamma\bigl(1-\frac{\Rcharge}{2}+\I r \rho(\sigma) -\frac{\rho(\flux)}{2}\bigr)} \,
	\frac{\Gamma\bigl(1-\frac{\Rcharge}{2}+\I r \rho(\sigma) + \frac{\rho(\flux)}{2}\bigr)}
	{\Gamma\bigl(\frac{\Rcharge}{2}-\I r \rho(\sigma) + \frac{\rho(\flux)}{2}\bigr)}
	= \prod_\rho (-1)^{\rho(\flux)} \;.
\end{equation}
We have used $\Gamma(x+\flux/2)/\Gamma(1-x+\flux/2)=(-1)^{\flux} \Gamma(x-\flux/2)/\Gamma(1-x-\flux/2)$ which is a consequence of Euler's identity $\Gamma(y)\Gamma(1-y)=\pi/\sin\pi y$. Since weights sum to zero for simple factors of $G$, the sign simply shifts theta angles of some $U(1)$ gauge factors by $\pi$.

The one-loop determinant of a vector multiplet can be recast as that of a collections of chiral multiplets of R-charge $\Rcharge=2$ and gauge charges equal to the roots $\alpha$ of $G$. This can be understood in terms of the Higgs mechanism. If the theory had an extra chiral multiplet of R-charge 0 in the adjoint representation, we could give a VEV to its diagonal components without breaking the R-symmetry. The VEV would break the gauge group to its maximal torus $U(1)^{\rank G}$, and give a mass both to vector and chiral multiplet components along the roots $\alpha$. Taking into account the observation above, we have schematically the relations: $Z_\text{gauge} = 1/\prod_\alpha Z_\text{chiral}^{(\alpha),\, q=0} = \prod_\alpha Z_\text{chiral}^{(\alpha),\, q=2}$.
This fact implies that $Z_{S^2}$ of a non-Abelian GLSM is the specialization of $Z_{S^2}$ of an associated ``Cartan theory'' which has gauge group $U(1)^{\rank G}$, has $\rank G$ parameters $z_{\ell}$, and has one chiral multiplet for each weight $w$ of $\repr$ and one for each root $\alpha$ of $G$. The original partition function is retrieved by setting $z_{\ell}=1$ for each FI parameter that does not correspond to a $U(1)$ factor of the original theory.

\subsection{The A-system}
\label{BLssec:sphere-pde}

Remarkably, the partition function \eqref{BLZCoulomb} obeys a system of differential equations
that are holomorphic in the $z_{\ell}$ and a similar system for $\bar{z}_{\ell}$.
We first review the results of \cite{BLHalverson:2013eua},
which apply to abelian GLSMs and to non-abelian GLSMs through their associated Cartan theory.
Set $r=1$ for brevity.
Consider a GLSM with abelian gauge group $G=U(1)^{N}$ and chiral multiplets of charges~$Q^{\ell}_I$ under the $\ell$-th gauge group factor.  Its partition function is
\begin{equation}\label{BLZS2abelian}
  Z_{S^2}
  =
	\sum_{\flux\in\mathbb{Z}^N} \int_{\mathbb{R}^N} \frac{\dd[^N]{\sigma}}{(2\pi)^N}
  \prod_{\ell=1}^N \biggl[ z_\ell^{\I\sigma_\ell+\frac{\flux_\ell}{2}} \bar{z}_\ell^{\I\sigma_\ell-\frac{\flux_\ell}{2}} \biggr]
  \prod_{I} \biggl[
	\frac{\Gamma\bigl(\frac{\Rcharge_I}{2}-\I\twmass_I-\frac{\flavorflux_I}{2}-Q_I^{\ell} (\I\sigma_{\ell}+\frac{\flux_{\ell}}{2})\bigr)}
	{\Gamma\bigl(1-\frac{\Rcharge_I}{2}+\I\twmass_I-\frac{\flavorflux_I}{2}+Q_I^{\ell} (\I\sigma_{\ell}-\frac{\flux_{\ell}}{2})\bigr)} \biggr]
\end{equation}
with an implicit summation over~$\ell$.  If we shift the summation on $\flux$ by any $\fluxshift\in\mathbb{Z}^N$ and the contour for each $\sigma_{\ell}$ by $-\I\fluxshift_\ell/2$ (the contour encounters no pole), then the classical action is multiplied by $z_{\ell}^{\fluxshift_{\ell}}$ and the arguments of gamma functions are shifted by $-Q_I^\ell \fluxshift_\ell$ in the numerator. Extracting these shifts from the gamma function arguments yields some factors linear in $\I\sigma_{\ell}+\flux_{\ell}/2$ which can be reproduced by acting on~$Z_{S^2}$ with the holomorphic differential operators $z_{\ell}\partial/\partial z_{\ell}$.  We find that for any $\fluxshift\in\mathbb{Z}^N$,
\begin{equation}\label{BLGKZ}
  \begin{aligned}
  & \prod_{I,\ Q_I^\ell \fluxshift_\ell>0}
	\biggl( \frac{\Rcharge_I}{2}-\I\twmass_I-\frac{\flavorflux_I}{2}-Q_I^{\ell} \frac{\partial}{\partial \log z_{\ell}}\biggr)_{Q_I^\ell \fluxshift_\ell}
	\, Z_{S^2}
	\\
	& =
	\Biggl( \prod_{\ell=1}^N z_{\ell}^{\fluxshift_\ell} \Biggr)
  \prod_{I,\ Q_I^\ell \fluxshift_\ell<0}
	\biggl( \frac{\Rcharge_I}{2}-\I\twmass_I-\frac{\flavorflux_I}{2}-Q_I^{\ell} \frac{\partial}{\partial \log z_{\ell}}\biggr)_{-Q_I^\ell \fluxshift_\ell}
	\, Z_{S^2}
	\end{aligned}
\end{equation}
in terms of Pochhammer symbols $(x)_n = \prod_{i=0}^{n-1} (x+i) = \Gamma(x+n)/\Gamma(x)$.  The same system with $z_{\ell}\to(-1)^{\sum_I Q_I^{\ell}}\bar{z}_{\ell}$ and $\flavorflux\to-\flavorflux$ holds.

This ``A-system'' of equations---a slight generalization of the GKZ (Gel'fand, Kapranov, Zelevinski) A-hypergeometric systems---is highly redundant: the equation for $\fluxshift+\fluxshift'$ is a consequence of those for $\fluxshift$ and~$\fluxshift'$, at least if $Q_I^\ell \fluxshift_\ell$ and $Q_I^\ell \fluxshift'_\ell$ have the same sign for all $I$.  The space of holomorphic solutions to \eqref{BLGKZ} is typically finite-dimensional, so $Z_{S^2}$~is a linear combination of holomorphic times antiholomorphic solutions:
\begin{equation}\label{BLZCFFsplit}
  Z_{S^2} = \sum_k C_k \, \mathcal{F}_k(z) \, \widetilde{\mathcal{F}}_k(\bar{z}) \;.
\end{equation}
For GLSMs that flow in the infrared to NLSMs on Calabi-Yau manifolds, the differential equations are the well-known Picard-Fuchs differential equations and the functions $\mathcal{F}_k$ are periods of the mirror Calabi-Yau.

For non-Abelian examples it turns out that many solutions to the associated Cartan theory's A-system are absent from explicit factorizations of $Z_{S^2}$ into \eqref{BLZCFFsplit}. This suggests the existence of more stringent differential equations whose set of solutions would capture exactly the (anti)holomorphic dependence of~$Z_{S^2}$.
Let us focus for concreteness on SQCD, namely $G=U(N)$ with $N_f$~fundamental and $N_f$~antifundamental chiral multiplets.  We replace the single FI parameter~$z$ by $z_{\ell}$, $1\leq\ell\leq N$ as in the associated Cartan theory, but we write the vector multiplet one-loop determinant as a differential operator rather than as a chiral multiplet determinant.  Concretely:
\begin{equation}\label{BLZSQCDasSQED}
  Z_{S^2}^{\text{SQCD}}(z,\bar{z})
  = \frac{1}{N!}
	\Biggl[
	\prod_{k<j}^{N} \Bigl(\I z_k\frac{\partial}{\partial z_k}-\I z_j\frac{\partial}{\partial z_j}\Bigr)
  \Bigl(\I \bar{z}_k\frac{\partial}{\partial \bar{z}_k}-\I \bar{z}_j\frac{\partial}{\partial \bar{z}_j}\Bigr)
	\prod_{\ell=1}^{N} Z_{S^2}^{N=1}(z_{\ell},\bar{z}_{\ell})
	\Biggr]_{\substack{z_{\ell}=(-1)^{N-1}z\\\bar{z}_{\ell}=(-1)^{N-1}\bar{z}}} \;.
\end{equation}
The sign $(-1)^{N-1}$ comes from $(-1)^{\alpha(\flux)}$ in the vector multiplet one-loop determinant (see \autoref{BLtab:sign}).
The partition function~$Z_{S^2}^{N=1}$ of SQED (an Abelian theory) obeys the A-system, which reduces in this case to a single equation ($\fluxshift=1$):
\begin{equation}\label{BLSQEDdiff}
  \Biggl[
  \prod_{I=1}^{N_f}
	\biggl( \frac{\Rcharge_I}{2}-\I\twmass_I-\frac{\flavorflux_I}{2}-\frac{\partial}{\partial \log z_{\ell}}\biggr)
	- z_{\ell}
	\prod_{I=1}^{N_f}
	\biggl( \frac{\Rcharge_I}{2}-\I\twmass_I-\frac{\flavorflux_I}{2}+\frac{\partial}{\partial \log z_{\ell}}\biggr)
	\Biggr] Z_{S^2}^{N=1} = 0 \;.
\end{equation}
The space of holomorphic solutions to this ${N_f}^\text{th}$ order differential equation is spanned by $N_f$ functions $\mathcal{F}_I(z)$ for $I=1,\ldots,N_f$.
The antiholomorphic counterpart $z\to\bar{z}$ also holds, so in an appropriate basis
$Z_{S^2}^{N=1}(z,\bar{z}) = \sum_I C_I \mathcal{F}_I(z) \mathcal{F}_I(\bar{z})$.
Expand each SQED partition function $Z_{S^2}^{N=1}$ in~\eqref{BLZSQCDasSQED} as such a sum, so as to get $(N_f)^N$ factorized terms in total.  Note that the holomorphic differential operator in~\eqref{BLZSQCDasSQED} is antisymmetric in the~$z_k$, hence only $\binom{N_f}{N}$~terms remain.\footnote{\label{BLfoot:susybreaks}For $N_f<N$, localization gives a vanishing result for~$Z_{S^2}$: this is due to supersymmetry breaking.}

We now show that these $\binom{N_f}{N}$ terms are solutions of an ordinary differential equation of order~$\binom{N_f}{N}$ in $z\partial/\partial z$, rather than a system of differential equations in $z_{\ell}\partial/\partial z_{\ell}$. By expanding the differential operator in~\eqref{BLZSQCDasSQED} into a sum of monomials which are products of derivatives acting on individual factors $Z_{S^2}^{N=1}(z_j,\bar{z}_j)$, then setting $z_j=(-1)^{N-1} z$ and $\bar{z}_j=(-1)^{N-1}\bar{z}$ as indicated in~\eqref{BLZSQCDasSQED}, one writes the SQCD partition function as~\eqref{BLFpzzbar} below, for $p_j=j$.  Consider more generally
\begin{equation}\label{BLFpzzbar}
  F_{p}(z,\bar{z})
	=
	\varepsilon_{\ell_1\cdots\ell_N}
	\prod_{j=1}^{N}\Biggl[
	\biggl(\I z\frac{\partial}{\partial z}\biggr)^{p_j-1}
	\biggl(\I\bar{z}\frac{\partial}{\partial\bar{z}}\biggr)^{\ell_j-1}
	Z_{S^2}^{N=1}(z,\bar{z})
  \Biggr]
\end{equation}
for integers $p_j\geq 1$.  Reordering the~$p_j$ only affects signs, and $F_p$ vanishes if any $p_i=p_j$.
The $\I z\partial/\partial z$ derivative of~$F_{p}$ is a sum of $N$~such functions (each with one $p_j\to p_j+1$).
On the other hand, the GKZ equation~\eqref{BLSQEDdiff} for $Z_{S^2}^{N=1}(z,\bar{z})$ expresses its ${N_f}^\text{th}$ derivative as a linear combination of its lower derivatives.  Thus $F_{p}$ with any~$p_j>N_f$ can be written as a sum of terms with lower~$p$.
All derivatives of $Z_{S^2}^{\text{SQCD}}=F_{1,2,\ldots,N}$ are hence linear combinations (with known holomorphic coefficients) of the $\binom{N_f}{N}$ functions $F_{p}$ for which all $1\leq p_j\leq N_f$.  This establishes the existence of a holomorphic differential equation of order~$\binom{N_f}{N}$ obeyed by the SQCD partition function (coefficients can be made polynomial in~$z$).

This proof extends to quiver gauge theories, and gives bounds on the number of terms needed in the factorization of $Z_{S^2}$ that are stronger than those deduced from the GKZ system of the associated Cartan theory. On the other hand the proof is not constructive; no closed form expression for the differential equation is known at present.

\subsection{Higgs branch localization}
\label{BLssec:sphere-higgs}

As we have just seen, the sphere partition function of a GLSM can be factorized into holomorphic times antiholomorphic functions of its complexified FI parameters.  We now interpret the factors physically as being due to vortices at the poles, by localizing the path integral directly into this form.  After this ``Higgs branch'' localization was found for 2d $\cN{=}(2,2)$ theories \cite{BLBenini:2012ui, BLDoroud:2012xw}, it was used to explain a similar factorization in 3d $\cN{=}2$ theories \cite{BLFujitsuka:2013fga, BLBenini:2013yva} and 4d $\cN=1$ theories \cite{BLPeelaers:2014ima, BLYoshida:2014qwa}.

Because the Coulomb branch result \eqref{BLZCoulomb} is analytic in $\frac{\Rcharge_I}{2}-\I r \twmass_I$, we can work with $\Rcharge_I=0$.  We also ignore fluxes for external vector multiplets for simplicity, and set $r=1$. Of course, we assume that the theory has $U(1)$ gauge factors as otherwise the factorization property is vacuously true.

We localize using in addition to $\mathcal{L}_{\text{v.m.}}+\mathcal{L}_{\text{c.m.}}$ the deformation term \cite{BLBenini:2012ui}
\begin{equation}
  \mathcal{L}_{\text{Higgs}} = \delta_{\supercharge}\Tr\Bigl[ -\I\bigl(\phi\bar{\phi}-\chi\bigr)\delta_{\supercharge}\sigma\Bigr] \;.
\end{equation}
For each $U(1)$ gauge factor, it includes a parameter~$\chi$ which will play the role of an FI parameter.  The trace denotes the natural pairing between $\bar{\phi}$ and $(\delta_{\supercharge}\sigma)\phi$ on the one hand, and the projection onto each $U(1)$ gauge factor with coefficients~$\chi$ on the other hand.  The bosonic part of this deformation term includes $\I D(\phi\bar{\phi}-\chi)$, and leads to the on-shell value $D=-\I(\phi\bar{\phi}-\chi)$ up to unimportant coefficients.  After integrating out~$D$, saddle points of the deformation term are exactly the $\supercharge$-invariant configurations analyzed in \autoref{BLssec:sphere-bps}:
\begin{align}
\eta(\theta,\varphi) &= e^{\I\varphi[k^{\pm},\,\cdot\, ]} \eta(\theta,0) \;,\qquad&
\phi(\theta,\varphi) &= e^{\I\varphi k^{\pm}}\phi(\theta,0) \;, \\
\label{BLdthetaeta}
\partial_\theta \eta &= f(\theta)\sin\theta \, (\phi\bar{\phi}-\chi) \;,\qquad&
\sin\theta \, \partial_\theta \phi &= f(\theta) \, \eta \phi \;,
\end{align}
where $k^+= \eta(0)$ is used in the simply-connected region $0\leq\theta<\pi$ while $k^-=-\eta(\pi)$ in the region $0<\theta\leq\pi$.  The remaining fields are a constant~$\sigma$ which commutes with all $\eta(\theta,\varphi)$ and such that $(\sigma+\twmass_I)\phi_I=0$ for all flavors $I$, and $A=(k^\pm - \cos\theta\,\eta)\dd{\varphi}$ whose flux $\flux=k^+-k^-$ is GNO quantized.  We have also seen that any non-zero~$\phi$ lies among non-negative integer eigenspaces of $k^{\pm}$.  Since $(\sigma,k^\pm)$ commute pairwise, a constant gauge transformation diagonalizes them.

The partition function localizes to an integral over all solutions to these equations, and one should compute one-loop determinants.  One technique described in \autoref{BLssec:sphere-det} to compute one-loop determinants involves localization to fixed points of~$\supercharge^2$, namely the poles.  One-loop determinants are then a product of contributions from each pole that only depend on values of the $\supercharge$-invariant field configuration at those points.  The one-loop determinants \eqref{BLZ1l} computed in the Coulomb branch factorize as functions of $\I\sigma\pm\frac{\flux}{2}$, which we associated to the North and South poles by considering correlators of the $\supercharge$-invariant operators $\I\sigma\pm\eta$ in \autoref{BLssec:sphere-coulomb}.  We deduce that the one-loop determinant is, more generally,
\begin{equation}\label{BLZ1lhig}
  Z_{\text{1-loop}}
  = \prod_{\alpha>0} (-1)^{\alpha(\flux)} \, \alpha\bigl(\I \sigma+\eta(0)\bigr) \, \alpha\bigl(\I \sigma-\eta(\pi)\bigr) \;
  \prod_{I,w}
	\frac{\Gamma\bigl(-\I \twmass_I-w(\I \sigma+\eta(0))\bigr)}
	{\Gamma\bigl(1+\I \twmass_I+w(\I \sigma-\eta(\pi))\bigr)} \;.
\end{equation}
Both $k^+ \mp k^- = \eta(0)\pm \eta(\pi)$ are integrals over the squashed sphere.  One is the flux:
\begin{equation}
  \eta(0) + \eta(\pi)
	= \frac{1}{2\pi} \int_0^\pi \dd{\theta} \int_0^{2\pi} \dd{\varphi} \partial_\theta(-\eta\cos\theta)
	= \frac{1}{2\pi} \int F
  = \flux
\end{equation}
and fluxes through each hemisphere are $k^+ = \eta(0)$ and $-k^- = \eta(\pi)$.
The other is the integral of the $D$-term equation with the volume form:
\begin{equation}\label{BLintegratedDterm}
  \eta(0) - \eta(\pi)
  = \int_0^{\pi} \dd{\theta} \, (- \partial_\theta \eta)
  = \frac{1}{2\pi} \int \bigl(\chi-\phi\bar{\phi}\bigr) \, \dvol_2 \equiv \Delta \;.
\end{equation}
For a fixed flux $\flux$, each ratio of Gamma functions in \eqref{BLZ1lhig} has the asymptotics
\begin{equation}\label{BLhiggsdetasymp}
  \abs*{\frac{\Gamma\bigl(-\I \twmass_I - w(\I \sigma+\flux/2+\Delta/2)\bigr)}
	  {\Gamma\bigl(1+\I \twmass_I+ w(\I \sigma-\flux/2+\Delta/2)\bigr)}}
	\stackrel{w(\Delta) \to\pm\infty}{=}
	e^{- \left( \rule{0pt}{.6em} 2w(\Delta)+1 \right) \left(  \rule{0pt}{.6em} \log\abs{w(\Delta)}-1 \right)+O(1)} \;.
\end{equation}
Taking the product over weights~$w$, we expect the one-loop determinant to be suppressed for large~$\Delta$.  We shall prove for a class of theories that in the appropriate limit $\chi\to\pm\infty$ the one-loop determinant indeed is suppressed for all saddle points except those for which the $D$-term $\phi\bar{\phi}-\chi$ is small throughout the sphere.  The path integral localizes in this limit to solutions to $\phi\bar{\phi}=\chi$ with vortices at the poles.

Let us focus for concreteness on $U(N)$ SQCD with $N_f$ fundamental and $N_a\leq N_f$ antifundamental chiral multiplets of generic twisted masses $\twmass_I$ and $\anti{\twmass}_I$ respectively.  The case $N_a\geq N_f$ is obtained by charge conjugation.  We assume $N_f\geq N$ to avoid supersymmetry breaking.  The results extend to $\prod_\ell U(N_\ell)$ quiver gauge theories by considering gauge groups one by one, and perhaps to more general matter contents.

Consider a smooth $\supercharge$-invariant configuration, and diagonalize $\sigma$ by a constant gauge transformation.  At most $N$ chiral multiplets are non-zero: in each eigenspace of $\sigma$, at most one chiral multiplet is non-zero because of $(\sigma+\twmass_I)\phi_I=0$ for fundamentals and $\anti{\phi}_I (-\sigma+\anti{\twmass}_I)=0$ for antifundamentals.  Focus first on a non-zero fundamental $\phi_I\neq 0$.  The trace of \eqref{BLdthetaeta} in the eigenspace $\sigma=-\twmass_I$ implies (with no summation on $I$ in this paragraph)
\begin{equation}
  \sin\theta \, \partial_\theta (\bar{\phi}_I \phi_I) = 2f(\theta)\bar{\phi}_I\eta\phi_I \,,\qquad
  \partial_\theta (\bar{\phi}_I \eta \phi_I) = f(\theta) \sin\theta \, (\bar{\phi}_I \phi_I - \chi) \bar{\phi}_I \phi_I \;.
\end{equation}
We prove by contradiction that $\bar{\phi}_I\phi_I\leq\chi$ for any such non-zero solution (in particular $\chi>0$). Consider the infimum $\theta_0$ of angles at which $\bar{\phi}_I\phi_I\geq\chi$ (if it exists).  The derivative of $\bar{\phi}_I\phi_I$ must be non-negative, thus $\bar{\phi}_I\eta\phi_I\geq 0$.  These two inequalities continue to hold for $\theta\in[\theta_0,\pi]$ since derivatives are non-negative.  However, we know that $\phi_I$ is among non-positive eigenspaces of $\eta(\pi)=-k^-$, thus the second inequality must be saturated everywhere, hence $\eta\phi_I=0$ and we find constant $\phi_I$ solutions with $\abs{\phi_I}^2=\chi$.  Similarly, solutions where an antifundamental $\anti{\phi}_I\neq 0$ is non-zero obey $\abs{\anti{\phi}_I}^2\leq-\chi$ hence require $\chi<0$.  Collecting these results, we deduce in particular that the $D$-term $\chi\unit_N-\phi_I\bar{\phi}_I+\bar{\anti{\phi}}_I\anti{\phi}_I$ is positive semidefinite for $\chi>0$ and negative semidefinite for $\chi<0$.  Furthermore, its trace is at least $\abs{\chi}$ (in absolute value) unless $N$ chiral multiplets are non-zero, fully Higgsing the gauge group (fixing $\sigma$).

The (diagonal) matrix $\Delta$ is the integral \eqref{BLintegratedDterm} of this semidefinite matrix.  One consequence is that $\Tr\Delta\to\pm\infty$ (with the same sign as $\chi$) in the limits $\chi\to\pm\infty$, except for saddle points for which $N$ chiral multiplets are non-zero and the $D$-term equation is approximately obeyed throughout the sphere.  Another consequence is that all eigenvalues of $\Delta$ have the same sign as $\Tr\Delta$; then taking the product of \eqref{BLhiggsdetasymp} over all weights we find that the full one-loop determinant is suppressed as $\Tr\Delta\to+\infty$ (for $N_a>N_f$ instead it would be suppressed as $\Tr\Delta\to-\infty$ while for $N_a=N_f$ it is suppressed in both limits).  Therefore, the only saddle points that contribute as $\chi\to+\infty$ are those with $N$ non-zero fundamental chiral multiplets.  The requirement that $\phi$ belongs to non-negative integer eigenspaces of $k^\pm$ at poles then forces all eigenvalues of $k^\pm$ to be non-negative integers.

At the (South) North pole $\phi_I$ obeys BPS (anti)vortex equations.  The eigenvalues of $k^\pm$ control the asymptotics $\phi_I\sim (e^{\I\varphi} \sin\theta)^{k^{\pm}}\phi_I^{\circ}$ at the poles thus counting (anti)vortices there.  Since the integral of $D$ is $k^+ + k^-$, the distance over which $\abs{\phi_I}^2$ goes from $0$ at the poles to $\chi$ must scale like $k^\pm/\sqrt{\chi}$: vortices become point-like as $\chi\to\infty$.

Altogether, the path integral is a sum over $\binom{N_f}{N}$ Higgs branches $H$, namely choices of $N$ flavors with $\phi_I\neq 0$, and over vorticities $k^\pm_j$ ($j=1,\ldots,N$).  The classical action for each such configuration is evaluated using that the gauge flux and integrated $D$-term are $k^+ \mp k^-$, and the result factorizes: contributions from (anti)vortices at the (South) North pole depend (anti)holomorphically on $z$.  After some massaging,
\begin{gather}\label{BLZHiggs-sqcd}
  Z_{S^2}^{\text{SQCD}}
	=
	\sum_{\substack{H\subset\{1,\ldots,N_f\}\\\#H=N}} \Biggl[
  \prod_{J\in H} \frac{(z\bar{z})^{-\I \twmass_J}\prod_{I\not\in H} \gamma(-\I \twmass_I + \I \twmass_J)}{\prod_{I=1}^{N_a} \gamma(1+\I \anti{\twmass}_I+\I \twmass_J)} \, f_{H}(z) \, f_{H}\bigl((-1)^{N_f-N_a}\bar{z}\bigr) \Biggr]
	\\
  f_{\{J_1,\ldots,J_N\}}(x) = \sum_{k_1,\ldots,k_N\geq 0} \; 
	\prod_{j=1}^{N}
  \frac{ x^{k_j} \prod_{I=1}^{N_a} (-\I \anti{\twmass}_I-\I \twmass_{J_j})_{k_j} }
	{ \prod_{i=1}^{N} (\I \twmass_{J_i}-\I \twmass_{J_j}-k_i)_{k_j} \prod_{I\not\in\{J_i\}} (-\I \twmass_I+\I \twmass_{J_j}-k_j)_{k_j}}
\end{gather}
where $\gamma(y)=\Gamma(y)/\Gamma(1-y)$ and $z$ is the renormalized value \eqref{BLzrenorm}.  Up to a relabeling of parameters, the function $f_H(z)$ coincides with previously known vortex partition functions computed in the Omega background.  This is unsurprising as point-like vortices are unaffected by the precise IR regulator (sphere or Omega background).  The sign difference $(-1)^{N_f-N_a}$ between vortices and antivortices is explained by noting that $\supercharge^2$ rotates counterclockwise around one pole but clockwise around the other, hence one should map the rotation parameter $1/r\to-1/r$, and by noting that the renormalized $z$ includes a factor $r^{N_f-N_a}$.

We have glossed over a technical difficulty: the one-loop determinant \eqref{BLhiggsdetasymp} is singular for chiral multiplets $\phi_I$ that acquire a non-zero value in a given Higgs branch.  The corresponding zero-mode of $\phi_I$ is removed by the $D$-term equation, in other words integrating out $D$ makes $\phi_I$ massive.  To derive \eqref{BLZHiggs-sqcd} we have eliminated these zero-modes by taking the appropriate residues, but fixing signs is not straightforward in this approach.  On the other hand, we know that Coulomb branch and Higgs branch localization must yield the same result.  One can derive the Higgs branch expression \eqref{BLZHiggs-sqcd} from the Coulomb branch integral (if $N_a<N_f$ or $N_a=N_f$ and the true FI parameter $\zeta>0$) by closing the integration contours and writing the integral as a sum of residues.  From the Coulomb branch integral, $k^\pm$ appear as the flux $\flux=k^+-k^-$ and integers $k^+_j+k^-_j\geq 0$ labeling poles of gamma functions. Factorization is due to the fact that the integrand in the Coulomb branch expression is a product of a function of $z$ and $\I\sigma+\flux/2$ by a function of $\bar{z}$ and $\I\sigma-\flux/2$.

In more general theories, the Higgs branch localization result takes the form
\begin{equation}\label{BLZHiggs}
  Z_{S^2} = \sum_{\text{Higgs branches}} Z_{\text{cl}} \, Z'_{\text{1-loop}} \, Z_{\text{vortex}} \, Z_{\text{antivortex}} \;.
\end{equation}
The sum ranges over constant solutions to $(\sigma+\twmass_I)\phi_I=0$ and to the $D$-term equation $\phi\bar{\phi}=\chi$, which form a discrete set for generic twisted masses.%
\footnote{In theories with a complicated matter content, this should be checked explicitly.}
The factors are a classical contribution from these constant solutions, a one-loop determinant with poles removed as outlined above, and (anti)holomorphic contributions from point-like (anti)vortices at the poles.  The detailed expression can in principle be obtained by localizing onto the Higgs branch as we have just done for SQCD, but solutions to the BPS equations have not been investigated in general.

A simpler approach to obtain \eqref{BLZHiggs} is to start from the Coulomb branch integral and close contours: as for SQCD, the residues organize themselves into a factorized form.  This leads to vortex partition functions which were also later computed in the Omega background \cite{BLFujimori:2015zaa}. The relevant sets of poles are described
in \ifvolume{\autoref{MOsec:residues}}{subsection 6.2 of \volcite{MO}}.
We apply this technique to compare partition functions of dual
theories in \autoref{BLssec:dual-seiberg}.

\subsection{One-loop determinants}
\label{BLssec:sphere-det}

The first step in computing one-loop determinants around a $\supercharge$-invariant configuration $\Phi_0$ is to write the quadratic Lagrangian for fluctuations~$\delta\Phi$, including a gauge fixing term.  The Lagrangian takes the form $\delta\Phi \, \Delta[\Phi_0] \, \delta\Phi$ for some operator~$\Delta$ whose bosonic part~$\Delta_{\bosonic}$ is essentially a Laplacian, and whose fermionic part~$\Delta_{\fermionic}$ is essentially a Dirac operator.  The one-loop determinant reads
\begin{equation}\label{BLoneloopbla}
  Z_{\text{1-loop}}[\Phi_0] = \frac{\det\Delta_{\fermionic}[\Phi_0]}{\det\Delta_{\bosonic}[\Phi_0]} \;,
\end{equation}
where we omit the usual square root by considering $\Delta_{\bosonic}$ as a complex operator rather than a real operator on the same space of fields.
Additionally, the operators $\Delta_{\bosonic}$ and $\Delta_{\fermionic}$ split into direct sums of contributions from the vector and chiral multiplets, which decompose further into individual roots or weights.
Three techniques are commonly used to evaluate these determinants.

The most pedestrian approach (for the round sphere only) is to decompose fields into spherical harmonics.
In this decomposition $\Delta_{\bosonic}$ and~$\Delta_{\fermionic}$ are block diagonal, with blocks involving a finite number of modes.
The determinant of each block is straightforward to evaluate, and in their product, contributions of many bosonic and fermionic modes cancel.

A second approach harnesses the cancellation by constructing two eigenmodes of $\Delta_{\fermionic}$ for each eigenmode of $\Delta_{\bosonic}$, and viceversa.  The pairing fails to be exactly 2-to-1 for some modes, which thus contribute to the ratio of determinants~\eqref{BLoneloopbla}.  Only these eigenvalues need to be computed, instead of the whole spectrum of $\Delta$.

The third approach is more systematic.
To begin with, find a basis $(X,X')$ of the fluctuation fields such that $\supercharge X=X'$ and $\supercharge X'= \cR X$ where $\cR=\supercharge^2$ is bosonic.
Separate pairs $(X_0,X_0')$ with $X_0$ bosonic and $X_0'$ fermionic from pairs $(X_1,X_1')$ with opposite statistics,
and write down the part of~$V$ quadratic in fluctuations as
\begin{equation}
  V^{(2)} = X_0' D_{00} X_0 + X_1 D_{10} X_0 + X_0' D_{01} X_1' + X_1 D_{11} X_1' \;.
\end{equation}
The operators $\Delta_{\bosonic}$ and $\Delta_{\fermionic}$ are read from $\supercharge V^{(2)}$.  After some linear algebra, the constraint $\supercharge^2 V^{(2)}=0$ implies that%
\footnote{For a non-degenerate deformation term, $\Delta_\text{f}$ and $\Delta_\text{b}$ have no zero-modes transverse to the localization locus. $\cR_0$ and $\cR_1$ can have such zero-modes, which should be omitted from $\det \cR_1/\det \cR_0$, however these are transverse to $\coker D_{10}$ and $\ker D_{10}$.}
\begin{equation}
  \frac{\det \Delta_{\fermionic}}{\det \Delta_{\bosonic}}
  = \frac{\det' \cR_1}{\det' \cR_0}
  = \frac{\det_{\coker D_{10}} \cR_1}{\det_{\ker D_{10}} \cR_0}
  = \prod_{i} \cR(i)^{-m_{i}} \;,
\end{equation}
where $i$ indexes eigenvalues $\cR(i)$ of $\cR$, and $m_{i}$
is the multiplicity of $\cR(i)$ in $\ker D_{10}$ minus
that in $\coker D_{10}$.  These eigenvalues and
multiplicities are read from the $\cR$-equivariant index
\begin{equation}
  \ind_{\cR} (D_{10})
  = \Tr_{\ker D_{10}} e^{t\cR} - \Tr_{\coker D_{10}} e^{t\cR}
  = \sum_{i} m_{i} \, e^{t\cR(i)} \;,
\end{equation}
itself computed as a sum over fixed points of $\cR$, thanks to the
Atiyah-Bott-Berline-Vergne equivariant localization formula \cite{BLAtiyah:1984px, BLBerline:1982cl}.

Each of these methods requires lengthy calculations for which we refer to appendices of \cite{BLBenini:2012ui, BLDoroud:2012xw, BLGomis:2012wy}.
To clarify a sign that was originally missed (in the vector multiplet one-loop determinant) we must describe some salient points.
By continuity, squashing cannot affect signs, so we focus for simplicity on the round sphere $f(\theta)=r$ and let $r=1$.

As a warm-up before the vector multiplet, consider the one-loop determinant for fluctuations of a chiral multiplet along a particular weight~$w$ of~$\repr$.
Denote by $w(\flux)$ and $w(\sigma)$ the (on-shell) components of the vector multiplet when acting on this weight.
One finds the eigenvalues of $\Delta_{\bosonic}$ by expanding fields in (spin) spherical harmonics: $\big( J+ \frac12 \big)^2 - \big( \I w(\sigma) + \frac{1-\Rcharge}2 \big)^2$ with multiplicity $2J+1$ for $J- \frac{\abs{w(\flux)}}2 \in \mathbb{Z}_{\geq 0}$.  The fermionic Lagrangian is
\begin{equation}\label{BLLpsi}
  \mathcal{L}^{\text{c.m.}(w)}_{\fermionic}
	= \I\bar{\psi} \Delta^{\text{c.m.}(w)}_{\fermionic} \psi
	= \I\bar{\psi}\bigl(-\slashed{D}- w(\flux)/2+(\I w(\sigma)-\Rcharge/2)\gamma^3\bigr)\psi \;,
\end{equation}
where $\slashed{D}=\gamma^i D_i$ involves a flux $w(\flux)$ through the sphere.  Note that we extracted a factor of $\I$ from $\Delta_{\fermionic}$: this multiplies the determinant by an overall constant phase which only affects the normalization of the partition function.
The operator $\slashed{D}$ has $\abs{w(\flux)}$ zero-modes of chirality $\sign w(\flux)$, and it has, for each $J-\abs{w(\flux)}/2-1/2\in\mathbb{Z}_{\geq 0}$, $2J+1$ pairs of modes with eigenvalues $\pm\I\sqrt{(J+1/2)^2- w(\flux)^2/4}$ interchanged by $\gamma^3$.  The operator $\Delta_{\fermionic}$ thus decomposes into blocks with the following determinant (the matrix is expressed in a basis of eigenmodes of $\slashed{D}$)
\begin{equation}
  \begin{aligned}
	& \det \begin{pmatrix}
	  -\I\sqrt{(J+1/2)^2- w (\flux)^2/4}-w(\flux)/2 & -\I w(\sigma)+\Rcharge/2 \\
		-\I w(\sigma) + \Rcharge/2 & \I\sqrt{(J+1/2)^2- w(\flux)^2/4}-w(\flux)/2
	\end{pmatrix}
	\\
	& \qquad =\bigl(J+\tfrac{1-\Rcharge}{2}+\I w(\sigma) \bigr)\bigl(J+\tfrac{1+\Rcharge}{2}-\I w(\sigma) \bigr) \;,
	\end{aligned}
\end{equation}
as well as $\abs{w(\flux)}$ eigenvalues $-w(\flux)/2+ ( \I w(\sigma) - \Rcharge/2 ) \sign w(\flux)$.  Combining these ingredients,
\begin{align}
  \frac{\det\Delta^{\text{c.m.}(w)}_{\fermionic}}{\det\Delta^{\text{c.m.}(w)}_{\bosonic}}
	& = \begin{aligned}[t]
	  & \big( -w(\flux)/2 - (\Rcharge/2-\I w(\sigma) )\sign w(\flux) \big)^{\abs{w(\flux)}} \\
	  & \times \frac{\prod_{J=(\abs{w(\flux)}+1)/2}^{\infty} \big( J+1/2-\Rcharge/2+\I w(\sigma) \big)^{2J+1} \big( J+1/2+\Rcharge/2-\I w(\sigma) \big)^{2J+1}}
				 {\prod_{J=\abs{w(\flux)}/2}^{\infty} \big( J+1-\Rcharge/2+\I w(\sigma) \big)^{2J+1} \big( J+\Rcharge/2-\I w(\sigma) \big)^{2J+1}}
	  \end{aligned} \nonumber \\
  \label{BLcm1looppre}
	& = \big( -\sign w(\flux) \big)^{\abs{w(\flux)}}
	\prod_{J=\abs{w(\flux)}/2}^{\infty}\frac{J+1-\Rcharge/2+\I w(\sigma)}{J+\Rcharge/2-\I w(\sigma)} \;.
\end{align}
The infinite product is divergent and we apply zeta-function regularization, namely replace $\prod_{k\geq 0} (x+k)$ by $\sqrt{2\pi}/\Gamma(x)$. Combining the contributions from all weights $w$ of the chiral multiplet representation, we get the chiral multiplet one-loop determinant
\begin{equation}\label{BLcm1loop}
  Z^{\text{c.m.}}_{\text{1-loop}}
	= \prod_{w}
	\frac{ \big( -\sign w(\flux) \big)^{\abs{w(\flux)}} \Gamma \big( \frac{\Rcharge}{2}-\I w(\sigma) + \frac{\abs{w(\flux)}}{2} \big) }{ \Gamma \big( 1-\frac{\Rcharge}{2}+\I w(\sigma) + \frac{\abs{w(\flux)}}{2} \big)}
	= \prod_{w}
	\frac{\Gamma \big( \frac{\Rcharge}{2}-\I w(\sigma) - \frac{w(\flux)}{2} \big) }{ \Gamma \big( 1-\frac{\Rcharge}{2}+\I w(\sigma) - \frac{w(\flux)}{2} \big)} \;.
\end{equation}
The last equality can be proven using $\Gamma(x+1)=x\Gamma(x)$.

We now move on to the vector multiplet, with an emphasis on signs rather than the precise factors.  The quadratic action, hence the one-loop determinant, splits into contributions from each root~$\alpha$ of~$G$.  Again we denote by $\alpha(\flux)$ and $\alpha(\sigma)$ the relevant on-shell components of the vector multiplet.  The bosonic operator $\Delta^{\alpha}_{\bosonic}$, taking into account ghosts, is positive definite as for the chiral multiplet hence will not affect signs in the final result.  The fermionic action is closely related to the action~\eqref{BLLpsi} of an adjoint chiral multiplet of R-charge $\Rcharge=0$.  It is
\begin{equation}
  \mathcal{L}^{\text{v.m.}(\alpha)}_{\fermionic}
	= -\I\bar{\lambda} \Delta^{\text{v.m.}(\alpha)}_{\fermionic} \lambda
	= -\I\bar{\lambda}\bigl(-\slashed{D}+\tfrac{1}{2}[w(\flux), \cdot]+\I\gamma^3[w(\sigma),\cdot]\bigr)\lambda \;,
\end{equation}
which differs from its chiral multiplet counterpart for the same weight $\alpha$ in two respects: we extracted a different overall factor $-\I$ instead of~$\I$, and more importantly the sign in front of $w(\flux)$ changed.  Keeping track of the effect of this change on the eigenvalues found previously gives
\begin{equation}\label{BLdelvmalpha}
  \det\Delta^{\text{v.m.}(\alpha)}_{\fermionic}
	= \big( -\sign\alpha(\flux) \big)^{\abs{\alpha(\flux)}}
  \big( \tfrac{\abs{\alpha(\flux)}}2 + \I\alpha(\sigma) \big)^{\abs{\alpha(\flux)}}
  \prod_{J= \frac{ \abs{\alpha(\flux)}+1}2 }^{\infty} \big( (J+ \tfrac12)^2 + \alpha(\sigma)^2 \big)^{2J+1} \;.
\end{equation}
Upon taking the product over the roots~$\alpha$, the signs for positive and negative roots combine:
\begin{equation}
  \prod_{\alpha} \big( - \sign\alpha(\flux) \big)^{\abs{\alpha(\flux)}}
  = \prod_{\alpha>0} (-1)^{\alpha(\flux)}
	= e^{2\pi\I \delta(\flux)}
\end{equation}
where $\delta$ is the Weyl vector, half-sum of the positive roots, and $2\delta(\flux) \in \mathbb{Z}$ may be odd.
The other contributions in~\eqref{BLdelvmalpha} combine into positive factors, and most are cancelled by bosonic factors.
Altogether, the vector multiplet one-loop determinant is
\begin{equation}\label{BLvm1loop}
  Z^{\text{v.m.}}_{\text{1-loop}}
	= e^{2\pi\I\delta(\flux)}
  \prod_{\alpha>0} \biggl[ \frac{\alpha(\flux)^2}{4} + \alpha(\sigma)^2 \biggr] \;.
\end{equation}
More precisely, the one-loop determinant omits factors for roots with $\alpha(\flux)=0$.  We include these factors nevertheless: they arise as Vandermonde determinants when replacing the integral over all scalars $\sigma$ in the Lie algebra of $G$ commuting with $\flux$ by an integral over the Cartan subalgebra only.

Had we kept the radius explicitly, it would appear as $1/r^2$ in each factor of~\eqref{BLvm1loop} and $1/r$ in each factor of~\eqref{BLcm1looppre}. The zeta-function regularized form of $\prod_{k\geq 0} (x+k/r)$ is $\sqrt{2\pi}r^{rx-1/2}/\Gamma(rx)$, thus the one-loop determinant listed above are multiplied altogether by
\begin{equation}
  r^{-\dim G+\rank G} \prod_{w} r^{1-\Rcharge+2i r w(\sigma)}
	= r^{c/3+\rank G+2ir \sum_w w(\sigma)} \;,
\end{equation}
with $c/3=\sum_w (1-\Rcharge) - \dim G$ as in \eqref{BLcover3}.  This power of $r$, together with the power $r^{-\rank G}$ due to integrating over $r\sigma$ rather than $\sigma$, yields the overall power $r^{c/3}$ and the renormalization \eqref{BLzrenorm} of FI parameters in \autoref{BLssec:sphere-coulomb}.


\section{Other \texorpdfstring{\ifvolume{$\cN{=}(2,2)$}{$\boldsymbol{\cN{=}(2,2)}$}}{N=(2,2)} curved-space results}
\label{BLsec:curved}

In this section we will briefly present some of the main directions in which the simple computation of the $S^2$ partition function has been developed.

\subsection{Local operator insertions}
\label{BLssec:operators}

Besides the computation of pure Euclidean partition functions, localization is also extremely powerful in computing expectation values and correlators of BPS operators: both local and non-local, both order and disorder.%
\footnote{By ``order'' operators we mean standard polynomial functions of the fundamental fields in the Lagrangian, while ``disorder'' operators are defined as singular boundary conditions for the fields in the path-integral at points or submanifolds  \cite{BL'tHooft:1977hy}.}
In order to be computable, the operators must be invariant under the supercharge $\cQ$ used to localize, in other words they must be BPS. However the superalgebra on a curved manifold may be quite different from the flat-space one, therefore localization on $S^2$ (or other curved manifolds) grants us access to correlators that go beyond the standard ``chiral rings'' on flat space.

Let us discuss local operators on $S^2$.%
\footnote{Wilson line operators can be easily computed as well, see \cite{BLDoroud:2012xw}.}
It follows from \eqref{BLdeltalambda}-\eqref{BLdeltapsi} that the order operators invariant under $\cQ$ are the field-strength twisted chiral operators $\Sigma$---whose bottom components are the complex scalars $(\sigma - i\eta)$---at the North pole $x_\text{N}$, and their conjugates $\bar\Sigma$ at the South pole $x_\text{S}$ (since $\cQ^2 = J + R/2$, $\cQ$-closed local operators must sit at fixed points of $J$). For order operators the localization prescription is simply to insert the operator into the integrand, in this case $\sigma \mp i\fm/2r$ for $\Sigma$ and $\bar\Sigma$, respectively. Schematically and setting $r=1$, non-normalized correlators are given by
\begin{equation}\label{BLcurvedcorrelators}
\Big\langle \cO_\text{N}\big( \Sigma(x_\text{N}) \big) \, \cO_\text{S}\big( \bar\Sigma(x_\text{S}) \big) \Big\rangle_\text{n.n.} = \frac1{\big| \text{Weyl}\big|} \sum_\fm \int_\ft \frac{d^\text{rank}\sigma}{(2\pi)^\text{rank}} \; \cO_\text{N}\Big( \sigma - \frac{i\fm}2 \Big)\, \cO_\text{S}\Big( \sigma + \frac{i\fm}2 \Big) \, Z_\text{cl,1-loop}
\end{equation}
where $\cO_\text{N,S}$ are arbitrary gauge-invariant polynomial functions, while normalized correlators are further divided by $Z_{S^2}$.

There is a subtlety, though. Since we are forced to place all chiral operators at the same point (and similarly for anti-chirals), one might expect contact terms to show up. Such contact terms can be understood in terms of operator mixings \cite{BLGerchkovitz:2016gxx}. Indeed the correlators \eqref{BLcurvedcorrelators} do not satisfy the flat-space chiral ring relations. It turns out \cite{BLBenini:2014mia, BLClosset:2015rna} that they realize a sort of non-commutative deformation of the chiral ring. Let us present the simple example of $\bC\bP^{N-1}$, \ie{} the $U(1)$ GLSM with $N$ chiral multiplets of charge $1$. With a trick similar to the one used in \autoref{BLssec:sphere-pde}, one can show that the correlators satisfy
\begin{equation}
\label{BLdeformed ring CP}
\big\langle \Sigma^N \, \cO_\text{N}(\Sigma) \, \cO_\text{S}(\bar\Sigma) \big\rangle = \big(\tfrac ir\big)^N z \, \big\langle \cO_\text{N}\big( \Sigma - \tfrac ir \big) \,  \cO_\text{S}(\bar \Sigma) \big\rangle \;,
\end{equation}
where $\Sigma$ and $\bar\Sigma$ are inserted at the North and South pole, respectively.
In the limit $r\to \infty$ this reproduces the twisted chiral ring of the GLSM, namely $\Sigma^N = ( \frac ir )^N z $,%
\footnote{On flat space this should be written as $\Sigma^N = (iM_\text{UV})^N z_\text{UV}$, where the UV Fayet-Iliopoulos term is related to the renormalized one as in \eqref{BLzrenorm}.}
which coincides with the quantum cohomology of $\bC\bP^{N-1}$ \cite{BLWitten:1993xi}. To interpret the deformation, we rewrite (non-normalized) correlators of $\Sigma$ as derivatives of the partition function, $\Sigma \to \frac z{ir} \parfrac{}{z}$, which obviously do not commute with $z$. Then \eqref{BLdeformed ring CP} reduces to the A-system equation for $\bC\bP^{N-1}$, $( \parfrac{}{\log z} )^N - (-1)^N z = 0$, and the non-commutative ring is the $\cD$-module obtained as quotient of the Weyl algebra by that equation.

An interesting class of local disorder operators is given by vortex operators.%
\footnote{They are the equivalent of vortex line operators in 3d, and of certain simple surface operators in 4d.}
They are defined by the singular behavior of the gauge field $A$ around a point,
\begin{equation}
\label{BLvortex-op-def}
A \,\sim\, \gamma \, d\varphi \;,
\end{equation}
where $\varphi$ is an angular coordinate around that point and $\gamma \in \fg$. The definition \eqref{BLvortex-op-def} is in a gauge where matter fields are regular, therefore it corresponds to a point-like insertion of magnetic flux: $F_{12} = 2\pi \gamma\, \delta^2(x)$. On $S^2$ we can insert BPS local vortex operators at the two poles, and the supersymmetric operator also involve the local source $D = 2\pi i \gamma\, \delta^2(x)$. In \cite{BLHosomichi:2015pia} the correlator of two vortex operators at the two poles, and labelled by $\gamma^\text{N,S}$, has been computed with localization techniques in $\cN{=}(2,2)$ simple gauge theories. A useful trick is the equivalence between gauge theories in the presence of (quantized) vortex operators and on orbifolds \cite{BLBiswas:1997}. In the Abelian case the result is that the correlators, as functions of $\gamma$, are piecewise constant with jumps at certain specific values. In the non-Abelian case, instead, there is a dependence on the Levi subgroup specified by $\gamma$, which is the reductive subgroup of $G$ commuting with $\gamma$. In both cases the partition function still factorizes as in \autoref{BLssec:sphere-higgs}.


\subsection{Twisted multiplets on the sphere}
\label{BLssec:curved-twisted}

So far we have considered the simplest type of $\cN{=}(2,2)$ gauge theories: those made of vector and chiral multiplets. Things become even more interesting when including other types of multiplets. The addition of twisted chiral multiplets has been studied in \cite{BLGomis:2012wy}. A twisted chiral multiplet $Y:(y, \chi, G)$ satisfies $\overline D_+ Y = D_- Y = 0$ and it comprises a complex scalar $y$, a Dirac fermion $\chi$ and a complex auxiliary scalar $G$. One can write Lagrangians for them on the sphere that preserve the $\mathfrak{su}(2|1)$ supersymmetry algebra of $S^2$, and apply localization to compute partition functions and correlators.%
\footnote{In fact, \cite{BLGomis:2012wy} generalizes the construction to squashed rotationally-invariant spheres which preserve only an $\mathfrak{su}(1|1)$ superalgebra. It turns out that the localization computations do not depend on the squashing.}
Twisted chiral multiplets must be neutral under vector multiplets, however they can couple to the field-strength twisted chiral multiplet $\Sigma$ through a twisted superpotential $\wt W$.

The supersymmetric action on $S^2$ follows from the Lagrangian
\begin{equation}
\cL_\text{t.c.m.} = D^i \bar y D_i y + i \bar\chi \gamma^i D_i \chi + \big| G + \tfrac\Delta r y \big|^2 \;.
\end{equation}
The parameter $\Delta$ is the Weyl weight of the twisted chiral multiplet. In the special case $\Delta=1$, the action is the same as for an Abelian vector multiplet, because with the identifications $y = \sigma - i\eta$, $\chi = \lambda$ and $G = D - i F_{12}$ we construct the field-strength twisted chiral multiplet.

The interaction term $\cL_{\wt W}$ has been already considered in \autoref{BLssec:sphere-theory}. In the context of mirror symmetry, as we will see in \autoref{BLssec:dual-mirror}, particularly important is the non-minimal coupling $\wt W = \Sigma Y$ between a vector multiplet and twisted chiral multiplets, where $Y$ plays the role of a dynamical FI term.

One can perform localization with respect to the same supercharge $\cQ$ used before \cite{BLGomis:2012wy}. From the BPS equations $\delta \chi = \delta \bar\chi = 0$, one obtains the conditions
\begin{equation}
y = \text{const.} \;,\qquad\qquad G + \tfrac\Delta r y = 0 \;,
\end{equation}
in other words the complex scalar $y$ can be an arbitrary constant on the sphere. The Lagrangian $\cL_\text{t.c.m.}$ is $\cQ$-exact and can be used as a localization term. The one-loop determinant is trivial, in the sense that it does not depend on $y$. Thus, for a system of twisted chiral multiplets $Y_I$ the localization formula reads
\begin{equation}\label{BLtwistedchiralZ}
Z_{S^2} = \int \Big( \prod\nolimits_I d^2y_I \Big) \; e^{-4\pi i r \wt W(y) - 4\pi i r \overline{\wt W}(\bar y)} \;,
\end{equation}
where the index $I$ runs over the twisted chiral multiplets.
As we will see in \autoref{BLssec:dual-mirror}, this expression can be used to confirm the mirrors \cite{BLHori:2000kt} of Hori and Vafa.

The superalgebra on $S^2$ we have considered so far---that we will call $\mathfrak{su}(2|1)_A$---is a subalgebra of the $\cN{=}(2,2)$ superconformal algebra whose bosonic part comprises $\mathfrak{su}(2)$ rotations of $S^2$ and the $\mathfrak{u}(1)$ vector-like R-symmetry. There exists, though, another inequivalent choice, $\mathfrak{su}(2|1)_B$, which instead contains the $\mathfrak{u}(1)$ axial R-symmetry \cite{BLDoroud:2013pka}. The two are swapped by the $\bZ_2$ mirror outer automorphism that exchanges the R-symmetries and exchanges multiplets with twisted multiplets. Those superalgebras contain supercharges $\cQ_A$ and $\cQ_B$, respectively. The charge $\cQ_A$ annihilates twisted chiral operators at the North pole and their conjugates at the South pole: it can be used to compute their correlators on $S^2$ (as we saw in \autoref{BLssec:operators}), as well as the Zamolodchikov metric on the K\"ahler moduli space of conformal fixed points. Those quantities are independent of complex structure moduli because chiral (superpotential) deformations are $\cQ_A$-exact. Likewise, $\cQ_B$ annihilates chiral operators at the North pole and their conjugates at the South pole, it can be used to compute their correlators on $S^2$ and the Zamolodchikov metric on the complex structure moduli space of fixed points \cite{BLDoroud:2013pka}. They will be independent of K\"ahler moduli because twisted chiral (twisted superpotential) deformations are $\cQ_B$-exact.

At this point, the easiest way to construct $\mathfrak{su}(2|1)_B$-invariant actions on $S^2$ and compute their path-integrals with localization is to exploit the $\bZ_2$ mirror automorphism. Thus, the $\mathfrak{su}(2|1)_B$ partition function of a gauge theory of vector and chiral multiplets is equal to the $\mathfrak{su}(2|1)_A$ partition function of a theory of twisted vector and twisted chiral multiplets. In this way, we can perform all computations in the $\mathfrak{su}(2|1)_A$ framework we have used to far, provided we study all types of twisted multiplets, in particular twisted vector multiplets.

A twisted vector multiplet is a real multiplet $\widetilde V$ subject to gauge redundancy by a twisted chiral multiplet: $\widetilde V \cong \widetilde V + \Lambda_t + \overline\Lambda_t$. It has the same components as a standard vector multiplet, $\widetilde V : (A_\mu, \sigma, \bar\sigma, \eta, \bar\eta, D)$, but the supersymmetry transformations are different. In particular, the field strength sits in the chiral multiplets $\widetilde\Sigma : (\sigma, \eta, D + iF)$ and its conjugate, the transformation of which we have already discussed. Twisted chiral multiplets can be minimally coupled to twisted vector multiplets; in Wess-Zumino gauge, the transformations of the former pick up a dependence on the latter. One can then write down supersymmetric actions on $S^2$ \cite{BLDoroud:2013pka}, in particular the Yang-Mills Lagrangian for $\widetilde V$ equals the kinetic Lagrangian for the chiral multiplet $\widetilde\Sigma$. The FI term sits in a linear superpotential term for $\widetilde\Sigma$.

An important point to stress is that charged twisted chiral multiplets contribute to the gauge-$U(1)_R$ anomaly. Since $U(1)_R$ is part of the supersymmetry algebra on $S^2$ (rather than being an outer automorphism as on flat space), such an anomaly would spoil supersymmetry. Therefore, only theories for which the sums of the charges of twisted chirals under Abelian twisted vectors vanish can preserve supersymmetry on $S^2$ quantum mechanically.%
\footnote{The same statement is true if we try to use the B-type topological twist \cite{BLWitten:1991zz}.}

Both kinetic actions are $\cQ$-exact and can be used for localization \cite{BLDoroud:2013pka}. The bosonic part of the Lagrangian reads
\begin{multline}
\cL_\text{kin} \big|_\text{bos} = \frac1{2g^2} \Tr \Big( |D_\mu\sigma|^2 + \frac14 [\sigma, \bar\sigma]^2 + F^2 + \tilde D^2 \Big) \\
+ |D_\mu y|^2 + |G|^2 + \frac12 \big( |\sigma y|^2 + |\bar\sigma y|^2 \big) + \frac{g^2}2 (y\bar y - \chi)^2 \;,
\end{multline}
where $\tilde D \equiv D + i g^2 (y\bar y - \chi)$ and $\chi$ is the matrix of FI terms that commutes with the gauge generators. Under the standard reality conditions, that Lagrangian is semipositive definite. The path-integral localizes%
\footnote{As always, one should be careful to include the gauge-fixing sector. The details can be found in the reference.}
to its zeros modulo gauge transformations, that is the manifold
\begin{equation}
\cM = \Big\{ y \,\big|\, y = \text{const.} \,,\; y\bar y - \chi = 0 \Big\} / G \;=\; \bC^{|\fR|} /\!\!/_\chi G
\end{equation}
with all other fields vanishing. Therefore, $\cM$ is a K\"ahler quotient of $\bC^{|\fR|}$ at levels $\chi$ of the moment map, where $|\fR|$ is the dimension of the matter representation.

Let us consider the Abelian case discussed in \cite{BLDoroud:2013pka}: the gauge group is $U(1)^{N_c}$ and there are $N_f$ twisted chiral multiplets of charges $Q^{a=1 \ldots N_c}_{I = 1 \ldots N_f}$ subject to $\sum_I Q^a_I = 0$. The one-loop determinant turns out to be
\begin{equation}
Z_\text{1-loop} = \det (M^\dag M) \;,
\end{equation}
where $M$ is the $N_f \times N_c$ matrix $M\du{I}{a} = Q^a_I y_I$. Obviously it must have $N_f \geq N_c$, otherwise the gauginos have fermionic zero-modes and the determinant vanishes. After some algebra, the partition function can be written as
\begin{equation}\label{BLZAGLSM}
Z_{S^2} = \int \frac{d^{N_f} y \wedge d^{N_f} \bar y}{(2\pi)^{N_c}} \, \det(M^\dag M) \, \prod_{a} \delta \big( 2\mu_a + \chi_a \big) \; e^{-4\pi i r\wt W(y) - 4 \pi i r \overline{\wt W}(\bar y)} \;,
\end{equation}
where the functions
\begin{equation}
\mu_a = - \frac12 \sum\nolimits_I Q^a_I |y_I|^2
\end{equation}
are the moment maps for the gauge action. Twisted chiral and antichiral operators are easily inserted at the North and South pole, respectively, by including $\cO_\text{NP}(Y)$ and $\cO_\text{SP}(\bar Y)$ in the integrand.

An interesting check performed in \cite{BLDoroud:2013pka} in models with a low-energy geometric description as Calabi-Yau NLSMs, is that the partition function assumes the form
\begin{equation}
Z_{S^2} = i^{\dim \cM} \int_\cM \Omega \wedge \bar\Omega = e^{- \cK_\text{C}}
\end{equation}
in terms of the nonwhere vanishing holomorphic top form $\Omega$. Thus, the sphere partition function computes the K\"ahler potential on the complex structure moduli space.

There are other multiplets that represent the $\cN{=}(2,2)$ supersymmetry algebra, such as semichiral multiplets, semichiral vector multiplets and large vector multiplets \cite{BLBuscher:1987uw, BLLindstrom:2007vc}. The most general $\cN{=}(2,2)$ NLSM contains chiral, twisted chiral and semichiral multiplets, and has a non-K\"ahler target with bi-Hermitian \cite{BLGates:1984nk} (also known as generalized K\"ahler \cite{BLGualtieri:2003dx, BLGualtieri:2007ng}) geometry with torsion. Some of those models can be realized at the IR of GLSMs constructed out of the more general multiplets. For instance, the sphere partition function of GLSMs with semichiral multiplets has been computed in \cite{BLBenini:2015isa}.


\subsection{Localization on the hemisphere}
\label{BLssec:curved-hemi}

A very interesting development of the localization programme is to consider theories on manifolds with boundaries. This is an extremely rich and interesting problem in its own right, it allows to discuss domain walls and interfaces between different phases, it relates with lower-dimensional theories that may live along the boundary, and in the 2d case it makes contact with the physics of D-branes via string theory.

As a first step, \cite{BLSugishita:2013jca, BLHonda:2013uca, BLHori:2013ika} study 2d $\cN{=}(2,2)$ gauge theories on the hemisphere (topology of the disk $D_2$) and compute their partition function with localization. The result $Z_{D_2}(\cB)$ depends on the boundary conditions---or D-brane---at the boundary, and they mainly focus on B-branes \cite{BLWitten:1991zz, BLOoguri:1996ck, BLHori:2000ck, BLHerbst:2008jq}. The partition function depends holomorphically on twisted chiral parameters and it is independent of chiral parameters. In fact, it computes the inner product $\langle \cB | \unit \rangle$ between two states in the Ramond sector, one generated by the identity operator%
\footnote{More generally, one can compute $\langle \cB | \cO \rangle$ by inserting the twisted chiral operator $\cO$ at the pole.}
and the other by the boundary conditions, and this is called the \emph{central charge} of the D-brane \cite{BLOoguri:1996ck}. For theories that flow to NLSMs on K\"ahler manifolds, the large-volume limit reproduces the known geometric expression \cite{BLMinasian:1997mm}, however the localization formula contains all quantum corrections (to be compared with \cite{BLHori:2000kt, BLHori:2000ck}). This can help in identifying the precise correspondence between the original B-brane and the mirror A-brane.

The partition function takes the form of an integral of a meromorphic form, and the choice of contour is related to the choice of boundary conditions for vector multiplets. There is no a priori rule to decide the boundary conditions, but the convergence of the integral imposes strong constraints. Near the phase boundaries, a convergent contour can be found only for a very restricted class of branes, and this reproduces the \emph{grade restriction rule} found in \cite{BLHerbst:2008jq} in some specific cases, generalizing it to non-Abelian GLSMs and non-Calabi-Yau geometries.

Let us briefly describe the localization computation of $Z_{D_2}(\cB)$ \cite{BLSugishita:2013jca, BLHonda:2013uca, BLHori:2013ika}. The hemisphere is parametrized by $\theta$ in the range $\big[ 0, \frac\pi2\big]$ and the boundary breaks the superalgebra $\mathfrak{su}(2|1)_A$ to $\mathfrak{su}(1|1)_A$, whose bosonic $\mathfrak{u}(1)$ subalgebra is a combination of rotations and vector-like R-symmetry rotations. To construct the kinetic actions for vector and chiral multiplets, one can use the same $\cQ$-exact expressions as on $S^2$, however they will now involve boundary terms (we set $r=1$):
\begin{equation}\begin{aligned}
\delta_\cQ \delta_{\bar\epsilon}  \int \dvol_2 \, \Tr \big( \bar\lambda \lambda/2 - 2 D \sigma + \sigma^2 \big) &= \int \dvol_2 \; \cL_\text{v.m.} + \oint_{\theta = \frac\pi2} d\varphi \; \cL^\text{bd}_\text{v.m.} \\
\delta_\cQ \delta_{\bar\epsilon}  \int \dvol_2 \; \big( \bar\psi \psi - 2i \bar \phi \sigma\phi + (q-1) \bar\phi\phi \big) &= \int \dvol_2 \; \cL_\text{c.m.} + \oint_{\theta = \frac\pi2} d\varphi \; \cL^\text{bd}_\text{c.m.} \;.
\end{aligned}\end{equation}
Likewise, the FI and $\vartheta$-terms as well as more general twisted superpotential interactions are corrected by boundary terms to be supersymmetric. However the variation of the superpotential is a boundary term:
\begin{equation}
\delta \cL_W = i D_\mu \Big( \bar\epsilon \gamma^\mu \psi_i \, \partial_i W(\phi) \Big) + i D_\mu \Big( \epsilon \gamma^\mu \bar\psi_{\bar\imath}\, \partial_{\bar\imath} W(\bar\phi) \Big)
\end{equation}
called the Warner term \cite{BLWarner:1995ay}. This is not easily canceled by a boundary term as before.

To cancel the variation of the superpotential, we need to construct the so-called Chan-Paton boundary interaction \cite{BLHerbst:2008jq}. First we need a $\bZ_2$-graded vector space
\begin{equation}
\cV = \cV^e \oplus \cV^o \;.
\end{equation}
We can think of the space $\text{End}(\cV)$ as a superalgebra, and take the odd endomorphisms to anticommute with the fermionic fields. Then $\cV$ must furnish a unitary representation of $G \times G_F \times U(1)_R$, \ie{} its coordinates are assigned R-charges $q_*$ and a gauge/flavor representation $\rho_*$. Finally we should construct two polynomials $\fQ(\phi)$, $\bar\fQ(\bar\phi)$ (complex conjugate in Lorentzian signature) with values in $\text{End}(\cV)^o$, invariant under $G\times G_F$ and with R-charge $1$ and $-1$, respectively, such that
\begin{equation}
\fQ(\phi)^2 = W(\phi) \, \unit_\cV \;,\qquad\qquad \bar\fQ(\bar\phi)^2 = \bar W(\bar\phi) \, \unit_\cV \;.
\end{equation}
These equations are called a \emph{matrix factorization} of $W$, and $\fQ$ is called a \emph{tachyon profile}. With it we construct the super-connection
\begin{equation}
\cA_\varphi = \rho_* \big( A_\varphi + i \sigma \big) + \frac{R}{2} - \frac i2 \{\fQ, \bar \fQ\} + \frac12 (\psi_+ - \psi_-)^i \partial_i \fQ + \frac12 (\bar\psi_+ - \bar\psi_-)^i \partial_i \bar \fQ
\end{equation}
and the boundary interaction is given by the supertrace
\begin{equation}
\text{Str}_\cV \Big[ \text{Pexp} \Big( i \oint d\varphi\, \cA_\varphi \Big) \Big] \;.
\end{equation}
One can show that the SUSY variation of the Chan-Paton interaction cancels the Warner term \cite{BLHerbst:2008jq}. The term $\{\fQ,\bar \fQ\}$ represents a boundary potential.

At this point one should specify boundary conditions invariant under supersymmetry and compatible with the Euler-Lagrange equations. The issue is somehow delicate, and details can be found in \cite{BLSugishita:2013jca, BLHonda:2013uca, BLHori:2013ika}. The boundary conditions for vector multiplets include $D_\theta \sigma = 0$ and $F_{\mu\nu} = 0$, therefore---compared to the $S^2$ case---the BPS moduli space does not include the flux parameter $\fm$. For chiral multiplets there are two options: Neumann or Dirichlet. Neumann boundary conditions include $D_\theta \phi = D_\theta \bar\phi = 0$ and describe directions along the D-brane (in target space). Dirichlet boundary conditions include $\phi = \bar\phi = 0$ and describe directions perpendicular to the D-brane. The BPS equations fix $\sigma = D = \text{const}$. The one-loop determinants are
\begin{equation}\begin{aligned}
Z_\text{gauge} &= \prod_{\alpha>0} i\alpha (\sigma) \,  \sin \big( i \pi \alpha (\sigma) \big) \\
Z_\text{matter}^\text{Neu} &= \prod_{w\in\fR} \Gamma\big( -i w(\sigma) \big) \;,\qquad\qquad Z_\text{matter}^\text{Dir} = \prod_{w\in \fR} \frac{- 2\pi i \, e^{\pi w(\sigma) } }{ \Gamma\big(1 + iw(\sigma) \big)} \;.
\end{aligned}\end{equation}
As usual, twisted masses are introduced by including an external vector multiplet with scalar component $\tau$, and R-charges are accounted by the shift $\tau \to \tau + i q/2$. The final localization formula is
\begin{equation}
Z_{D_2}(\cB) = \frac1{|\text{Weyl}(G)|} \int \frac{d^N\sigma}{(2\pi)^N} \; z^{i\sigma} \; \text{Str}_\cV \big[ e^{-2\pi \sigma} \big] \; Z_\text{1-loop}
\end{equation}
where $z = e^{-2\pi \zeta + i\vartheta}$ is the exponential of the complexified FI parameter and $N = \rank G$. As we have discussed before, the integration contour should be defined with care: it should be a deformation of the real contour $\sigma \in \bR^N$ which ensures convergence, and only for a very restricted set of branes can this be achieved for all values of $z$ \cite{BLHori:2013ika}.

To clarify the role of the Chan-Paton interaction, let us give a simple example of D0-branes on $\bC^n$. The model has $n$ free chiral multiplets with flavor symmetry $G_F = U(n)$. To describe D0-branes one can simply impose Dirichlet boundary conditions in all directions, which does not break $U(n)$. There is no gauge sector and the partition function reads
\begin{equation}
Z_{D_2}(\text{D0}) = \prod_{j=1}^n \frac{ -2\pi i \, e^{\pi \tau_j}}{\Gamma(1+i\tau_j)}
\end{equation}
where $\tau_j$ are the twisted masses (equivariant parameters). On the other hand, we can impose Neumann boundary conditions and construct a boundary interaction. To construct $\cV$ we take fermionic oscillators $\{\eta_i, \bar\eta^j\} = \delta_i^j$ and a Clifford vacuum $|0\rangle$ such that $\bar\eta^j |0\rangle = 0$: then we identify $\cV$ with the fermionic Fock space. We choose the tachyon profile
\begin{equation}
\fQ(\phi) = \phi^i \eta_i \;,\qquad\qquad \bar\fQ(\bar\phi) = \bar\phi_j \bar\eta^j \;.
\end{equation}
Clearly $\fQ(\phi)^2 = \bar\fQ(\bar\phi)^2 = 0$. As $\eta_i$ generate the Fock space and transform in the antifundamental representation of $U(n)$, one gets $\text{Str}_\cV \big[ e^{-2\pi \tau} \big] = \prod_j (1- e^{2\pi \tau_j})$. Hence
\begin{equation}
Z_{D_2}(\text{D0)} = \prod_{j=1}^n \big( 1 - e^{2\pi \tau_j} \big) \, \Gamma(-i\tau_j) = \prod_{j=1}^n \frac{ -2\pi i \, e^{\pi \tau_j}}{\Gamma(1+i\tau_j)} \;.
\end{equation}
We have indicated the boundary conditions in the same way as before, namely as $\text{D0}$, because they realize the same D-brane in the IR, and indeed the central charges agree. The boundary interaction creates a potential $\{\fQ, \bar\fQ\} = \bar\phi^i \phi_i$ whose only minimum is at the origin, therefore at low energies it gives the same D-brane as by imposing Dirichlet boundary conditions. This is a simple example of tachyon condensation \cite{BLSen:1998sm}: one can describe all lower-dimensional branes using space-filling branes and a suitable boundary tachyon profile.

An important outcome of the localization computation is the expression for the D-brane central charge in large-volume geometric phases \cite{BLHori:2013ika}:
\begin{equation}
Z_{D_2}(\cB) = \int_X \hat\Gamma_X \; e^{B + i\omega/2\pi} \; \text{ch}(\cB) \;.
\end{equation}
Here $X$ is the K\"ahler target manifold, $\omega$ its K\"ahler class, $B$ is the B-field and $\text{ch}(\cB)$ is the Chern character of the complex of holomorphic vector bundles that specifies the brane (see \cite{BLHori:2013ika} for details). Moreover $\hat\Gamma_X$ is the \emph{Gamma class} of the holomorphic tangent bundle of $X$, defined in the standard way in terms of the Chern roots by the function
\begin{equation}
\hat\Gamma(x) = \Gamma\Big( 1 - \frac x{2\pi i} \Big) \;.
\end{equation}
The formula was already known to mathematicians \cite{BLHosono:2004jp, BLKatzarkov:2008hs, BLIritani:2009},%
\footnote{In particular notice that $\hat\Gamma_X$ appears in place of the A-roof class $\hat A_X$.}
and the appearance of $\hat\Gamma_X$ from the perturbative part of the NLSM path-integral has also been confirmed in \cite{BLHalverson:2013qca}. A similar analysis for orientifold planes, as opposed to D-branes, has been done in \cite{BLKim:2013ola} by studying the partition function of GLSMs on $\bR\bP^2$, and an expression for the central charge in large-volume geometric phases has been found.

\subsection[\texorpdfstring{$\Omega$}{Omega}-deformed A-twist on the sphere]{$\boldsymbol{\Omega}$-deformed A-twist on the sphere}
\label{BLssec:curved-omega}

The modern ``Coulomb branch localization'' framework initiated by Pestun \cite{BLPestun:2007rz} can be applied to well-studied setups to obtain new interesting results. In particular, the canonical way to preserve supersymmetry on a curved manifold is to perform the so-called \emph{topological twist} \cite{BLWitten:1988ze}: one turns on a background vector field that couples to the R-symmetry, equal and opposite to (some component of) the spin connection. On $S^2$, the A-type topological twist corresponds to one unit of magnetic flux for the vector R-symmetry $R$.%
\footnote{In the B-type twist one turns on one unit of flux for the axial R-symmetry. Since the two are equivalent, up to the $\bZ_2$ automorphisms that swaps multiplets with twisted multiplets, we shall consider the A-twist only.}
Thus, the A-twist \cite{BLWitten:1988xj} consists of a supersymmetric background for 2d $\cN{=}(2,2)$ theories---different from the one we have discussed so far---preserving an $\mathfrak{su}(1|1)_A$ subalgebra (the bosonic $\mathfrak{u}(1)$ is the vector-like $R$).

The interesting observables in the A-model \cite{BLWitten:1988xj, BLWitten:1991zz} are correlation functions of twisted chiral operators at non-coincident points (and their descendants). For gauge theories, they are gauge-invariant polynomials $\cO\big(\Sigma(p) \big)$ of the field-strength twisted multiplet at points $p$. Those operators form a chiral ring, and the correlations functions are independent of the insertion points $p$. For theories that flow to NLSMs, the correlators equal the structure constants of the quantum cohomology ring of the target.

Localization has been applied to topologically twisted theories since the beginning, however in a form more similar to the ``Higgs branch localization'' in which the path-integral localizes to holomorphic maps in A-twisted NLSMs, and to point-like vortices in A-twisted GLSMs (see \eg{}  \cite{BLWitten:1993yc, BLMorrison:1994fr}). It turns out that Coulomb branch localization can be applied as well, producing compact and easily-calculable expressions for correlators in A-twisted gauge theories on $S^2$ \cite{BLBenini:2015noa, BLClosset:2015rna} and yielding new results in non-Abelian theories.

The A-twisted background can be generalized into a sort of $\Omega$-deformed $S^2$ \cite{BLClosset:2014pda}:%
\footnote{A geometric way to understand this deformation is to start in 3d with an $S^2$-bundle over $S^1$ in which $S^2$ rotates by an angle $\epsilon_\Omega$ as we wind around $S^1$. Then reduce on $S^1$ to 2d \cite{BLBenini:2015noa}.}
the superalgebra is still $\mathfrak{su}(1|1)$, however the $\mathfrak{u}(1)$ bosonic subalgebra is a linear combination of $R$ and rotations of $S^2$ along an axis, controlled by a parameter $\epsilon_\Omega$. This deformation had already appeared in the mathematical literature, starting with the work of Givental \cite{BLGivental:1996}. Since rotations are part of the supersymmetry algebra, for $\epsilon_\Omega\neq 0$ local operators can only be inserted at the North and South poles.

Localization can be applied to the more general $\Omega$-deformed A-twist as well \cite{BLBenini:2015noa, BLClosset:2015rna}. The BPS configurations are parametrized by the Cartan part of the complex scalar $\sigma$ in the vector multiplet, as well as by diagonal magnetic fluxes $\fm$ in the coroot lattice $\Gamma_\ft$. For a theory with exponentiated complex FI term $z = e^{-2\pi \zeta +i \vartheta}$ the classical action is
\begin{equation}
Z_\text{cl} = z^{\Tr \fm} \;.
\end{equation}
The one-loop determinants for vector and chiral multiplets are
\begin{equation}
Z_\text{1-loop}^\text{gauge} = \prod_{\alpha>0} \bigg[ \alpha(\sigma)^2 - \frac{\alpha(\fm)^2 \epsilon_\Omega^2}4 \bigg] \;,\quad
Z_\text{1-loop}^\text{chiral} = \prod_{w \in \fR} \prod_{j = - \frac{|B_w|-1}2}^{\frac{|B_w|-1}2} \bigg( \frac1{w(\sigma) + j \,\epsilon_\Omega} \bigg)^{\sign B_w}
\end{equation}
where $B_w = w(\fm) - q_w + 1$. Here $q_w$ are the R-charges, which on the A-twisted background have to be chosen integral (because of Dirac quantization). Each operator insertion $\cO(\Sigma)$ brings an extra factor
$$
\cO\Big( \sigma \pm \epsilon_\Omega \frac\fm2 \Big) \;.
$$
The sign is $\pm$ for insertions at the North/South pole, while for $\epsilon_\Omega =0$ operators can be inserted at any point. Then the localization formula is
\begin{equation}
\label{BLA-twist formula}
\Big\langle \prod\nolimits_i \cO_i \Big\rangle_\Omega = \frac1{|\text{Weyl}(G)|} \sum_{\fm \in \Gamma_\ft} \Bigg[ \sum_{\sigma_* \in \fM_\text{sing}} \JKres_{\sigma = \sigma_*} \big( \sQ(\sigma_*), \eta \big) \; \prod\nolimits_i \cO_i \; Z_\text{cl} \, Z_\text{1-loop} + \text{bdry} \Bigg] \;.
\end{equation}
This formula has a similar flavor to the previous ones, however the integration is over a specific middle-dimensional contour in the complex $\sigma$-plane which effectively computes a weighted sum of residues of the integrand at the singular points $\sigma_* \in \fM_\text{sing}$. This particular contour is called the Jeffrey-Kirwan residue \cite{BLJeffreyKirwan}: its importance for recent localization computations has been recognized in \cite{BLBenini:2013xpa} in the context of the elliptic genus, and we will explain it in detail in \autoref{BLsec:torus}, including the notation used in \eqref{BLA-twist formula}. The JK residue depends on the choice of an auxiliary parameter $\eta \in \ft$, while the last term represents boundary contributions at infinity of the $\sigma$-plane. The sum of all contributions does not depend on $\eta$, however for certain choices the boundary contribution vanishes making the computation easier. More details can be found in \cite{BLBenini:2015noa, BLClosset:2015rna}.


\subsection{General supersymmetric backgrounds}

To study supersymmetric backgrounds more general than the $\cN{=}(2,2)$ massive superalgebras on the round $S^2$ or the A-twist, one can employ a systematic method developed by Festuccia and Seiberg in four dimensions \cite{BLFestuccia:2011ws}, and adapted to the two-dimensional case in \cite{BLClosset:2014pda} (the method is summarized in \volcite{DU}). The method consists in coupling the flat-space supersymmetric theory of interest to some off-shell supergravity,%
\footnote{This step has the technical limitation that we need an off-shell formulation both for supergravity and the quantum field theory. Thus, it becomes difficult to apply the method when the number of supercharges is large.}
giving an expectation value to the bosonic fields in the graviton multiplet (including the metric and the auxiliary fields), and then taking a rigid limit in which the Newton constant goes to zero but the background remains fixed. In this limit, supersymmetry of the background is guaranteed by the vanishing of gravitino variations (possibly including gaugino variations, when they are part of the graviton multiplet).

The two-dimensional Lorentzian $\cN{=}(2,2)$ superalgebra can have at most the R-symmetry group automorphism $U(1)_\text{left} \times U(1)_\text{right} \simeq U(1)_R \times U(1)_A$, \ie{} a vector and an axial part. We restrict to theories with a vector-like R-symmetry, then the algebra admits a complex central charge $Z$.%
\footnote{The $\cN{=}(2,2)$ superalgebra admits two complex central charges, each breaking one of the two R-symmetries. A superconformal theory cannot have central charges.}
On Euclidean flat space, the supersymmetry algebra is
\begin{equation}
\{Q_\alpha, \wt Q_\beta\} = \big[ 2\gamma^\mu P_\mu + 2i \bP_+ Z - 2i \bP_- \wt Z \big]_{\alpha\beta} \;,\qquad\qquad \{Q_\alpha,Q_\beta\}=\{\wt Q_\alpha, \wt Q_\beta\} = 0 \;,
\end{equation}
where $\bP_\pm$ are the projectors on positive/negative chirality spinors. Tilded quantities are complex conjugate in Lorentzian signature, but are independent complexified quantities in Euclidean signature.  To $Q_\pm$, $\wt Q_\pm$ we assign R-charges $-1$ and $+1$, respectively.

Theories with an R-symmetry have an $\cR$-multiplet%
\footnote{For a discussion of supercurrent multiplets in two dimensions see \cite{BLDumitrescu:2011iu}.}
containing the conserved operators
\begin{equation}
\cR_\mu \,:\quad  \big( T_{\mu\nu},\, S_{\alpha\mu},\, \wt S_{\alpha\mu},\,  j_\mu^R,\, j_\mu^Z,\, j_\mu^{\wt Z} \big) \;,
\end{equation}
namely the stress tensor, the supersymmetry currents, and the currents for R-symmetry and central charges. Correspondingly, there exists an off-shell 2d supergravity---dimensional reduction of new minimal 4d supergravity \cite{BLSohnius:1981tp, BLSohnius:1982fw, BLGates:1983nr}---whose graviton multiplet couples to the $\cR$-multiplet. In a Wess-Zumino gauge it contains the fields
\begin{equation}
\cG_\mu \,:\quad \big( g_{\mu\nu},\, \Psi_{\alpha\mu},\, \wt\Psi_{\alpha\mu},\, V_\mu,\, C_\mu,\, \wt C_\mu \big) \;,
\end{equation}
namely the metric, the gravitinos and the gauge fields coupling to R- and central charges. The gauge fields $V_\mu, C_\mu, \wt C_\mu$ appear in covariant derivatives, $D_\mu = \nabla_\mu - i q V_\mu + \frac z2 \wt C_\mu - \frac{\tilde z}2 C_\mu$, as well as through their field strengths. It is convenient to introduce the dual field strengths
\begin{equation}
\cH = - i \varepsilon^{\mu\nu} \partial_\mu C_\nu \;,\qquad\qquad \wt\cH = -i \varepsilon^{\mu\nu} \partial_\mu \wt C_\nu \;.
\end{equation}
The gravitino variations---which will be referred to as the generalized Killing spinor (GKS) equations---are:
\begin{equation}\begin{aligned}
\label{BLGKS equations 2d}
0 = \tfrac12\delta \Psi_\mu &= (\nabla_\mu - i V_\mu) \epsilon - \frac12 \mat{\cH & 0 \\ 0 & \wt\cH} \gamma_\mu \epsilon + \ldots \\
0 = \tfrac12 \delta \wt\Psi_\mu &= (\nabla_\mu + i V_\mu) \wt\epsilon - \frac12 \mat{ \wt\cH & 0 \\ 0 & \cH} \gamma_\mu \wt\epsilon + \ldots \;.
\end{aligned}\end{equation}
In the rightmost terms we used conventions in which the chirality matrix is $\gamma_3 = \smat{1 & 0 \\ 0 & -1}$. The two supersymmetry parameters $\epsilon$, $\wt\epsilon$ are complex Dirac spinors with R-charges $1,-1$ respectively, and no central charges. The dots represent terms that vanish when we set $\Psi_\mu = \wt\Psi_\mu = 0$. These equations have to be solved for the background fields as well as for $\epsilon$, $\wt\epsilon$, and the number of solutions for the latter is the number of preserved supercharges.

From the off-shell supergravity transformations of fields (which can be found in \cite{BLClosset:2014pda}) one deduces the deformed supersymmetry algebra on a background:%
\footnote{The contraction of spinor indices is $\psi \chi = \psi_\alpha \varepsilon^ {\alpha\beta} \chi_\beta$, namely the symbol $\psi\chi$ stands for $\psi^\sT \smat{0 & 1 \\ -1 & 0} \chi$ in standard matrix notation. Moreover in this subsection $\sigma$ indicates the full complex scalar in the vector multiplet, indicated by $\sigma - i\eta$ before.}
\begin{equation}
\label{BLdeformed SUSY algebra 2d}
\begin{aligned} \{ \delta_\epsilon, \delta_{\tilde\epsilon}\} &= i \cL_K - i \epsilon \bfQ \wt\epsilon \\
\{\delta_{\epsilon_1}, \delta_{\epsilon_2}\} &= \{ \delta_{\tilde\epsilon_1}, \delta_{\tilde\epsilon_2} \} = 0
\end{aligned} \qquad\qquad
\bfQ = \mat{z - \sigma - \frac r2 \cH & 0 \\ 0 & \wt z -\wt\sigma - \frac r2 \wt\cH}
\end{equation}
acting on a field $\varphi_{q,z,\tilde z}$ of fixed charges. The first term is a gauge-covariant Lie derivative, and it represents a translation along the vector field
\begin{equation}
K^\mu = \epsilon \gamma^\mu \wt\epsilon \;.
\end{equation}
It follows from the GKS equations that such a vector field is Killing, $\nabla_{(\mu} K_{\nu)} = 0$ and $\cL_K \cH = \cL_K \wt H = 0$, unless it vanishes. The second term is a mix of R-symmetry, $Z/\wt Z$-symmetry and gauge/global symmetry rotations.

Let us discuss some important solutions to (\ref{BLGKS equations 2d}) (the full set of solutions in presented in \cite{BLClosset:2014pda}). The first solution, known for a long time \cite{BLWitten:1988xj}, is the \emph{topological A-twist}:
\begin{equation}
V_\mu = -\frac14 \omega_\mu^{ab} \varepsilon_{ab} \;,\qquad \epsilon = \mat{0 \\ \epsilon_-} \;,\qquad \wt\epsilon = \mat{\wt\epsilon_+ \\ 0} \;,\qquad \cH = 0 \;,\qquad \wt\cH = 0
\end{equation}
where $\omega_\mu^{ab}$ is the spin connection and $\epsilon_-, \wt\epsilon_+$ are constant.%
\footnote{More generally, $\wt\cH$ could be an arbitrary function; this does not affect the supersymmetry algebra. We cannot turn on holonomies for $V_\mu$ because $\epsilon, \tilde\epsilon$ (and the supercharges) would no longer be periodic and there would not be solutions.}
This solution exists on any orientable Riemann surface $\Sigma$. There are two Killing spinors of opposite R-charge and chirality. On a compact Riemann surface of genus $g$, the background has $(g-1)$ units of R-symmetry flux: $\frac1{2\pi} \int_\Sigma dV = g-1$. In particular the R-charge of all gauge-invariant operators should be quantized:
\begin{equation}
q \, (g-1) \in \bZ \;.
\end{equation}
The deformed supersymmetry algebra, in the absence of central charges, is simply%
\footnote{With central charges one finds the $\mathfrak{su}(1|1)$ superalgebra $\{\delta_\epsilon, \delta_{\tilde\epsilon}\} = i \epsilon_- \wt\epsilon_+ (z-\sigma)$.}
\begin{equation}
\delta_\epsilon^2 = \delta_{\tilde\epsilon}^2 = 0 \;,\qquad\qquad \{\delta_\epsilon, \delta_{\tilde\epsilon}\}=0 \;.
\end{equation}
Similarly, there is the $\overline{\text{A}}$-twist with $1-g$ units of R-symmetry flux. For $g>1$ these are the only two solutions.

For $g=0$ (topologically $S^2$), if the metric has a rotational symmetry around an axis, there exists a one-parameter family of deformations called the $\Omega$-background 
in \cite{BLClosset:2014pda}. Let $K^\mu$ be the Killing vector field, which for definiteness we can take as $K = \partial_\varphi$ in the coordinates \eqref{BLcoordinatesS2}. Then
\begin{equation}
V_\mu = - \frac12 \omega_\mu^{12} \;,\quad \epsilon = \mat{ \epsilon_\Omega K_{\hat z} \\ 1}\epsilon_- \;,\quad \wt\epsilon = \mat{ 1 \\ - \epsilon_\Omega K_{\hat{\bar z}} } \wt\epsilon_+ \;,\quad \cH = - \frac i2 \epsilon_\Omega \varepsilon^{\mu\nu} \partial_\mu K_\nu \;,\quad \wt\cH = 0 \;,
\end{equation}
where $\epsilon_\Omega$ is a complex parameter. We have used the flat complex index that follows from $e^{\hat z} = e^{\hat 1} + i e^{\hat 2}$. In practice we can identify $C_\mu = \frac{\epsilon_\Omega}2 K_\mu$. The background preserves two complex supercharges of opposite R-charge, and the deformed supersymmetry algebra is
\begin{equation}
\{ \delta_\epsilon, \delta_{\tilde\epsilon}\} = i \epsilon_\Omega \cL_K^{s'} + i \epsilon_- \wt\epsilon_+ Z \;,
\end{equation}
where $\cL_K^{s'}$ is a Lie derivative covariant with respect to gauge/flavor rotations, not with respect to $R$, $Z$ and $\wt Z$, but where the spin $s$ is replaced by $s' = s + \frac q2$. This is the background that we considered in \autoref{BLssec:curved-omega}.

On $S^2$ there is a second class of interesting solutions with no net R-symmetry flux: they give \emph{untwisted} backgrounds. The simplest case is that of a round $S^2$, then
\begin{equation}
V_\mu = 0 \;,\qquad \cH = \wt\cH = \frac ir \;,\qquad \nabla_\mu \epsilon = \frac i{2r} \gamma_\mu \epsilon \;,\qquad \nabla_\mu \wt\epsilon = \frac i{2r} \gamma_\mu \wt\epsilon \;.
\end{equation}
The spinors solve the Killing spinor equation: on the round $S^2$ there are four solutions---two for $\epsilon$ and two for $\wt\epsilon$---so the number of preserved supersymmetries is maximal. With no central charges the deformed supersymmetry algebra is
\begin{equation}
\{\delta_\epsilon, \delta_{\tilde\epsilon}\} = i \cL_K - \frac{\epsilon\wt\epsilon}{2r} \, R \;,
\end{equation}
and the Killing vectors $K^\mu$ generate the $\mathfrak{so}(3)$ isometry algebra of $S^2$. The full superalgebra is $\mathfrak{su}(2|1)$. Notice that the background is not the analytic continuation of a real background in Lorentzian signature, and this in general breaks reflection positivity. However, if the theory is superconformal, the auxiliary fields $\cH$, $\wt\cH$ couple to redundant operators and reflection positivity is recovered.

As described in \autoref{BLssec:sphere-theory}, on more general topological spheres with only $U(1)$ isometry one can still preserve two supercharges of opposite R-charge without twisting. Considering for definiteness the metric $ds^2 = f(\theta)^2 d\theta^2 + r^2 \sin^2\theta\, d\varphi^2$, the background is
\begin{equation}
V = \frac12 \Big( 1 - \frac rf \Big) d\varphi \;,\quad \epsilon = e^{\frac i2 \theta \gamma_1} e^{\frac i2 \varphi} \mat{ \epsilon_0 \\ 0} \;,\quad \wt\epsilon = e^{\frac i2 \theta \gamma_1} e^{- \frac i2 \varphi} \mat{ 0 \\  \wt \epsilon_0} \;,\quad \cH = \wt\cH = \frac if
\end{equation}
with constant $\epsilon_0$, $\wt\epsilon_0$.

One last example we want to mention is that of $\cN{=}(2,2)$ theories with both vector and axial R-symmetries, placed on a flat $T^2$. In this case one can turn on a flat connection for, say, the left-moving R-symmetry $U(1)_\text{left}$. As a result the left-moving supercharges are lifted, however there remain the two right-moving supercharges with opposite charge under $U(1)_\text{right}$. This case is discussed in \autoref{BLsec:torus}.

Supersymmetric actions on the curved backgrounds are constructed in a way similar to flat space: the supersymmetry variations of gauge-invariant D-terms and F-terms are total derivatives, therefore their spacetime integrals are supersymmetric invariants. The top D-term component of a neutral general supermultiplet with $q=0$ can be used to construct the super-Yang-Mills and matter kinetic actions, $\cL_D = D$. For instance from the D-term of $-\frac12 \wt\Phi e^{-2\cV} \Phi$ one obtains the kinetic action of a chiral multiplet $\Phi$:
\begin{equation}\begin{aligned}
\cL_\Phi &= D_\mu \wt\phi D^\mu \phi + \wt\phi D \phi + \frac12 \big( \tfrac q2 R_s + \cH \wt z + \wt\cH z \big) \wt\phi \phi + \frac12 \wt\phi \{ \bfQ, \wt\bfQ \} \phi \\
&\quad - \wt F F + i \wt\psi \gamma^\mu D_\mu \psi + i \wt\psi \wt\bfQ \psi + i \sqrt2\, \wt\psi \wt\lambda \phi + i \sqrt2\, \wt\phi \lambda \psi
\end{aligned}\end{equation}
where $R_s$ is the scalar curvature.
For NLSMs one uses the D-term of the K\"ahler potential $K(\wt\Phi, \Phi)$. From the D-term of a gauge-invariant multiplet whose lowest component is $\frac12 \Tr \wt\sigma\sigma$ one obtains the SYM action
\begin{equation}\begin{aligned}
\cL_\cV &= \frac12 \Big( F_{12} - \frac12 \wt\cH \sigma + \frac12 \cH \wt\sigma \Big)^2 + \frac12 D_\mu \wt\sigma D^\mu \sigma + \frac18 [\sigma,\wt\sigma]^2 \\
&\quad + i \wt\lambda \gamma^\mu D_\mu \lambda - i \wt\lambda \mat{ [\wt\sigma,\cdot] & 0 \\ 0 & [\sigma,\cdot] } \lambda - \frac12 \Big( D + \frac12 \wt\cH\sigma + \frac12 \cH \wt\sigma \Big)^2
\end{aligned}\end{equation}
with trace implicit. The top F-term component of a neutral chiral multiplet with $q=2$, $z= \wt z =0$ gives superpotential interactions,
\begin{equation}
\cL_W = F_W + \wt F_W \;,
\end{equation}
where $F_W$ is the F-term component of the superpotential $W(\Phi)$:%
\footnote{Integrating out $F_i$, $\wt F_i$ one obtains the real positive potential $\sum_i \big| \parfrac{W}{\phi_i} \big|^2$. Alternatively, to keep the fields $F_i$, $\wt F_i$ one should redefine them with an $i$, then the kinetic action is positive-definite and the superpotential action is imaginary.}
\begin{equation}
F_W = \parfrac{W}{\phi_i} F_i - \frac12 \parfrac{^2W}{\phi_i \partial \phi_j} \psi_j \psi_i \;,\qquad\qquad \wt F_W = \parfrac{\wt W}{\wt \phi_i} \wt F_i + \frac12 \parfrac{^2 \wt W}{\wt \phi_i \partial \wt \phi_j} \wt\psi_j \wt\psi_i \;.
\end{equation}
The top component of a twisted chiral multiplet with $q=z=\wt z=0$ (called twisted F-term or G-term) can be corrected to give supersymmetric actions, since $\delta(G - i \wt\cH \omega)$ is a total derivative (here $\omega$ is the lowest component of a twisted chiral multiplet). The twisted superpotential action is then
\begin{equation}
\cL_\cW = G_\cW - i \wt\cH \cW(\omega) + \wt G_\cW + i \cH \wt\cW(\wt\omega) \;,
\end{equation}
where
\begin{equation}
G_\cW = \parfrac{\cW}{\omega_i} G_i - \frac12 \parfrac{^2 \cW}{\omega_i \partial\omega_j} \eta_j \eta_i
\end{equation}
is the G-term of the twisted superpotential $\cW$.


\section{Elliptic genera of \texorpdfstring{\ifvolume{$\cN{=}(2,2)$}{$\boldsymbol{\cN{=}(2,2)}$}}{N=(2,2)} and \texorpdfstring{\ifvolume{$\cN{=}(0,2)$}{$\boldsymbol{\cN{=}(0,2)}$}}{N=(0,2)} theories}
\label{BLsec:torus}

We discuss now the Euclidean path-integral of two-dimensional $\cN{=}(2,2)$ supersymmetric theories on a torus $T^2$. This quantity, called the \emph{elliptic genus}, was first introduced in the physics literature in \cite{BLSchellekens:1986yi, BLSchellekens:1986yj, BLPilch:1986en} in the context of free orbifolds and in \cite{BLWitten:1986bf, BLWitten:1987cg} in the context of non-linear sigma models. As usual, the path-integral on a circle computes the trace over a Hilbert space of states, and we start from such a Hamiltonian definition:
\begin{equation}
Z_{T^2}(\tau, z, u) = \Tr_\text{RR} \, (-1)^F q^{H_L} \bar q^{H_R} y^J \prod\nolimits_a x_a^{K_a} \;.
\end{equation}
The trace is over the Ramond sector of the Hilbert space of the theory on a spatial circle, \ie{} one takes periodic boundary conditions for fermions. Then $F$ is the fermion number, we define the parameters
\begin{equation}
q = e^{2\pi i \tau} \;,\qquad\qquad y = e^{2\pi i z} \;,\qquad\qquad x_a = e^{2\pi i u_a} \;,
\end{equation}
and $q$ specifies the complex structure of a torus $w \cong w+1 \cong w + \tau$, with $\tau = \tau_1 + i\tau_2$. $H_L$ and $H_R$ are the left- and right-moving Hamiltonians respectively, defined in Euclidean signature in terms of Hamiltonian and momentum as $2H_L = H + iP$, $2H_R = H - iP$. We assume that the theory has a left-moving $U(1)$ R-symmetry $J$ (which might be discrete if the theory is not conformal) and a flavor group $K$ (with Cartan generators $K_a$). Their fugacities are $y$ and $x_a$. Given a charge vector $\rho^a$, we use the notation $x^\rho = \prod\nolimits_a x_a{}^{\rho^a} = e^{2\pi i \rho^a u_a}$. We also write $\rho(u)=\rho^a u_a$, considering $\rho\in \fk^*$ and $u\in \fk$, where $\fk$ is the Cartan algebra of the flavor symmetry group $K$. The elliptic genus with $u_a\neq 0$ is sometimes called the equivariant elliptic genus, while setting $z=u_a=0$ the elliptic genus reduces to the Witten index. The $q\to0$ limit of the elliptic genus is called the $\chi_y$ genus.

Physically, the elliptic genus is interesting because it detects spontaneous supersymmetry breaking: if supersymmetry is broken, then the Witten index is zero (although the opposite is not necessarily true) \cite{BLWitten:1982df}. Moreover, if the theory is superconformal the operators $H_L, H_R, J$ equal the zero-mode generators $L_0, \bar L_0, J_0$ of the superconformal algebra%
\footnote{When not uniquely fixed, \eg{} by the superpotential, the superconformal R-symmetries can be determined through the $c$-extremization principle of \cite{BLBenini:2012cz, BLBenini:2013cda}.}
and the elliptic genus is equal to the \emph{superconformal index}, which counts superconformal primary operators in flat space.

Mathematically, the elliptic genus of a NLSM with complex target manifold $X$ is a topological invariant---related to the elliptic cohomology of $X$ \cite{BLWitten:1986bf}---equal to the Euler characteristic of a specific infinite-dimensional formal vector space $E_{q,y}$. The Witten index equals the Euler number of $X$. See \eg{} \cite{BLGritsenko:1999fk}.

If the theory has a discrete spectrum (at least in the equivariant sense), then the elliptic genus is a holomorphic function of $q$ because the contributions from states with $H_R \neq 0$ cancels between pairs of states with opposite values of $(-1)^F$. The genus has very interesting modular transformation properties as well. Since the spectrum of the Ramond sector is invariant under charge conjugation:
\begin{equation}
Z_{T^2}(\tau,z,u) = Z_{T^2}(\tau,-z,-u) \;.
\end{equation}
When the R-symmetry is non-anomalous and the theory flows to an IR fixed point, the modular transformations of the elliptic genus are:
\begin{equation}
Z_{T^2} \bigg( \frac{a\tau + b}{c\tau + d} \,,\, \frac z{c\tau + d} \,,\, \frac{u}{c\tau + d} \bigg) = \exp\bigg[ \frac{\pi i c}{c\tau + d} \Big( \frac{c_L}3 z^2 - 2\cA^a_L u_a z \Big) \bigg] \, Z_{T^2}(\tau,z,u)
\end{equation}
for $\smat{a & b \\ c & d} \in SL(2,\bZ)$. Here $c_L$ is the left-moving IR central charge, proportional to the 't~Hooft anomaly of $J$, while $\cA^a_L$ is the 't~Hooft anomaly between $J$ and $K_a$:
\begin{equation}
c_L = -3 \sum_\text{fermions} \gamma_3 J^2 \;,\qquad\qquad \cA^a_L = \sum_\text{fermions} \gamma_3 J K_a \;.
\end{equation}
The sums are taken over all fermions in the theory, and $\gamma_3$ is the chirality matrix that we take positive (negative) on right (left) movers. For a NLSM on a Calabi-Yau manifold $X$ of complex dimension $d = c_L/3$, the elliptic genus is a Jacobi form of weight zero and index $d/2$. 

Later on, the elliptic genus of Gepner models was  computed using the known characters of $\cN{=}2$ superconfomal algebras \cite{BLEguchi:1988vra, BLOoguri:1989fd}. Then it was realized that the elliptic genus of Landau-Ginzburg models can be computed by localization \cite{BLWitten:1993jg}, which led to a formula for the elliptic genus of Gepner models using the orbifold Landau-Ginzburg description \cite{BLBerglund:1993fj, BLKawai:1993jk, BLKawai:1994fm, BLBerglund:1994zg}. In this review we will be mostly concerned with the more recent computation of the elliptic genus of gauge theories \cite{BLGadde:2013ftv, BLBenini:2013nda, BLBenini:2013xpa} with localization techniques. The resulting formula agrees with that of Landau-Ginzburg models in case the gauge group is trivial, and with known mathematical results of the elliptic genus of complete intersections in toric varieties \cite{BLMaZhou, BLGuoZhou} when the theory has a smooth geometric phase.

It turns out that one can equally well consider the elliptic genus of theories with only $\cN{=}(0,2)$ supersymmetry. Mathematically, they describe more general bundles than the tangent bundle on $X$. Their equivariant elliptic genus is defined as
\begin{equation}
Z_{T^2}(\tau,u) = \Tr_\text{R} \, (-1)^F q^{H_L} \bar q^{H_R} \prod\nolimits_a x_a^{K_a} \;.
\end{equation}
If the theory has a low-energy description as a NLSM with target a holomorphic vector bundle over a compact complex manifold, as in the models in \cite{BLDistler:1993mk}, the elliptic genus encodes the Euler characteristic of that vector bundle \cite{BLKawai:1994np}.%
\footnote{Setting $u_a=0$, the equivariant elliptic genus reduces to the partition function of the chiral CFT associated to the half-twisted model.}

If the theory has discrete spectrum, the elliptic genus is a holomorphic function of $\tau$ with the following modular transformation properties:
\begin{equation}
Z_{T^2} \bigg( \frac{a\tau + b}{c\tau + d} \,,\, \frac{u}{c\tau + d} \bigg) = \epsilon(a,b,c,d)^{c_R-c_L} \; \exp\bigg[ - \frac{\pi i c}{c\tau + d} \, \cA^{ab} u_a u_b \bigg] \, Z_{T^2}(\tau,u)
\end{equation}
for $\smat{a & b \\ c & d} \in SL(2,\bZ)$. The multiplier system $\epsilon(a,b,c,d)$ is a phase, independent of $u_a$, universally defined by
\begin{equation}
\frac{\eta\big( \frac{a\tau+b}{c\tau + d} \big)}{\theta_1 \big(\frac{a\tau+b}{c\tau+d} \,\big|\, \frac{u}{c\tau+d})} = \epsilon(a,b,c,d) \; e^{-\frac{i\pi c}{c\tau+d} x^2} \; \frac{\eta(\tau)}{\theta_1(\tau|u)} \;.
\end{equation}
It is through $\epsilon$ that the gravitational anomaly shows up. In the theories under consideration, $c_R - c_L$ equals three times the number of right-moving minus left-moving fermions, while $\cA^{ab}$ are the flavor 't~Hooft anomalies:
\begin{equation}
c_R - c_L = 3 \sum_{\text{fermions}} \gamma_3 \;,\qquad\qquad \cA^{ab} = \sum_\text{fermions} \gamma_3 K_a K_b \;.
\end{equation}

One could also consider alternative elliptic genera where the trace is taken in the Neveu-Schwartz sector (on one or both sides), as in \cite{BLGadde:2013ftv}. The resulting functions are all equivalent, because easily related by spectral flow.

\paragraph{The Jeffrey-Kirwan residue.} The main focus in this review is the computation of the elliptic genera of gauge theories, through localization techniques. Such a computation has been done in \cite{BLGadde:2013ftv, BLBenini:2013nda, BLBenini:2013xpa} for the simplest theories, building on \cite{BLWitten:1993jg, BLGrassi:2007va}, and then generalized in many ways, for instance in \cite{BLHarrison:2013bya, BLMurthy:2013mya, BLAshok:2013pya, BLIsrael:2015aea}. In particular, the expression found in \cite{BLBenini:2013nda, BLBenini:2013xpa} involves a particular type of higher-dimensional residue operation called the \emph{Jeffrey-Kirwan} (JK) \emph{residue} defined in \cite{BLJeffreyKirwan} and motivated by \cite{BLWitten:1992xu}. Schematically:
$$
Z_{T^2} = \sum_{u_*} \JKres_{u=u_*} Z_\text{1-loop}(u)
$$
where $Z_\text{1-loop}$ is a meromorphic top-form on an $r$-dimensional complex manifold ($r$ is the rank of the gauge group), and the sum is over all singular points. The JK residue depends on a choice of vector $\eta$, however the total sum does not (and the parameter $\eta$ is just auxiliary). The details are presented below.

Such a residue operation arises from a careful treatment of the bosonic and fermionic zero-modes in the problem. It has then been realized that very similar systems of zero-modes arise in many other localization contexts, for instance for the Witten index in one dimension \cite{BLHori:2014tda, BLCordova:2014oxa}, for the Nekrasov partition function \cite{BLNekrasov:2002qd} in four \cite{BLNakamura:2015zsa} and five \cite{BLHwang:2014uwa} dimensions, for the topological twist in two dimensions \cite{BLBenini:2015noa, BLClosset:2015rna, BLClosset:2015ohf} and its higher-dimensional generalizations \cite{BLBenini:2015noa}. In all these cases, the result of localization for gauge theories takes the form of a JK residue, possibly with extra boundary contributions.

\subsection{Multiplets, Lagrangians and supersymmetry}
\label{BLssec:torus-theory}

In the path-integral formulation, the elliptic genus equals the Euclidean path-integral of the theory on a flat $T^2$, in the presence of flat connections $A^\text{R}$ and $A^\text{flav}$ for the R- and flavor symmetries, respectively, coupled to the R- and flavor symmetry currents:
\begin{equation}
z = \oint_t A^\text{R} - \tau \oint_s A^\text{R} \;,\qquad\qquad u_a =\oint_t A^\text{$a$-th flav} - \tau \oint_s A^\text{$a$-th flav} \;,
\end{equation}
where $t,s$ are the temporal and spatial cycles.%
\footnote{Choosing a constant connection $A^\text{R}_\mu$, we have $z = (-2i \tau_2)\, A^\text{R}_{\bar w}$ and similarly for the flavor holonomies.}
This is equivalent to specifying non-trivial boundary conditions twisted by the R- and flavor charges, along both the spatial and temporal cycles. Since the background is flat, the actions are just the standard ones. In the $\cN{=}(2,2)$ case we have already discussed multiplets, supersymmetry transformations and actions for gauge theories in \autoref{BLsec:sphere} and \autoref{BLsec:curved}. So, let us move the $\cN{=}(0,2)$ case.

We are interested in $\cN{=}(0,2)$ gauge theories. The reader can consult \cite{BLWitten:1993yc, BLDistler:1993mk, BLDistler:1995mi} for more details.
Using the complex coordinate $w$, the supersymmetry parameters satisfy $\gamma^w \epsilon = \gamma^w \bar\epsilon = 0$. It is convenient to write spinors in components, in particular the SUSY parameters are $\epsilon^+, \bar\epsilon^+$. We consider theories formulated in terms of chiral, Fermi and vector multiplets.

First we have a chiral multiplet $\Phi = (\phi, \bar\phi, \psi^-, \bar\psi^-)$ with variations
\begin{equation}\begin{aligned}
\label{BLSUSY var chiral}
\delta \phi &= -i \bar\epsilon^+ \psi^- \qquad\qquad\qquad & \delta \psi^- &= 2i \, \epsilon^+ D_{\bar w} \phi \\
\delta \bar\phi &= -i \epsilon^+ \bar\psi^- & \delta\bar\psi^- &= 2i \, \bar\epsilon^+D_{\bar w} \bar\phi \;.
\end{aligned}\end{equation}
Second we have a Fermi multiplet $\Lambda = (\psi^+, \bar\psi^+, G, \bar G)$ with variations
\begin{equation}\begin{aligned}
\label{BLSUSY var Fermi}
\delta \psi^+ &= \bar\epsilon^+ G + i \epsilon^+ E \qquad\qquad & \delta G &= 2\, \epsilon^+ D_{\bar w} \psi^+ - \epsilon^+ \psi_E^- \\
\delta \bar\psi^+ &= \epsilon^+ \bar G + i \bar\epsilon^+ \bar E & \delta \bar G &= 2\, \bar\epsilon^+ D_{\bar w} \bar\psi^+ - \bar\epsilon^+ \bar\psi_E^- \;.
\end{aligned}\end{equation}
Here $\cE(\Phi_i) = (E, \bar E, \psi_E^-, \bar\psi_E^-)$ is a chiral multiplet, holomorphic function of the fundamental chiral multiplets in the theory, and it is part of the definition of $\Lambda$. Notice that \mbox{$E = E(\phi_i)$} and its fermionic partner is $\psi_E^- = \sum_i \psi_i^- \, \partial E/\partial \phi_i$. Third we have a vector multiplet \hbox{$V = (A_\mu, \lambda^+, \bar\lambda^+, D)$} with variations
\begin{equation}\begin{aligned}
\label{BLSUSY var vector}
\delta A_w &= \tfrac 12 \big( \epsilon^+ \bar\lambda^+ - \bar\epsilon^+ \lambda^+ \big) \qquad &
\delta \bar\lambda^+ &= \bar\epsilon^+ (-D-iF_{12}) \qquad &
\delta (-D-iF_{12}) &= 2\, \epsilon^+ D_{\bar w} \bar\lambda^+ \\
\delta A_{\bar w} &= 0 &
\delta \lambda^+ &= \epsilon^+ (-D+iF_{12}) &
\delta (-D+iF_{12}) &= 2\, \bar\epsilon^+ D_{\bar w} \lambda^+ \;.
\end{aligned}\end{equation}
Comparing with \eqref{BLSUSY var Fermi}, notice that the fields in the second and third column form a Fermi multiplet $\Upsilon = (\bar\lambda^+, \lambda^+, -D-iF_{12}, -D + iF_{12})$ with $\cE = 0$.

The supersymmetric action for chiral multiplets comes from the Lagrangian
\begin{equation}
\cL_\Phi = D_\mu \bar\phi D^\mu \phi + i \bar\phi D \phi + 2\, \bar\psi^- D_w \psi^- - \bar\psi^- \lambda^+ \phi + \bar\phi \bar\lambda^+ \psi^- \;,
\end{equation}
for Fermi multiplets we have
\begin{equation}
\cL_\Lambda = - 2\, \bar\psi^+ D_{\bar w} \psi^+ + \bar E E + \bar G G + \bar\psi^+ \psi_E^- - \bar\psi_E^- \psi^+ \;,
\end{equation}
and for vector multiplets we have
\begin{equation}
\cL_\Upsilon = \Tr \Big[ F_{12}^2 + D^2 - 2\, \bar\lambda^+ D_{\bar w} \lambda^+ \Big] \;.
\end{equation}
The last one equals the Lagrangian for the Fermi multiplet $\Upsilon$. Interactions are specified by holomorphic functions $J^a(\phi)$ of the chiral multiplets (and anti-holomorphic functions $\bar J^a(\bar\phi)$ of their partners), where $a$ parametrizes the Fermi multiplets in the theory:
\begin{equation}
\cL_J = \sum\nolimits_a \big( i G_a J^a - \psi_a^+ \psi_J^{-a} \big) \;,\qquad\qquad \cL_{\bar J} = \sum\nolimits_a \big( i \bar G_a \bar J^a - \bar\psi_a^+ \bar\psi_J^{-a} \big) \;.
\end{equation}
Their supersymmetry variation is a total derivative as long as
\begin{equation}
\sum\nolimits_a E_a(\phi) J^a(\phi) = 0 \;.
\end{equation}

It turns out that all these actions are $\cQ$-exact. This reflects the fact that the elliptic genus is a ``topological invariant'', unaffected by continuous deformations of the parameters in the theory. Defining the anticommuting supercharge $\cQ$ by using commuting spinor parameters and choosing them $\epsilon^+ = \bar\epsilon^+ = 1$, the action of $\cQ$ is then immediately read off from the supersymmetry variations and one finds, up to total derivatives:
\begin{equation}\begin{aligned}
\cL_\Phi &= \cQ \big( 2i\bar\phi D_w \psi^- - i \bar\phi \lambda^+ \phi \big) \;,\qquad\qquad &
\cL_\Lambda &= \cQ \big( \bar\psi^+ G - i \bar E \psi^+ \big) \\
\cL_J &= \cQ \big( {\textstyle \sum_a} i \psi_a^+ J^a \big) \;,\qquad\qquad &
\cL_\Upsilon &= - \cQ \, \Tr \big( \lambda^+ (D + iF_{12}) \big) \;.
\end{aligned}\end{equation}

The $\cN{=}(2,2)$ theories can be regarded as special cases of $\cN{=}(0,2)$, in which the left-moving R-symmetry appears as a flavor symmetry. To reduce from $(2,2)$ to $(0,2)$ supersymmetry, we define projectors $P_\pm = (1 \pm \gamma_3)/2$. Then the chiral multiplet $\Phi_{(2,2)}$ splits into a chiral multiplet $\Phi = (\phi, \bar\phi, P_-\psi, P_-\bar\psi)$ and a Fermi multiplet $\Lambda = (P_+\psi, P_+\bar\psi, F, \bar F)$. The vector multiplet $V_{(2,2)}$ splits into a vector multiplet $V$
and an adjoint chiral multiplet $\Sigma = (\sigma, \bar\sigma, P_-\lambda, P_-\bar\lambda)$. If $\Phi_{(2,2)}$ is charged under $V_{(2,2)}$, then its Fermi component $\Lambda$ has $\cE = \Sigma \Phi$ (where $\Sigma$ acts in the correct representation). Superpotential interactions $W(\Phi_{(2,2)})$ become interactions $J^a(\phi) = \partial W/\partial \phi_a$. A twisted chiral multiplet $Y_{(2,2)}$ (which must be neutral) splits into a chiral and a Fermi multiplet. In particular the twisted chiral multiplet $\Sigma_{(2,2)}$ constructed out of $V_{(2,2)}$ splits into $\Upsilon$ and $\Sigma$. A twisted superpotential $\widetilde W(\Sigma_{(2,2)})$ becomes an interaction $J^\Upsilon(\sigma) = \partial \widetilde W / \partial \sigma$, and a complexified Fayet-Iliopoulos term is simply a constant $J^\Upsilon = \frac\theta{2\pi} + i\zeta$.

\subsection{The localization formula and the JK residue}
\label{BLssec:torus-loc}

We will first present the formula for the elliptic genus obtained from localization in \cite{BLBenini:2013nda, BLBenini:2013xpa}, and then outline its derivation in the next subsection.

The localization computation proceeds along the same steps as in \autoref{BLsec:sphere}. First, the moduli space of BPS configurations is parametrized by flat $G$-connection on $T^2$ modulo gauge transformations, where $G$ is the gauge group. For simplicity, we will assume that the non-Abelian part of $G$ is connected and simply-connected.%
\footnote{Non-simply-connected and disconnected groups can be treated as well, see \eg{} \cite{BLBenini:2013nda}.}
Let $\fh$ be the Cartan algebra of $G$, then the Cartan torus of $G$ can be identified with $\fh/ \Gamma_\fh$ where $\Gamma_\fh$ is the coroot lattice. We define
\begin{equation}
\label{BLmoduli space}
\fM =\fh_\bC /(\Gamma_\fh + \tau \Gamma_\fh) \;,
\end{equation}
then the moduli space is $\fM/\text{Weyl}(G)$. We parametrize the complexified Cartan algebra $\fh_\bC$ by $u_a$, then $\fM$ is the product of $r$ copies of $T^2$ where $r=\rank G$. Similarly, we introduce variables $\xi_b$ on the complexified Cartan algebra of the flavor group $K$ and fugacities $\zeta_b = e^{2\pi i \xi_b}$.

Second, the one-loop determinants produce a meromorphic $(r,0)$-form $Z_\text{1-loop}(\tau, z, u, \xi)$. In the $\cN{=}(2,2)$ case, for a gauge theory with chiral multiplets transforming in the (possibly reducible) representation $\fR$, such a form is
\begin{equation}
\label{BL1-loop (2,2)}
Z_\text{1-loop} = \Bigg( \frac{2\pi \eta(q)^3 }{ \theta_1(q, y^{-1}) } \Bigg)^r \prod_{\alpha \,\in\, G} \frac{\theta_1(q, x^\alpha) }{ \theta_1(q, y^{-1} x^\alpha) } \; \prod_{\rho \,\in\, \fR} \frac{ \theta_1 (q, y^{R_\rho/2 -1} \zeta^{K_\rho} x^\rho) }{ \theta_1 (q, y^{R_\rho/2} \zeta^{K_\rho} x^\rho)} \; du_1 \cdots du_r \;.
\end{equation}
The first product is over the roots $\alpha$ of $G$, while the second one is over the weights $\rho$ of $\fR$. The elliptic Dedekind and Jacobi functions are defined as
\begin{equation}
\eta(q) = q^\frac1{24} \prod_{n=1}^\infty (1-q^n) \;,\qquad \theta_1(q,y) = -i q^\frac18 y^\frac12 \prod_{n=1}^\infty (1-q^n)(1-yq^n)(1-y^{-1}q^{n-1}) \;.
\end{equation}
Finally, $R_\rho/2$ is the left-moving R-charge (if the axial R-charge is zero, then $R_\rho$ is the vector R-charge) and $K_\rho$ is the flavor weight of the chiral multiplet associated to the weight $\rho$. These charges are constrained by superpotential interactions, and this is the only place where the superpotential appears. If extra (neutral) twisted chiral multiplets $\Sigma_c$ with axial R-charge $R^{(A)}_c$ are present, one should also include the factor
$$
\prod_c \frac{\theta_1 \big(q, y^{- {R^{(A)}_c/2} \,+\, 1} \big) }{ \theta_1 \big(q, y^{-R^{(A)}_c/2}\big) } \;.
$$
In the $\cN{=}(0,2)$ case, for a theory with chiral multiplets in representation $\fR_\text{chiral}$ and Fermi multiplets in representation $\fR_\text{Fermi}$, the meromorphic form is
\begin{equation}
\label{BL1-loop (0,2)}
Z_\text{1-loop} = \Bigg( \frac{2\pi \eta(q)^2}{i} \Bigg)^r \! \prod_{\alpha\,\in\, G} \! i \frac{\theta_1(q,x^\alpha)}{\eta(q)} \! \prod_{\rho \,\in\, \fR_\text{chiral}} \! \frac{i \, \eta(q)}{\theta_1(q, \zeta^{K_\rho} x^\rho)} \! \prod_{\rho \,\in\, \fR_\text{Fermi}} \!\!\! i \frac{\theta_1(q, \zeta^{K_\rho} x^\rho)}{\eta(q)} \; du_1 \cdots du_r \;.
\end{equation}
Note that the only difference between $u$ and $\xi$ is that $u$ will be integrated over. We will sometimes keep $\xi$ implicit in the following formul\ae.

The meromorphic form $Z_\text{1-loop}$ has poles in $u$, along hyperplanes corresponding to all chiral and off-diagonal vector multiplets for $\cN{=}(2,2)$, and to chiral multiplets for $\cN{=}(0,2)$. Each of those multiplets introduces a singular hyperplane $H_i \subset \fM$. We will use the index $i$ for them, and call $Q_i \in \fh^*$ the weight of the multiplet under the gauge group. For the different types of multiplets we have:
\begin{equation}
\label{BLHi}
\begin{array}{rcr@{\ }ll}
\text{vector}_{(2,2)}: & H_i = \Big\{& - z + Q_i(u) &= 0
\pmod{\bZ+ \tau\bZ } \Big\} \;, \qquad\qquad & Q_i = \alpha \\[.5em]
\text{chiral}_{(2,2)}:  & H_i = \Big\{& \frac{R_i}2 z + Q_i(u) + K_i(\xi) &= 0 \pmod{ \bZ+ \tau\bZ } \Big\} \;, &Q_i = \rho \\[.5em]
\text{chiral}_{(0,2)}:  & H_i = \Big\{& Q_i(u) + K_i(\xi) &= 0
\pmod{ \bZ + \tau\bZ } \Big\} \;, &Q_i = \rho \;.
\end{array}
\end{equation}
Note that a single $H_i$ can contain multiple parallel disconnected hyperplanes. We denote by $\sQ = \{Q_i\}$ the set of all charge covectors. Then we define
\begin{equation}
\fM_\text{sing} = \bigcup\nolimits_i H_i
\end{equation}
in $\fM$, and we denote by $\fM_\text{sing}^* \subset \fM_\text{sing}$ the set of isolated points in $\fM$ where at least $r$ linearly-independent hyperplanes meet:
\begin{equation}
\label{BLM sing star}
\fM_\text{sing}^* = \big\{ u_* \in \fM \,\big|\, \text{at least $r$ linearly independent $H_i$'s meet at } u_* \big\} \;.
\end{equation}
Given $u_*\in \fM_\text{sing}^*$, we denote by $\sQ(u_*)$ the set of charges of the hyperplanes meeting at $u_*$: \begin{equation}
\sQ(u_*) = \{  Q_i \,\big|\, u_* \in H_i \} \;.
\end{equation}
Next, one has to choose a generic%
\footnote{Denote by $\Cone_\text{sing}(\sQ) \subset \fh^*$ the union of the cones generated by all subsets of $\sQ$ with $r-1$ elements. Then each connected component of $\fh^* \setminus \Cone_\text{sing}(\sQ)$ is called a \emph{chamber}. By a ``generic'' covector we mean an $\eta \not\in \Cone_\text{sing}(\sQ)$: such $\eta$ identifies a chamber in $\fh^*$.}
non-zero $\eta\in\fh^*$. Then, the elliptic genus is given by the formula:%
\footnote{For a technical reason, one has to assume the following condition on the gauge theory charges: \emph{For any $u_*\in\fM_\text{sing}^*$, the set $\sQ(u_*)$ is  contained in a half-space of $\fh^*$}. A hyperplane arrangement with this property at $u_*$ is called \emph{projective} \cite{BLSzenesVergne}. Notice that if the number of hyperplanes at $u_*$ is exactly $r$, the arrangement is automatically projective. When the condition is not met, one needs to relax the constraints on R- and flavor charges coming from the superpotential, resolve $u_*$ into multiple singularities which are separately projective, and eventually take a limit where the charges are the desired ones.
If at every $u_*$ the number of hyperplanes meeting at $u_*$ is exactly $r$, we call the situation \emph{non-degenerate}.
\label{BLfoot:projective}
}
\begin{equation}
\label{BLeg main formula}
Z_{T^2}(\tau,z,\xi) = \frac1{|\text{Weyl}(G)|} \sum_{u_* \,\in\, \fM_\text{sing}^*} \JKres_{u=u_*}\!\big(\sQ(u_*),\eta \big) \;\; Z_\text{1-loop}(\tau,z,u,\xi) \;.
\end{equation}
Here $\JKres$ is the Jeffrey-Kirwan residue operation, which is explained in detail below. It is locally constant as a function of $\eta$, but it can jump as $\eta$ crosses from one chamber to another. Nonetheless the sum on the right hand side is independent of $\eta$.

\subsubsection{The Jeffrey-Kirwan residue}
\label{BLsssec:torus-JK}

The Jeffrey-Kirwan residue operation has been introduced in \cite{BLJeffreyKirwan}; there are several equivalent formulations available in the literature, and we follow \cite{BLSzenesVergne}. We define the residue at $u_* = 0$; for generic $u_*$ one just shifts the coordinates. Consider $n$ hyperplanes meeting at $u=0\in \bC^r$:
\begin{equation}
H_i = \big\{ u \in \bC^r \,\big|\, Q_i(u)=0 \big\}
\end{equation}
for $i=1,\ldots, n$ and with $Q_i \in (\bR^r)^*$. Here we indicate the set of charges $\sQ(u_*) = \{Q_i\}$ simply by $\sQ_*$: the charges define the hyperplanes $H_i$ and give them an orientation. The set $\sQ_*$ defines a hyperplane arrangement (for further details on hyperplane arrangements see \eg{} \cite{BLOrlikTerao}). The coefficients defining the hyperplanes are all real, \ie{} we are dealing with a complexified central arrangement. A residue operation is a linear functional on the space of meromorphic $r$-forms that are holomorphic on the complement of the arrangement, such that it annihilates exterior derivatives of  rational $(r-1)$-forms.

Take a  meromorphic $r$-form $\omega$ defined in a neighborhood $U$ of $u=0$, and holomorphic on the complement of $\bigcup_i H_i$. When $n=r$, we can define the residue of $\omega$ at $u=0$ by its integral over $\prod_{i=1}^r \cC_i$, where each $\cC_i$ is a small circle around $H_i$ (and the overall sign depends on the order of the $H_i$'s). This stems from the fact that the homology group $H_r \big( U \setminus \bigcup_{i=1}^r H_i, \bZ \big) = \bZ$, and therefore there is a natural generator defined up to a sign. When $n>r$ however, $H_r \big( U \setminus \bigcup_{i=1}^n H_i,\bZ \big) = \bZ^{c_{n,r}}$ with $c_{n,r}>1$, and it is imperative to specify the precise cycle to choose.

For a projective arrangement and given an $\eta\in (\bR^r)^*$, the Jeffrey-Kirwan residue is the linear functional defined by the conditions:
\begin{equation}
\label{BLSVcondition}
\JKres_{u=0}(\sQ_*,\eta) \, \frac{\dd Q_{j_1}(u)}{Q_{j_1}(u)} \wedge \cdots \wedge \frac{\dd Q_{j_r}(u)}{Q_{j_r}(u)} =
\begin{cases}
\sign \det (Q_{j_1} \dots Q_{j_r}) & \text{if } \eta\in \Cone(Q_{j_1} \ldots Q_{j_r}) \\
0 & \text{otherwise}
\end{cases}
\end{equation}
where $\Cone$ denotes the cone spanned by the vectors in the argument. We can rewrite it as
\begin{equation}
\label{BLSVcondition2}
\JKres_{u=0}(\sQ_*,\eta) \,  \frac{\dd u_1 \wedge \dots \wedge \dd u_r}{Q_{j_1}(u) \cdots Q_{j_r}(u)} =
\begin{cases}
\dfrac1{|\det (Q_{j_1} \dots Q_{j_r})|} & \text{if } \eta\in \Cone(Q_{j_1} \ldots Q_{j_r}) \\[1em]
0 & \text{otherwise}
\end{cases}
\end{equation}
after choosing coordinates $u_a$ on $\fh$.
The definition \eqref{BLSVcondition}-\eqref{BLSVcondition2} is in general vastly over-determined since there are many relations between the forms $\bigwedge_{\alpha=1}^r \dd Q_{j_\alpha}/Q_{j_\alpha}$, but it has been proven in \cite{BLBrionVergne} that \eqref{BLSVcondition} is consistent, and it is given by an integral over an explicit cycle.
A constructive definition of the JK residue, as a sum of iterated standard residues, has been given in \cite{BLSzenesVergne} and reviewed in \cite{BLBenini:2013xpa}.

In the simplest case of $r=1$, applying~\eqref{BLSVcondition2} one finds
\begin{equation}
\JKres_{u=0}\big(\{q\}, \eta\big) \; \frac{\dd u}u = \begin{cases} \sign(q) & \text{if } \eta q > 0 \;, \\ 0 & \text{if } \eta q < 0 \;. \end{cases}
\end{equation}
Substituting into~\eqref{BLeg main formula}, one finds that the elliptic genus in the rank-1 case is given by
\begin{equation}
\label{BLeg rank-1 formula}
Z_{T^2} = \frac1{|\text{Weyl}(G)|} \sum_{u_+ \,\in\, \fM_\text{sing}^+} \frac1{2\pi i} \oint_{u=u_+} Z_\text{1-loop}
= - \frac1{|\text{Weyl}(G)|} \sum_{u_- \,\in\, \fM_\text{sing}^-} \frac1{2\pi i} \oint_{u=u_-} Z_\text{1-loop}
\end{equation}
by choosing $\eta = 1$ and $\eta = -1$ respectively, where $\fM_\text{sing}^{+(-)}$ is the subset of singularities with positive (negative) associated charge. In other words, one should sum (with sign) the residues coming from fields of either positive or negative charge. As the sum of all residues vanishes, the result is independent of the choice of $\eta$.

\subsection{The derivation}
\label{BLssec:torus-calc}

Let us briefly sketch how the formula~\eqref{BLeg main formula} is derived. The formula and its derivation was first obtained in \cite{BLBenini:2013nda, BLBenini:2013xpa} and then extended in \cite{BLHori:2014tda, BLBenini:2015noa, BLClosset:2015rna, BLClosset:2015ohf}.

The standard localization procedure reduces the path-integral to an integral over the BPS supermanifold of zero-modes, schematically
\begin{equation}
Z_{T^2} = \int_{\cM_\text{BPS}} \cD\varphi_0 \; e^{-S[\varphi_0]} \; Z_\text{1-loop}[\varphi_0] \;.
\end{equation}
In the present case, it turns out that $\cM_\text{BPS}$ contains fermionic zero-modes as well as singular loci with extra bosonic zero-modes. With a suitable regulator, the two problems solve each other and one is left with a contour integral within the bosonic component of $\cM_\text{BPS}$.

Solving the bosonic BPS equations---read off from \eqref{BLSUSY var chiral}, \eqref{BLSUSY var Fermi} and \eqref{BLSUSY var vector}---chiral and Fermi multiplets are set to zero and one finds the moduli space of flat connections on $T^2$ modulo gauge transformations. In the simplest case of a gauge group $G$ with connected and simply-connected non-Abelian part (and arbitrary Abelian part):
\begin{equation}
\cM_\text{BPS} \Big|_\text{bos} = \fM / \text{Weyl}(G) \;,\qquad\qquad \fM = \fh_\bC / (\Gamma_\fh + \tau \Gamma_\fh) \;.
\end{equation}
Here $\fM$ is $r$ copies of $T^2$ and we parametrize it by the variables $u$.

Around each of the BPS configurations, besides the bosonic zero-modes that parametrize $\cM_\text{BPS} \big|_\text{bos}$ there are also fermionic zero-modes and together they form complete supermultiplets. Each bosonic zero-mode is paired with a fermionic zero-mode coming from the Cartan gaugini (in $\cN{=}(0,2)$ notation). The right-moving Cartan gaugini $\lambda$ are not lifted because they are charged only under the R-symmetry, and we cannot turn on a flat connection for the R-symmetry without breaking supersymmetry. In a gauge where $A_\mu$ is constant, we can identify (up to unimportant coefficients) $u = A_{\bar w}$, then the fermionic zero-modes are constant $\lambda^+$, $\bar\lambda^+$. We can close the supersymmetry algebra ``off-shell'' if we introduce an auxiliary bosonic zero-mode $D_0$, which is the constant profile of $D$. The supersymmetry algebra follows from~\eqref{BLSUSY var vector}:
\begin{equation}\begin{aligned}
\label{BLzeromodealgebra}
Q u &= 0 \;,\qquad& Q \bar u &= \tfrac12 \bar\lambda^+ \;,\qquad& Q \bar\lambda^+ &= 0 \;,\qquad& Q \lambda^+ &= -D_0 \;,\qquad& QD_0 &= 0 \\
\wt Q u &=0 \;,\qquad& \wt Q \bar u &= - \tfrac12 \lambda^+ \;,\qquad& \wt Q \lambda^+ &=0 \;,\qquad& \wt Q \bar\lambda^+ &= - D_0 \;,\qquad& \wt Q D_0 &= 0 \;.
\end{aligned}\end{equation}
We used that the flux is zero and all modes are constant on $T^2$.

The one-loop determinants, keeping the dependence on $D_0$ which serves as a regulator of the final expression, are easy to compute using the Hamiltonian definition. For instance, for an $\cN{=}(2,2)$ chiral multiplet of charge weight $\rho$ one finds
\begin{equation}
\label{BL1loopDexample}
\cZ^{(2,2)\text{ chiral}}(u, \bar u,D_0) = \prod_{m,n} \frac{\big( m+n\tau+(1- \tfrac R2 ) z - \rho(u) \big) \big( m + n \bar\tau + \frac R2 \bar z + \rho(\bar u) \big)}{ \big| m + n\tau + \frac R2 z + \rho(u)  \big|^2 + i \rho(D_0)} \;.
\end{equation}
The determinants reduce to $Z_\text{1-loop}$ in \eqref{BL1-loop (2,2)} and~\eqref{BL1-loop (0,2)} for $D_0=0$. Then they develop singularities on $\fM$ along the hyperplanes $H_i$ because extra bosonic zero-modes appear, but there are no divergences for generic $D_0 \neq 0$.

The last step is to integrate over the moduli space of BPS configurations. For simplicity, let us restrict to the case that $G$ has rank $1$. Since all action terms are $\cQ$-exact, we perform localization simply by sending to zero all interactions. The singular hyperplanes---which in this case are just points---arise because we take the limit $\se \to 0$, where $\se$ is the gauge coupling. In the weakly-interacting theory the contribution from the neighborhood of a singular point $u_* \in \fM_\text{sing}$ where there are $M$ quasi-zero-modes $\phi_i$---whose charges $Q_i$ have the same sign by the assumption in \autoref{BLfoot:projective}---is roughly:
\begin{equation}
I = \int d^{2M}\phi\, \exp\bigg[ - \sum\nolimits_i |Q_i(u - u_*)|^2 |\phi_i|^2 - \frac{\se^2}2 \Big( \zeta - \sum\nolimits_i Q_i |\phi_i|^2 \Big)^2 \bigg] \;,
\end{equation}
where $\zeta$ is the FI term. The second term comes from the D-term potential and it ensures that the integral is convergent, even at $u=u_*$. By rescaling $\phi_i \to \big| Q_i \se \big|^{-1/2} \phi_i$ we can find an upper bound $|I| \lesssim C/\se^M$ for some constant $C$. Therefore, we can split the integral over $\fM$ into two pieces, removing from $\fM$ an $\varepsilon$-neighborhood $\Delta_\varepsilon$ of $\fM_\text{sing}$. The integral over $\Delta_\varepsilon$ is bounded by $\varepsilon^2/\se^M$ up to constants, 
therefore in a scaling limit $\se, \varepsilon \to 0$ in which $\varepsilon^2/\se^M\to0$ as well, it does not contribute. We thus have
\begin{equation}
\label{BLZmStartingExp}
Z_{T^2} = \frac1{|\text{Weyl}(G)|} \lim_{\se, \varepsilon\to 0} \int_{\fM\setminus\Delta_\varepsilon} \hspace{-1em} d^2u \int_{\bR + i \eta} \hspace{-1em} dD_0 \int d\lambda^+ d\bar\lambda^+ \; \cZ(u,\bar u, \lambda^+ , \bar\lambda^+ , D_0) \;.
\end{equation}
We have restored the dependence on the zero-mode $D_0$ of the auxiliary field, since it will be used as a regulator momentarily. Then $\cZ$ is the effective partition function obtained by integrating out all massive modes. Setting $\lambda^+ = \bar\lambda^+ = 0$ and in the $\se\to0$ limit it gives what we have written in~\eqref{BL1loopDexample}, and further setting $D_0=0$ it gives the one-loop determinants in~\eqref{BL1-loop (2,2)} and~\eqref{BL1-loop (0,2)}. The function $\cZ$ is holomorphic in $D_0$ around the origin as long as $u\not\in \Delta_\varepsilon$. Therefore we have the freedom to shift the real integration contour on the complex $D_0$-plane along the imaginary direction, as long as this shift is small: in~\eqref{BLZmStartingExp} we have called $\eta$ such a shift.

The partition function $\cZ(u,\bar u, \lambda^+, \bar\lambda^+, D_0)$ depends on the gaugino zero-modes because of the Lagrangian couplings $\lambda\psi \phi$ to the matter fields we have integrated out. As noticed in \cite{BLBenini:2015noa, BLClosset:2015rna}, the dependence is fixed by supersymmetry. From the algebra~\eqref{BLzeromodealgebra} it follows
\begin{equation}
0 = \wt Q\cZ = - \frac{\lambda^+}2 \, \parfrac{\cZ}{\bar u} - D_0 \parfrac{\cZ}{\bar\lambda^+} \qquad\Rightarrow\qquad
\parfrac{^2 \cZ}{\lambda^+ \partial \bar\lambda^+} = - \frac1{2D_0} \, \parfrac\cZ{\bar u} \Big|_{\lambda^+ = \bar\lambda^+ = 0} \;.
\end{equation}
The integral over the fermionic zero-modes produces a total derivative, and by Stoke's theorem we obtain the contour integral expression
\begin{equation}
Z_{T^2} = \frac1{|\text{Weyl}(G)|} \lim_{\se,\varepsilon\to0} \int_{\partial\Delta_\varepsilon} \hspace{-.8em} du \int_{\bR + i\eta} \hspace{-.2em} \frac{dD_0}{D_0} \, \cZ(u,\bar u, D_0) \;.
\end{equation}
Consider a component of $\partial\Delta_\varepsilon$ around a point $u_* \in \fM_\text{sing}$. Suppose that we have chosen $\eta>0$. From the unregularized expression in~\eqref{BL1loopDexample} of the chiral one-loop determinant, we see that the poles in the complex $D_0$-plane are in the half $\rho \im D_0 > 0$. If $\rho<0$ the poles are in the negative half-plane. As $\varepsilon\to0$ they collapse towards $D_0=0$, because the term in absolute value is of order $\varepsilon$ on the contour $\partial\Delta_\varepsilon$, however the contour $\bR + i \eta$ is safely far from them. The $D_0$-integral remains finite as $\varepsilon\to0$, and then the $u$-integral vanishes because its contour shrinks. On the contrary, if $\rho>0$ the poles are in the upper $D_0$-half-plane and, as $\varepsilon \to 0$, they would cross the contour $\bR + i \eta$. To avoid that, we shift the contour to $\bR - i \eta$ and we collect minus the residue at $D_0 = 0$. As before, the integral along $\bR - i\eta$ does not yield any contribution as $\varepsilon \to 0$. Minus the residue at $D_0 = 0$, though, gives
$$
\lim_{\se,\varepsilon\to0} \int_{\partial\Delta_\varepsilon} \hspace{-.2cm} du\; \cZ(u, \bar u, 0) =  \Res_{u=u_*} Z_\text{1-loop}(u) \;,
$$
up to constants that can be fixed in one known example. Had we chosen $\eta<0$ instead, a similar argument goes through and one obtains minus the residue at $u=u_*$ if $\rho<0$, zero if $\rho>0$.

We reach the conclusion that for $\eta>0$ we collect the residues of $Z_\text{1-loop}(u)$ at the points $u_* \in \fM_\text{sing}^+$ corresponding to chiral fields with positive charges, while for $\eta < 0$ we collect minus the residues at the points $u_* \in \fM_\text{sing}^-$ corresponding to chiral fields with negative charges, reproducing~\eqref{BLeg rank-1 formula}. The generic higher-rank case is much more intricate, but conceptually very similar, and it leads to the JK residue. We refer the reader to the references for details.

\subsection{Extensions and applications}
\label{BLssec:torus-more}

The localization formula we presented has been generalized in many ways. The authors of \cite{BLMurthy:2013mya, BLAshok:2013pya} have considered gauge theories with St\"uckelberg fields, which in the IR may realize non-compact sigma models a prototype of which is the $SL(2,\bR)/U(1)$ (cigar) coset. The resulting genus is a Jacobi-like form that is non-holomorphic in the modular parameter $\tau$ of the torus, with mock modular behavior. The reason is that those models have a continuous spectrum above a threshold, even when equivariant parameters are turned on, and the density of states of bosons and fermions in the continuum need not be equal.

The prototype $\cN{=}(2,2)$ GLSM is the one introduced by Hori and Kapustin in \cite{BLHori:2001ax}. The model has a $U(1)$ vector field $V$, a chiral multiplet $\Phi$ of charge $1$, and a chiral St\"uckelberg field $P$ that transforms as
\begin{equation}
P \,\to\, P + i \Lambda
\end{equation}
under super-gauge transformations $V \to V - i \Lambda + i \bar\Lambda$. Moreover the imaginary part of $p$ (the scalar in $P$) is periodic, $p \sim p + 2\pi i$. The action is
\begin{equation}
S = \frac1{4\pi} \int d^2w\, d^4\theta \Big[ \bar\Phi e^V \Phi + \frac k4 \big( P + \bar P + V \big)^2 - \frac1{2\se^2} \bar \Sigma \Sigma \Big] \;,
\end{equation}
where $k>0$ and $\Sigma$ is the field-strength twisted chiral multiplet. When written in components, the imaginary part of $p$ acts as a standard St\"uckelberg field that gives the photon a mass, while the real part is a dynamical FI term. Integrating out the massive photon and after some RG flow, one obtains the $SL(2,\bR)_k/U(1)$ SCFT with central charge $c= 3 + \frac 6k$, which in the large $k$ limit has a description as the NLSM with cigar target
\begin{equation}
ds^2 = 2k \big( du^2 + \tanh^2 u \, d\psi^2 \big)
\end{equation}
with $\psi \cong \psi + 2\pi$, and a non-trivial background dilaton \cite{BLWitten:1991yr}.

When computing the elliptic genus with localization, the presence of the field $P$ introduces many differences. First, the action involving $P$ is naively $\cQ$-exact but in fact it gives rise to a non-trivial boundary term in field space \cite{BLHori:2001ax}, therefore the final answer will depend on $k$. Luckily, $P$ appears quadratically in the action and so its path-integral can be computed exactly.

Second, the field $P$ has fermionic zero-modes $\chi^-$, $\bar\chi^-$ coupled to the gaugino zero-modes $\lambda^+$, $\bar\lambda^+$. Therefore, when integrating over the fermionic zero-modes they are absorbed at tree level and we simply generate a constant factor---instead of a total derivative as before. The integration over the bosonic zero-modes of $P$ and the massive modes is standard, and the details can be found in \cite{BLAshok:2013zka, BLMurthy:2013mya, BLAshok:2013pya}. A point to note is that the imaginary part of $p$ is a periodic scalar which admits winding modes: $p_2(w + 1) = p_2(w) + 2\pi n$, $p_2(w+\tau) = p_2(w) + 2\pi m$. The path-integral over $P$ then gives
\begin{equation}
Z_P = \frac{k}{D_0 \tau_2} \, \frac{\theta_1(q,y)}{\eta(q)^3} \sum_{m,n\in\bZ} \exp \bigg[ - \frac{\pi k}{\tau_2} \Big( m + n\tau + u + \frac zk \Big)\Big( m + n\bar\tau + \bar u + \frac zk \Big) \bigg] \;.
\end{equation}
The one-loop determinant of the chiral multiplet is also modified. The UV left-moving R-symmetry $J_\mu$ is anomalous and the IR conserved superconformal R-symmetry is obtained by mixing with the gauge-invariant quantity $A_\mu + \partial_\mu p_2$. The R-symmetry background experienced by $\Phi$ thus depends on the winding numbers $(m,n)$:
\begin{equation}
Z_\Phi = y^{\rho n} \frac{\theta_1(q, y^{R/2-1} x^\rho) }{ \theta_1(q, y^{R/2} x^\rho) } \;,
\end{equation}
where $\rho$ is the gauge charge.
Putting everything together, and generalizing to multiple chiral multiplets $\Phi_i$, one finds the formula:
\begin{equation}
\label{BLeg Stuckelberg}
Z_{T^2} = k \int_{T^2} \frac{d^2u}{\tau_2} \prod_i \frac{\theta_1(q, y^{\frac{R_i}2-1} x^{\rho_i} ) }{ \theta_1(q, y^{\frac{R_i}2} x^{\rho_i})} \sum_{m,n\in\bZ} y^{n \sum_i \rho_i} \, e^{-\frac{\pi k}{\tau_2} \left( m + n\tau + u + \frac zk \right) \left( m + n \bar\tau + \bar u + \frac zk \right)} \;.
\end{equation}
The cigar coset theory corresponds to a single field with $\rho=1$ and $R=0$.
Notice that in the presence of multiple chiral multiplets with coincident poles, the integral is divergent and it should be regularized with care.

The expression above is not holomorphic in $\tau$: it is the product of a usual Jacobi form and an Appell-Lerch sum. The latter are intimately related to a very interesting class of functions called mock modular forms \cite{BLZwegers, BLZagier2007, BLDabholkar:2012nd}. Their key feature is that they transform as modular forms, but they suffer from a holomorphic anomaly.

The example above can obviously be generalized to more Abelian groups and matter contents. As an interesting application, \cite{BLHarvey:2014nha} studies the equivariant elliptic genera of a class of gravitational instantons that are given by hyper-K\"ahler four-manifolds of asymptotically locally flat (ALF) type. The simplest of these spaces is the Taub-NUT manifold, while more general constructions yield the multi-center $A_k$ ALF spaces of Gibbons and Hawking \cite{BLGibbons:1979zt} and other spaces. Taub-NUT can be obtained from an $\cN{=}(4,4)$ GLSM consisting of a $U(1)$ vector multiplet, a hypermultiplet of charge $1$ and a neutral St\"uckelberg hypermultiplet \cite{BLTong:2002rq}.%
\footnote{In $\cN{=}(2,2)$ notation, the model has a vector multiplet $V$, two neutral chiral multiplets $\Phi$, $\Psi$, two chiral multiplets $Q$, $\wt Q$ of charges $\pm1$, and a St\"uckelberg chiral multiplet $P$. There is also a superpotential $W = \wt Q \Phi Q + \Phi \Psi$.}
The model has $SU(2)^3$ R-symmetry and $U(1)_f$ flavor symmetry (one $SU(2)$ corresponds to rotations of $\bR^3$, while $U(1)_f$ rotates the circle). The coupling $k$ controls the size of the asymptotic circle, and in the $k\to\infty$ limit the model turns into $\bC^2$. Localization in this case gives%
\footnote{In this model the R-symmetry current is not anomalous, therefore the St\"uckelberg field has no R-charge.}
\begin{equation}
\label{BLeg Taub-NUT}
Z_{T^2} = k \int_\bC \frac{d^2u}{\tau_2} \; \frac{\theta_1( q, x y \zeta_1) \, \theta_1(q, x y^{-1} \zeta_1) }{ \theta_1(q, x \zeta_1\zeta_2) \, \theta_1(q, x \zeta_1 \zeta_2^{-1}) } \, e^{- \frac{k\pi}{\tau_2} \left| u \right|^2} \;,
\end{equation}
where we have used the sum over winding sectors to ``unfold'' the integral over the whole complex plane. Here $\zeta_1$, $\zeta_2$, $y$ are fugacities for $U(1)_f$ and for two left-moving R-symmetries, respectively. In the $k\to\infty$ limit one obtains the equivariant elliptic genus of $\bC^2$, which equals the integrand of~\eqref{BLeg Taub-NUT} evaluated at $u=0$ and it is holomorphic in $\tau$.

Physically, the equivariant deformations produce a potential roughly proportional to the length of the orbits of the associated $U(1)$ actions. In the case of Taub-NUT the potential produced by $\zeta_1 = e^{2\pi i \xi_1}$ is
\begin{equation}
V = \frac{\xi_1^2}{\frac1k + \frac1{\left| \vec x \right|^2}} \;,
\end{equation}
where $\vec x$ is a coordinate on $\bR^3$. Around the origin ($\vec x=0$) the potential is quadratic and it gives a discrete IR spectrum, while at large $|\vec x|$ it is a constant allowing for a continuous spectrum of scattering states. The latter are responsible for the loss of holomorphy in $\tau$.

One can also generalize the elliptic genus to the \emph{twining genera}: in the minimal $\cN{=}(0,2)$ case, given a theory with a discrete symmetry $g$ that commutes with the right-moving supersymmetry algebra, they are defined as
\begin{equation}
Z_g(\tau,u) = \Tr_\text{R} \, (-1)^F \, g \, q^{H_L} \, \bar q^{H_R} \, \prod\nolimits_a x_a^{K_a}
\end{equation}
with an insertion of $g$ into the trace. These objects decompose into characters of the discrete symmetry group, and therefore contain valuable information about the spectrum of the theory. As shown in \cite{BLHarrison:2013bya}, it is easy to extend the localization computation to the twining genera. If the symmetry element $g$ acts on chiral and Fermi multiplets as
\begin{equation}
g\, \Phi_i = e^{2\pi i \alpha_i} \Phi_i \;,\qquad\qquad g\, \Lambda_i = e^{2\pi i \beta_i} \Lambda_i \;,
\end{equation}
the corresponding one-loop determinants are modified to
\begin{equation}
Z_{\Phi_i} = \frac{i \, \eta(q) \, e^{\pi i \alpha_i} }{ \theta_1(q, e^{2\pi i \alpha_i} x^{\rho_i}) } \;,\qquad\qquad Z_{\Lambda_i} = \frac{i \, \theta_1(q, e^{2\pi i \beta_i} x^{\rho_i}) }{ \eta(q) \, e^{\pi i \beta_i} } \;.
\end{equation}
Then one takes the JK residue at the (possibly shifted) poles. The formula has been used for instance in \cite{BLHarrison:2013bya} to study the twining genera of $\cN{=}(0,4)$ NLSM with K3 target under the action of elements of $M_{24}$, the largest Mathieu group, in connection with the so-called ``moonshine conjectures''.

For $\cN{=}(0,2)$ gauge theories, the elliptic genus is one of the few quantities that can be computed non-pertubatively via localization.%
\footnote{For deformations of $\cN{=}(2,2)$ gauge theories, one can also compute correlators on $S^2$ \cite{BLClosset:2015ohf}.}
Therefore it constitutes a crucial test of conjectured IR dualities among different gauge theories. One example are the trialities of \cite{BLGadde:2013lxa}, which will be discussed in \autoref{BLssec:dual-02}.

To conclude, we mention two recent applications of the elliptic genus of gauge theories. One is in the context of ``AGT correspondences'' \cite{BLAlday:2009aq} (see also \volcite{TA}). Through the compactification of M5-branes on $T^2 \times \cM_4$, where $\cM_4$ is an arbitrary four-manifold, one can try to relate the elliptic genus of certain $\cN{=}(0,2)$ gauge theories (which depend on $\cM_4$) and the Vafa-Witten partition function \cite{BLVafa:1994tf} of 4d $\cN{=}4$ SYM on $\cM_4$. This program has been initiated in \cite{BLGadde:2013sca}. Another application is to interpret the partition functions of certain 6d SCFTs as the generating functions of the elliptic genera of their BPS strings \cite{BLHaghighat:2013gba} (see \volcite{KL}).


\section{Dualities}
\label{BLsec:dual}

The supersymmetric observables exactly computed in previous sections provide powerful means to test dualities.  We review in this section a few such applications of the sphere partition function and the elliptic genus. First we consider mirror symmetry in \autoref{BLssec:dual-mirror}, comparing sphere partition functions of 2d $\cN{=}(2,2)$ GLSMs which flow to mirror Calabi-Yau manifolds. In \autoref{BLssec:dual-seiberg} we discuss Seiberg-like dualities between 2d $\cN{=}(2,2)$ gauge theories with $U(N)$ gauge groups and (anti)fundamental matter. Then in \autoref{BLssec:dual-variants} we list generalizations, while \autoref{BLssec:dual-02} describes 2d $\cN{=}(0,2)$ dualities which have been checked using the elliptic genus.

\subsection{Mirror symmetry}
\label{BLssec:dual-mirror}

Numerous NLSMs with Calabi-Yau target spaces can be realized as the low-energy limit of gauged linear sigma models. The Calabi-Yau moduli space coincides with the conformal manifold of the NLSM, which is typically spanned by exactly marginal (chiral and twisted chiral) operators of the GLSM.  Twisted chiral operators alter the (complexified) K\"ahler structure of the Calabi-Yau, while chiral operators alter its complex structure.  Both the moduli space of K\"ahler structure deformations and that of complex structure deformations are K\"ahler manifolds, whose metric derives from a K\"ahler potential.

As reviewed in \volcite{MO}, these K\"ahler potentials can be efficiently computed from GLSM sphere partition functions \cite{BLJockers:2012dk, BLGomis:2012wy, BLGerchkovitz:2014gta, BLGomis:2015yaa}. Placing a GLSM on the sphere preserves either the vector or the axial R-symmetry.  These two choices lead to distinct partition functions $Z_A$ and $Z_B$, computed in \autoref{BLsec:sphere} and \autoref{BLssec:curved-twisted}.  Both are independent of the superrenormalizable gauge coupling hence are invariant under the RG flow.  The first one depends only on twisted chiral parameters (FI parameters, theta angles, twisted masses, vector R-charges) and gives the K\"ahler potential $\mathcal{K}_K$ on the moduli space of K\"ahler structure deformations, while the second one depends only on chiral parameters (the superpotential) and gives the K\"ahler potential $\mathcal{K}_C$ on the moduli space of complex structure deformations:
\begin{equation}
  Z_A = e^{-\mathcal{K}_K}
  \qquad\text{and}\qquad
  Z_B = e^{-\mathcal{K}_C} \;.
\end{equation}
This streamlines the extraction of genus zero Gromov-Witten invariants from $\mathcal{K}_K$ (see \cite{BLJockers:2012dk, BLSharpe:2012ji, BLHonma:2013hma, BLHalverson:2013qca}).  Remarkably, most known Calabi-Yau manifolds are paired such that the moduli space of complex structure deformations of one is identical to the moduli space of K\"ahler structure deformations of the other and viceversa. The manifolds are called mirrors of each other.  The interchange of $\mathcal{K}_K$ and $\mathcal{K}_C$ can be shown by proving that $Z_A$ of one GLSM is equal to $Z_B$ of the other.

Mirror symmetry generalizes to GLSMs whose low-energy limit is an NLSM on a K\"ahler manifold (with non-negative first Chern class) rather than a Calabi-Yau manifold: a large class of GLSMs have Landau-Ginzburg models as their mirrors \cite{BLHori:2000kt}.  In this section, we prove following~\cite{BLGomis:2012wy,BLDoroud:2013pka} (an example was worked out in \cite{BLBenini:2012ui}) that $Z_A$ of a GLSM is equal to $Z_B$ of the Landau-Ginzburg mirror proposed by Hori and Vafa.  In fact, to avoid switching back and forth between the backgrounds that preserve $\mathfrak{su}(2|1)_A$ and $\mathfrak{su}(2|1)_B$, we apply to the Landau-Ginzburg model the involution of the superconformal algebra which exchanges these two subalgebras and exchanges vector/chiral multiplets with twisted vector/chiral multiplets.  As a result, we wish to write $Z_A$ of a GLSM (computed in \autoref{BLsec:sphere}) as $Z_A$ of a Landau-Ginzburg model of twisted chiral multiplets (computed in \autoref{BLssec:curved-twisted}).

The key mathematical identity is (omitting irrelevant constant factors)%
\footnote{The Hankel contour goes around the cut of $s^{-b-1}$ along the negative real axis, at a constant distance $\epsilon>0$ above and below it. To prove the second equality, either redefine $Y = -\log t$ and $\bar Y = -\log s$, or decompose $Y = x + iy$ and recognize that the last expression is the Fourier transform $2\pi \int_{-\infty}^{+\infty} dx\, e^{(b-a)x} J_{a+b}(2 e^{-x})$ of the Bessel function of the first kind.}
\begin{equation}
  \frac{\Gamma(a)}{\Gamma(1+b)}
  = \int_0^{\infty} \dd{t} e^{-t} t^{a-1} \int_{\text{Hankel}} \dd{s} e^{s} s^{-b-1}
  = \int_{-\infty}^{\infty} \int_{-\pi}^{\pi} \dd[^2]{Y} \exp\Bigl( -e^{-Y} - a Y + e^{-\bar{Y}} + b \bar{Y} \Bigr) \,,
\end{equation}
where $Y \cong Y + 2\pi i$ is periodic and $a+b \in \mathbb{Z}$. This identity applies readily to the one-loop determinant \eqref{BLZ1l} of a chiral multiplet:
\begin{equation}
  Z_\text{1-loop}^\text{chiral}
  = \prod_{w} \frac{\Gamma\bigl(\frac{\Rcharge}{2}-\I r\twmass-\frac{\flavorflux}{2}-\I r w(\sigma) -\frac{w(\flux)}{2}\bigr)}{\Gamma\bigl(1-\frac{\Rcharge}{2}+\I r\twmass-\frac{\flavorflux}{2}+\I r w(\sigma) -\frac{w(\flux)}{2}\bigr)}
  = \prod_{w} \int \dd[^2]{Y_w} e^{-\frac{\Rcharge}{2}(Y_w+\bar{Y}_w)} \, e^{-4\pi\I r\twsupo_w-4\pi\I r\overline{\twsupo}_w}
\end{equation}
where the twisted superpotential~$\twsupo$ involves the bottom component $\Sigma=r\sigma-\I\flux/2$ of the field strength twisted chiral multiplet:
\begin{equation}
  4\pi\I r\twsupo_w = e^{-Y_w} - \bigl( \I r\twmass + \tfrac{\flavorflux}{2} + \I w(\Sigma) \bigr) Y_w \;.
\end{equation}
For non-zero R-charges, the weight factor $\exp [ -\frac{\Rcharge}{2}(Y_w+\bar{Y}_w) ]$ is naturally absorbed in the integration measure (up to a constant) by using the variables $\widetilde{X}_w = \exp (-\frac{\Rcharge}{2}Y_w )$.

Translating the full partition function of a GLSM to the variables $Y$ and~$\Sigma$ yields
\begin{equation}
\label{BLZGLSMasLG}
  Z_\text{GLSM}
  = \int \dd[^2]{\Sigma}
  \prod_{\alpha>0} \biggl[ e^{\pi\alpha(\Sigma-\bar{\Sigma})} \alpha(\Sigma) \, \alpha(\bar{\Sigma}) \biggr]
  \prod_{I,w} \biggl[ \int \dd[^2]{Y^I_w} e^{-\frac{\Rcharge_I}{2}(Y^I_w+\bar{Y}^I_w)} \biggr] \,
  e^{-4\pi\I r\twsupo-4\pi\I r \overline{\twsupo}}
\end{equation}
where $\Sigma$ has a GNO quantized imaginary part, we dropped a factor of $r^{c/3}$ and numerical constants, and where
\begin{equation}\label{BLmirrortwsupo}
  4\pi\I r\twsupo
  = \sum_{\ell}\Bigl[ \bigl(\vartheta_{\ell}+2\pi\I\zeta_{\ell}\bigr) \Tr_{\ell} \Sigma \Bigr]
  + \sum_{I,w} \Bigl[ e^{-Y^I_w} - \bigl( \I r\twmass_I + \tfrac{\flavorflux_I}{2} + \I w(\Sigma) \bigr) Y^I_w \Bigr] \;.
\end{equation}
This twisted superpotential was found much earlier \cite{BLHori:2000kt} to be generated by vortices.

Equation \eqref{BLZGLSMasLG} is the $Z_A$ partition function \eqref{BLtwistedchiralZ} of a Landau-Ginzburg model with twisted chiral multiplets $\Sigma$ and $Y$ in a certain target space (whose volume form leads to the non-trivial integration measure) and subject to the twisted superpotential \eqref{BLmirrortwsupo}, or equivalently the $Z_B$ partition function of a model with chiral multiplets.  We can go further in the case of Landau-Ginzburg models mirror to Abelian GLSMs: then~$\Sigma$ appears only linearly in the twisted superpotential hence plays the role of a Lagrange multiplier.  Integrating it out yields delta function constraints between the variables $Y^I$:
\begin{equation}
\label{BLYQeqtau}
  \I\vartheta_{\ell}-2\pi\zeta_{\ell} + \sum_I Y^I Q_I^{\ell} \,\in\, 2\pi\I\mathbb{Z} \;.
\end{equation}
Here $Q_I^{\ell}$~denotes the charge of the $I$-th chiral multiplet under the $\ell$-th gauge group and we have dropped the index~$w$ as each irreducible representation is one-dimensional.  Each of the $N$ constraints \eqref{BLYQeqtau} eliminates one of the $N_f$ twisted chiral multiplets $Y^I$.  The twisted superpotential reads
\begin{equation}
  4\pi\I r\twsupo
  = \sum_{I=1}^{N_f} \biggl[ e^{-Y^I} - \Bigl(\I r\twmass_I+\frac{\flavorflux_I}{2} \Bigr) Y^I \biggr]_{\eqref{BLYQeqtau}}
\end{equation}
and the integration measure reduces to (up to permutations of the $Y^I$'s)
\begin{equation}
  \prod_{I=1}^{N_f-N} \int \dd[^2]{Y^I} e^{-\frac{\Rcharge_I'}{2} \left( Y^I+\bar{Y}^I \right) }
\end{equation}
where $\Rcharge_I'$~are certain combinations of R-charges, most conveniently determined by mixing the R-symmetry with gauge symmetries so that R-charges of eliminated twisted chiral multiplets vanish.  Switching to variables $\widetilde{X}^I=\exp (-\frac{\Rcharge'}{2}Y^I )$ yields a Landau-Ginzburg model whose target space is flat, but with a conical singularity at $\widetilde{X}=0$.

Let us consider as an example the quintic hypersurface in $\mathbb{CP}^4$.  We start with a $U(1)$ GLSM with $5$ chiral multiplets $X_i$ of gauge charge $+1$ and R-charge $\Rcharge$ and a chiral multiplet $P$ of gauge charge $-5$ and R-charge $\Rcharge_P=2-5\Rcharge$, with a superpotential $\supo=P\,G_5(X)$ where $G_5$~is a generic homogeneous polynomial of degree~$5$.  The parameter $\Rcharge$ mixes the R-symmetry with the gauge symmetry.  Following the steps above, we introduce twisted chiral fields $Y_1,\ldots,Y_5,Y_P$ then integrate out $\Sigma$ and obtain the constraint \eqref{BLYQeqtau}, namely $5 Y_P = 2\pi\zeta-\I\vartheta + Y_1 + \ldots + Y_5 \pmod{2\pi\I}$.  For convenience, mix the R-symmetry with the gauge symmetry to set $\Rcharge\to 2/5$ so that $\Rcharge_P=0$.  The variables $\widetilde{X}_i=\exp(-Y_i/5)\in\mathbb{C}^*/\mathbb{Z}_5$ absorb the integration measure, at the cost of an orbifold singularity at the origin.  In those variables, the twisted superpotential of the Landau-Ginzburg model reads
\begin{equation}\label{BLmirrorquinticLGsupo}
  4\pi\I r\twsupo
  = \widetilde{G}_5(\widetilde{X})
  = \widetilde{X}_1^5 + \widetilde{X}_2^5 + \widetilde{X}_3^5 + \widetilde{X}_4^5 + \widetilde{X}_5^5
  + e^{(-2\pi\zeta+\I\vartheta)/5} \widetilde{X}_1 \widetilde{X}_2 \widetilde{X}_3 \widetilde{X}_4 \widetilde{X}_5 \;.
\end{equation}
This model does not depend on the $101$ parameters in the original superpotential $P\, G_5(X)$, hence it can only be the mirror of the quintic GLSM at a specific point in the moduli space.  Since the theories coincide at that point, their whole moduli spaces must be the same, but let us describe more precisely how this comes about.  Continuing from $Z_A$ of the Landau-Ginzburg model, we add a $U(1)$ twisted vector multiplet under which each $\widetilde{X}_i$ has charge $+1$ and a twisted chiral multiplet $\widetilde{P}$ of charge $-5$ and replace $\twsupo$ by $\widetilde{P} \, \widetilde{G}_5(\widetilde{X})$ to make it gauge invariant. The $Z_A$ partition function \eqref{BLZAGLSM} of this twisted GLSM coincides with that of the Landau-Ginzburg model.  The twisted GLSM is a special case of the quintic GLSM with multiplets replaced by twisted multiplets and $G_5\to\widetilde{G}_5$.  When its FI parameter $\chi$ is negative, the low-energy limit is the Landau-Ginzburg model \eqref{BLmirrorquinticLGsupo}: the D-term equation $\sum_i \abs{\widetilde{X}_i}^2 - 5\abs{\widetilde{P}}^2 = \chi$ forces $P\neq 0$ and the F-term equation then sets all $\widetilde{X}_i=0$.  When $\chi>0$ instead, the low-energy limit is an NLSM on the quintic hypersurface $\widetilde{G}_5=0$, orbifolded as described above.  This orbifold has fixed points and curves.  Blowing up the singularities requires a choice of $100$ volume parameters which combine with $\chi$ to reproduce the $101$ chiral parameters of the original GLSM.  The NLSM on the blown-up orbifold of $\{\widetilde{G}_5=0\}$ could also be described by a twisted $U(1)^{101}$ GLSM with $106$ twisted chiral multiplets, but this is somewhat cumbersome.

The partition functions $Z_A$ and $Z_B$ of mirror theories have also been proven equal for all complete intersections in products of weighted projective spaces.

\subsection{Seiberg duality with unitary groups}
\label{BLssec:dual-seiberg}

We now describe a 2d $\cN{=}(2,2)$ analogue of 4d $\cN{=}1$ Seiberg duality, explain how sphere partition functions and elliptic genera of the dual theories are compared, and deduce some variants of this duality in the next subsection. The first such $\cN{=}(2,2)$ dualities were analyzed in \cite{BLHori:2006dk}.

The duality states that two theories (named ``electric'' and ``magnetic'' theories) have the same infrared limit.  The electric theory has a $U(K)$ gauge group, $N_f$~fundamental chiral multiplets~$\phi_F$ ($1\leq F\leq N_f$), and $N_a$~antifundamental chiral multiplets~$\anti{\phi}_A$ ($1\leq A\leq N_a$).  We assume $N\leq\max(N_f,N_a)$, as otherwise supersymmetry is spontaneously broken (both the elliptic genus and the sphere partition function vanish, see \autoref{BLfoot:susybreaks}) and there is no duality.  The magnetic theory has the same field content with
$$
(K,N_f,N_a) \;\to\; (K',N_f',N_a') = \big( \max(N_f,N_a)-K \,,\, N_a \,,\, N_f \big)
$$
and additionally it has $N_a N_f$ gauge singlet chiral multiplets $M_{AF}$ with a cubic superpotential $\supo=M_{AF}\anti{\phi}'_{F} \phi'_{A}$.  The complexified FI parameters $t=2\pi\zeta+\I\vartheta$ and the vector R-charges are
\begin{gather}
  (-1)^{N_f'-K'} e^{-t'} = \bigl[ (-1)^{N_f-K} e^{-t} \bigr]^{-1}
  \text{ namely }
  \zeta' = -\zeta \,, \
  \vartheta' = -\vartheta + \min(N_f,N_a)\pi
  \\
  \label{BLseibergRch}
  \Rcharge'_A = 1-\anti{\Rcharge}_A \,,
  \qquad
  \anti{\Rcharge}'_F = 1-\Rcharge_F \,,
  \qquad
  \Rcharge'^{(M)}_{AF} = \anti{\Rcharge}_A + \Rcharge_F \,.
\end{gather}
These R-charges give the superpotential R-charge $2$ as required by supersymmetry, and are also consistent with the matching of chiral rings: $M_{AF}$ is mapped to the meson $\anti{\phi}_A \phi_F$ of the electric theory. The flavor symmetry $S[U(N_f)\times U(N_a)]$ shared by the two theories can be coupled to a background vector multiplet to include twisted masses and flavor fluxes (the same in both theories), and $M$ transforms in the bifundamental representation of $U(N_f)\times U(N_a)$.

Charge conjugation lets us assume $N_a\leq N_f$.  It would in fact be enough to check the duality for $N_a=N_f$, then decouple chiral multiplets by giving them large twisted masses.

Let us compare elliptic genera of the two theories \cite{BLGadde:2013ftv, BLBenini:2013xpa}. Recall that the elliptic genus $Z_{T^2}(\tau,z,\xi)$ depends on the period $\tau$ of $T^2$, an R-symmetry holonomy $z$, and flavor symmetry holonomies $\xi$ in the flavor Cartan algebra. It is a sum of JK residues \eqref{BLeg main formula} of $Z_{\text{1-loop}}(\tau,z,u,\xi)$ at values of the gauge holonomies $u$ (in the gauge Cartan algebra) where this meromorphic $(\rank G,0)$-form has poles.  More precisely, each component of $u$ lies in a torus $\mathbb{C}/(\mathbb{Z}+\tau\mathbb{Z})$ and similarly for $\xi$. For the electric theory,%
\footnote{With a slight abuse of notation, we identify $\theta_1(\tau|z) \equiv \theta_1(q,y)$ where, as usual, $q = e^{2\pi i \tau}$ and $y=e^{2\pi i z}$.}
\begin{multline}\label{BLegSQCD1loop}
  Z_{\text{1-loop}}(\tau,z,u,\xi,\anti{\xi}) =
  \frac{1}{K!} \biggl(\frac{2\pi\eta(q)^3}{\theta_1(\tau|-z)}\biggr)^{K}
  \prod_{i\neq j}^K \frac{\theta_1(\tau|u_i-u_j)}{\theta_1(\tau|u_i-u_j-z)}
  \\
  \times \prod_{i=1}^K \Biggl(
  \prod_{F=1}^{N_f} \frac{\theta_1(\tau|u_i-\xi_F+(\frac{\Rcharge_F}{2}-1)z)}{\theta_1(\tau|u_i-\xi_F+\frac{\Rcharge_F}{2}z)}
  \prod_{A=1}^{N_a} \frac{\theta_1(\tau|-u_i+\anti{\xi}_A+(\frac{\tilde{\Rcharge}_A}{2}-1)z)}{\theta_1(\tau|-u_i+\anti{\xi}_A+\frac{\tilde{\Rcharge}_A}{2}z)}
  \Biggr) \dd[^K]{u} \;.
\end{multline}
In the following we omit R-charges by shifting $\xi_F\to\xi_F+(\Rcharge_F/2)z$ and $\anti{\xi}_A\to\anti{\xi}_A-(\anti{\Rcharge}_A/2)z$.

Consider first the case $N_a=N_f$.  The elliptic genus is a sum of residues of~\eqref{BLegSQCD1loop} at a set of poles which depends on a choice of auxiliary parameter~$\eta$ in the gauge Cartan algebra.  Choosing $\eta=(1,\ldots,1)$ selects poles due to fundamental chiral multiplets, at $u_i=\xi_{F_i}$ for $1\leq i\leq K$, with all $F_i$~distinct.  Altogether, poles which contribute are labelled by $K$-element subsets $\mathcal{I}$ of $\{1,\ldots,N_f\}$:
\begin{multline}\label{BLSeiberg-eg}
  Z_{T^2}^{U(K)}\bigl(\tau,z,\xi,\anti{\xi}\bigr) = \sum_{\mathcal{I}\in C(K,N_f)}
  \prod_{F\in\mathcal{I}} \Biggl(
  \prod_{E\not\in\mathcal{I}} \frac{\theta_1(\tau|\xi_F-\xi_E-z)}{\theta_1(\tau|\xi_F-\xi_E)}
  \prod_{A=1}^{N_a} \frac{\theta_1(\tau|-\xi_F+\anti{\xi}_A-z)}{\theta_1(\tau|-\xi_F+\anti{\xi}_A)}
  \Biggr)
  \\
  = \Biggl( \prod_{F=1}^{N_f} \prod_{A=1}^{N_a} \frac{\theta_1(\tau|-\xi_F+\anti{\xi}_A-z)}{\theta_1(\tau|-\xi_F+\anti{\xi}_A)} \Biggr) \;
  Z_{T^2}^{U(N_f-K)} \bigl(\tau,z,\, - \tfrac z2-\xi \,,\, \tfrac z2 - \anti{\xi}\bigr) \;.
\end{multline}
Besides straightforward rearrangements, the second line uses $\theta_1(\tau|-z)=-\theta_1(\tau|z)$.  Restoring the R-charges by $\xi_F \to\xi_F-(\Rcharge_F/2)z$ and $\anti{\xi}_A\to\anti{\xi}_A+(\anti\Rcharge_A/2)z$, we recognize the genus of the dual theory with R-charges \eqref{BLseibergRch}.  

For $N_a\neq N_f$ the left-moving $U(1)$ R-symmetry is anomalous and reduces to $\mathbb{Z}_{\abs{N_f-N_a}}$.  The R-symmetry fugacity $y$ must obey $y^{N_f-N_a}=1$, as $Z_{\text{1-loop}}$~is multiplied by $y^{N_f-N_a}$ upon shifting any component of~$u$ by~$\tau$.  Unfortunately, $Z_{\text{1-loop}}$~is ill-defined at $y=1$ so the localization calculation of the elliptic genus fails in that case (and whenever $y^K=1$).  However, we can introduce $N_f-N_a$ chiral multiplets~$P_j$ in the $\det^{-1}$ representation of $U(K)$ to cancel the R-symmetry anomaly and allow generic~$y$, then take the limit $y^{N_f-N_a}\to 1$.  Provided we choose R-charges of $P_j$ to be $\Rcharge_j = \Rcharge + 2j$ for some~$\Rcharge$, their one-loop determinant contributions to the elliptic genus cancel as $y^{N_f-N_a}\to 1$ (for $y=1$ a physical explanation is that one can turn on twisted masses for the $P_j$):
\begin{equation}\label{BLegSQCDextra}
  \Biggl(\prod_{j=1}^{N_f-N_a} \frac{\theta_1(q, y^{\Rcharge/2+j-1} x^{-1})}{\theta_1(q, y^{\Rcharge/2+j} x^{-1})}\Biggr)
  \xrightarrow{\ y^{N_f-N_a}\to 1\ } 1 \;.
\end{equation}
The elliptic genus of the theory enriched with the $P_j$ can be computed with $\eta=(1,\ldots,1)$ as above and yields exactly \eqref{BLSeiberg-eg} once one takes the limit \eqref{BLegSQCDextra}.  For the allowed values $(N_f-N_a)z\in\mathbb{Z}$, the above turns out to simplify to
\begin{equation}
  Z_{T^2}^{U(K)}\bigl(\tau,z,\xi,\anti{\xi}\bigr)_{y^{N_f-N_a}=1}
  = y^{-KN_a/2}\binom{N_f}{K}_y
  = y^{-KN_a/2} \, \frac{\prod_{j=N_f+1-K}^{N_f} \big( y^{j/2}-y^{-j/2} \big) }{ \prod_{j=1}^{K} \big( y^{j/2}-y^{-j/2} \big) } \;.
\end{equation}
The elliptic genus vanishes for $y^K\neq 1$: this could be derived by using $\eta=(-1,\ldots,-1)$ in the theory without~$P_j$.  Incidentally, we learn by setting $y=1$ that the theories have $\binom{N_f}{K}$~vacua.

The equality of elliptic genera implies that BPS states of the dual theories have identical flavor and R-symmetry charges, but does not fix the map of (complexified) FI parameters nor the superpotential.  These are fixed by comparing A-type and B-type sphere partition functions, respectively.

Next, we sketch the proof~\cite{BLBenini:2012ui, BLBenini:2014mia, BLGomis:2014eya} that A-type sphere partition functions $Z_{S^2}^A$ of the two theories coincide.  This probes their twisted chiral rings (these have also been proven isomorphic).  As described in \autoref{BLssec:sphere-higgs} the partition function can be localized to \eqref{BLZHiggs-sqcd}: a ``Higgs branch configuration'' (labeled by $\mathcal{I}$ below) in the bulk of the sphere with vortices at the North pole and antivortices at the South pole. In detail,
\begin{equation}\label{BLSeibergZS2}
  Z_{S^2}^{A}
  = \sum_{\mathcal{I}\in C(K,N_f)} Z_{0}^{\mathcal{I}} \; Z_{+}^{\mathcal{I}}(e^{-t}) \; Z_{-}^{\mathcal{I}}\bigl(e^{-\bar{t}+\I\pi(N_f-N_a)}\bigr)
\end{equation}
where the semiclassical contribution~$Z_0^{\mathcal{I}}$ and the (anti)vortex contributions~$Z_{\pm}^{\mathcal{I}}$ are expressed in terms of the combinations $\Sigma_F^{\pm} = \Rcharge_F/2+\I r\twmass_F\pm\flavorflux_F/2$ of R-charge, twisted mass and flavor flux of fundamental chiral multiplets and similarly $\anti{\Sigma}_A^{\pm} = -\anti{\Rcharge}_A/2+\I r\anti{\twmass}_A\pm\anti{\flavorflux}_A/2$ for antifundamentals:
\begin{align}
  Z_{0}^{\mathcal{I}}
  &= \prod_{F\in\mathcal{I}} \Biggl(e^{-t\Sigma_F^+-\bar{t}\Sigma_F^-}
  \prod_{E\not\in\mathcal{I}} \frac{\Gamma(\Sigma_E^+ - \Sigma_F^+)}{\Gamma(1-\Sigma_E^- + \Sigma_F^-)}
  \prod_{A=1}^{N_a} \frac{\Gamma(\Sigma_F^+ - \anti{\Sigma}_A^+)}{\Gamma(1-\Sigma_F^- + \anti{\Sigma}_A^-)} \Biggr)
  \\
  Z_{\pm}^{\mathcal{I}}(x)
  &= \sum_{(k_F\geq 0)_{F\in\mathcal{I}}}
  \prod_{F\in\mathcal{I}}
  \frac{x^{k_F} \prod_{A=1}^{N_a} \bigl(\Sigma_F^{\pm}-\anti{\Sigma}_A^{\pm}\bigr)_{k_F}}
  {\prod_{E\in\mathcal{I}} \bigl(-\Sigma_E^{\pm}+\Sigma_F^{\pm}-k_E\bigr)_{k_F} \prod_{E\not\in\mathcal{I}} \bigl(\Sigma_E^{\pm}-\Sigma_F^{\pm}-k_F\bigr)_{k_F}} \;.
\end{align}
Dual partition functions are compared term by term.  The semiclassical parts $Z_{0,\text{electric}}^{\mathcal{I}}$ and $Z_{0,\text{magnetic}}^{\{1,\ldots,N_f\}\setminus\mathcal{I}}$ are equal up to simple factors elaborated on below.  Terms of order $x^k$ for some $k\geq 0$ in the vortex partition functions can be recast as a $k$-dimensional contour integral such that the poles on one side of the contour are labelled by $(k_F\geq 0)_{F\in\mathcal{I}}$ with $\sum_F k_F = k$.  Provided $N_a\leq N_f-2$, there is no pole at infinity and the sum of residues is equal to a sum over poles on the other side of the contour, which reproduces the $k$-vortex partition function of the dual theory.  For $\abs{N_a-N_f}\leq 1$ the integrand is singular at infinity and more tedious calculations are needed. The result is
\begin{multline}\label{BLSeiberg-ZS2}
  Z^{N_f,N_a}_{U(K)} \bigl(\Sigma^{\pm},\anti{\Sigma}^{\pm};t\bigr)
  = a_+\Bigl(e^{-t-\I\pi K'}\Bigr) \; a_-\Bigl(e^{-\bar{t}+\I\pi K'}\Bigr)
	\\
  \times \prod_{F,A} \frac{\Gamma(\Sigma_F^+ - \anti{\Sigma}_A^+)}{\Gamma(1-\Sigma_F^- +\anti{\Sigma}_A^-)} \;
  Z^{N_a,N_f}_{U(K')} \Bigl(\anti{\Sigma}^{\pm}+\tfrac{1}{2},\Sigma^{\pm}-\tfrac{1}{2};t'\Bigr)
\end{multline}
where the factors $a_\pm$ depend on $\Sigma_F^\pm$ and~$\anti{\Sigma}_A^\pm$ and are (anti)holomorphic functions of~$t$:
\begin{equation}\label{BLseibergapm}
  a_\pm(z) = z^{-K'/2} \prod_{F=1}^{N_f} \Bigl[e^{\pm\I\pi K'} z\Bigr]^{\Sigma_F^\pm} \prod_{A=1}^{N_a} \Bigl[e^{\pm\I\pi K'}\Bigr]^{\anti{\Sigma}_A^{\pm}} G_\pm(z)
\end{equation}
and the last function is $G_\pm(z)=1$ for $N_a\leq N_f-2$, $G_\pm(z)=e^{\mp z}$ for $N_a=N_f-1$, and $G_\pm(z)=(1+z)^{K'-\sum_F\Sigma_F^{\pm}+\sum_A\anti{\Sigma}_A^{\pm}}$ for $N_a=N_f$.  The last two factors in \eqref{BLSeiberg-ZS2} are the partition function of the dual theory with its mesons: shifts of $\Sigma^{\pm}$ and $\anti{\Sigma}^{\pm}$ by $\tfrac{1}{2}$ realize the map of R-charges \eqref{BLseibergRch}.  The factors $a_\pm$ are ambiguities due to finite renormalization of the partition function, but in quiver gauge theories they become physical and have neat interpretations in terms of cluster algebras \cite{BLBenini:2014mia} or Liouville/Toda correlation functions \cite{BLGomis:2014eya}. The phase of $a_+a_-$ comes from a background twisted superpotential that depends on $t$ and on the background field strength (twisted chiral) multiplets incorporating twisted masses and flavor fluxes.  The absolute value is independent of these background fields and can be ignored for our purposes.  It comes from improving the R-symmetry current.  Ambiguities of the sphere partition function under multiplication by (anti)holomorphic functions of $t$ play an important role in \volcite{MO}.

We will not elaborate on the comparison of B-type sphere partition functions performed very recently in \cite{BLDoroud:2016}. It yields that if the electric theory is endowed with a superpotential $\supo=\supo(\anti{\phi}_A\phi_F)$, then the magnetic theory has the superpotential $\supo(M_{AF})+M_{AF}\anti{\phi}'_{F} \phi'_{A}$.  This is consistent with the matching of chiral rings $M_{AF}=\anti{\phi}_A\phi_F$ and $\anti{\phi}'_{F}\phi'_{A}=0$.

\subsection{Variants of Seiberg duality}
\label{BLssec:dual-variants}

We now turn to consequences and analogues of the $\cN{=}(2,2)$ Seiberg duality.

The $\cN{=}(2,2)$ $SU(K)$ gauge theory with $N_f$ fundamental chiral multiplets is dual to the theory with $K\to K'=N_f-K$, as described in \cite{BLHori:2006dk}.  Chiral rings are generated by baryons, which match provided R-charges are $\Rcharge'_F = -\Rcharge_F + \sum_F \Rcharge_F / K'$.  Chiral multiplets of the two theories are in the (anti)fundamental of an $SU(N_f)$ flavor symmetry and have charges $1/K$ and $1/K'$ under a $U(1)$ baryonic symmetry.  Elliptic genera are shown to match in \cite{BLGadde:2013ftv, BLBenini:2013xpa}.  A-type sphere partition functions are shown to match \cite{BLBenini:2012ui} by integrating partition functions of the analogous $U(K)$ and $U(N_f-K)$ theories:
\begin{equation}\label{BLseibergZSU}
  \begin{aligned}
    Z_{SU(K)}^{N_f}(\Rcharge_F)
    & = \int_0^{2\pi} \frac{\dd{\vartheta}}{2\pi} \int_{-\infty}^{\infty} 4\pi\dd{\zeta}
    Z_{U(K)}^{N_f}(\Rcharge_F;\zeta,\vartheta)
    \\
    & = \int_0^{2\pi} \frac{\dd{\vartheta}}{2\pi} \int_{-\infty}^{\infty} 4\pi\dd{\zeta}
    Z_{U(N_f-K)}^{N_f}(\Rcharge_F';-\zeta,-\vartheta+\#\pi)
    = Z_{SU(N_f-K)}^{N_f}(\Rcharge_F').
  \end{aligned}
\end{equation}
We have used that partition functions of $U(K)$ and $U(N_f-K)$ theories are equal up to~\eqref{BLseibergapm}, namely powers of $e^{-t}$ and $e^{-\bar{t}}$ that can be absorbed by shifting the Coulomb branch parameter~$\sigma$ and the flux~$\flux$.  This shifts R-charges from $1-\Rcharge_F$ to $\Rcharge'_F$ given above, and affects twisted masses and flavor fluxes in the same way, compatible with flavor symmetries.

In the presence of $N_a\leq N_f-2$ additional antifundamental chiral multiplets, all steps of~\eqref{BLseibergZSU} go through ($\zeta$-dependent factors prevent the last step for $N_a=N_f-1$ and $N_a=N_f$) and yield
\begin{equation}
  Z_{SU(K)}^{N_f,N_a}(\Rcharge_F,\anti{\Rcharge}_A)
	= \prod_{F,A} \gamma(\anti{\Rcharge}_A/2+\Rcharge_F/2)
	Z_{SU(N_f-K)}^{N_a,N_f}\biggl(2-\anti{\Rcharge}_A+\frac{1}{K'}\sum_F\Rcharge_F, -\Rcharge_F+\frac{1}{K'}\sum_F\Rcharge_F\biggr) \,.
\end{equation}
At first this suggests that the two theories may be dual.  However, chiral rings do not match: the mesons $\anti{\phi}_A \phi_F$ and baryons $\phi_{F_1}\wedge\cdots\wedge\phi_{F_K}$ of the electric theory match with the singlets and baryons of the magnetic theory, but there is no chiral operator in the magnetic theory with the same R-charge as antibaryons $\anti{\phi}_{A_1}\wedge\cdots\wedge\anti{\phi}_{A_K}$.  The lack of duality is confirmed by noting that elliptic genera fail to match.  A similar situation was observed in \cite{BLHori:2013gga} where two GLSMs with equal $Z_A$ were shown to flow to different SCFTs, whose Calabi-Yau target spaces thus have the same quantum K\"ahler moduli space despite having different complex structure moduli spaces.

The same technique gives pairs of $U(K)$ and $U(K')$ theories that have equal sphere partition functions but are not all dual.  Start from the equality of partition functions of $U(K)$ and $U(N_f-K)$ theories with $N_f$~fundamentals and $N_a\leq N_f-2$ antifundamentals and additional singlets for the magnetic theory\footnote{To ease the parallel description of dual theories we have charge conjugated one theory.}.  Add $L$~singlets on both sides and gauge a $U(1)\subset S[U(N_f)\times U(N_a)]\times U(1)^L$ flavor symmetry, then shift its generator by that of $U(1)\subset U(K)$ or $U(1)\subset U(N_f-K)$.  Integrate over FI and theta parameters associated to the mixed $U(1)$.  This yields partition functions of $U(K)$ and $U(N_f-K)$ theories with matter in $N_f$~fundamental, $N_a$~antifundamental, and $L$~singlet representations of the $SU$ gauge group and with arbitrary $U(1)$ gauge charges.  As in dualities above, the magnetic theory has $N_fN_a$ additional singlets, now charged under the $U(1)$ gauge group.  Despite sphere partition functions being equal, the theories are not expected to be dual in general: their chiral rings typically do not match due to antibaryons dressed by singlets.  It would be interesting to find out which of these pairs of theories are indeed dual.  In \cite{BLHori:2006dk}, the duality was established for $N_a=0$, $L=1$, and with a superpotential $\supo=P\,G_d(B)$ where $P$~is the additional singlet in the $\det^{-d}$ representation of $U(K)$, and $G_d(B)$ is a degree~$d$ polynomial in the baryons~$B$.  See also \cite{BLUeda:2016wfa} for the case of $N_a=1$ multiplets in the $\overline{\square}\otimes\det^{-1}$ representation and other negative powers of $\det$ (then chiral rings contain no $U(1)$-invariant antibaryons).

Quiver gauge theories with $U(N_i)$ gauge and flavor symmetry factors and bifundamental chiral multiplets have multiple duals.  These are obtained by gauging part of the flavor symmetry $S[U(N_f)\times U(N_a)]$ in the duality between $U(K)$ and $U(\max(N_a,N_f)-K)$ theories above.  Denote by $a_{ij} \geq 0$ the number of chiral multiplets in the antifundamental representation of $U(N_i)$ and the fundamental of $U(N_j)$ for $i\neq j$.  Let $N_f(k) = \sum_i N_i a_{ik}$ and $N_a(k) = \sum_j a_{kj} N_j$ be the numbers of (anti)fundamental chiral multiplets for the node $U(N_k)$.  For any gauge factor $U(N_k)$ there exists a dual with
\begin{equation}
  N_k \to N_k'=\max \Bigl(N_f(k) \,,\, N_a(k) \Bigr)-N_k \,,
	\qquad
	a'_{ij} = \begin{cases}
	  a_{ij} + a_{ik} a_{kj} & \text{if $i\neq k$ and $j\neq k$} \\
		a_{ji} & \text{if $i=k$ or $j=k$}
	\end{cases}
\end{equation}
with cubic superpotential terms coupling the $N_j a'_{jk}=a_{kj} N_j$ fundamental and $a'_{ki}N_i=N_i a_{ik}$ antifundamental chiral multiplets of $U(N_k')$ with the corresponding $N_i a_{ik} a_{kj} N_j$ singlets of $U(N_k')$ while preserving the $U(N_i)$ and $U(N_j)$ symmetries.  The map of (complexified) FI parameters is more elaborate: for example in quivers with all $N_a(i)=N_f(i)$ so that FI parameters do not run
\begin{equation}
  z_k' = z_k^{-1}
	\quad \text{and for $k\neq i$,} \quad
	z_i' = z_i z_k^{a_{ki}} (z_k+1)^{a_{ik}-a_{ki}}
\end{equation}
in terms of the K\"ahler parameters $z_j=\exp\bigl(-t_j+\I\pi(N_f(j)-N_j)\bigr)$.
One can typically apply further Seiberg dualities to other nodes in the quiver, obtaining a web of dualities.  However, after dualizing the node $U(N_k)$, the quiver gauge theory involves adjoint matter if $a_{ik}\neq 0\neq a_{ki}$ for some~$i$: then one cannot apply Seiberg duality to $U(N_i)$.  This motivates the restriction to quivers such that for any of the dual descriptions, whenever $a_{ij}\neq 0\neq a_{ji}$ for some $i\neq j$, there exists a quadratic superpotential giving mass to $2\min(a_{ij},a_{ji})$ of these bifundamental chiral multiplets, which can thus be removed without affecting the infrared limit.  The condition is difficult to check, but has been proven for various classes of quivers.  Then all dual quivers can be taken to have $a_{ij}=0$ or $a_{ji}=0$ for all $i,j$, and the matter content is equally described by the antisymmetric matrix $B_{ij}=a_{ij}-a_{ji}$.  In \cite{BLBenini:2014mia} it was observed that $B$, together with the beta function of FI parameters, and the $z_j$, reproduce the structure of cluster seeds.  Dualities act on this data as cluster mutations.  The connection between cluster algebras and 2d $\cN{=}(2,2)$ quiver dualities is stronger than in higher dimensions, as it concerns not only the quiver described by~$B$ but also cluster coefficients and cluster variables.

Seiberg duality can also be realized as an explicit symmetry in a 2d CFT \cite{BLGomis:2014eya}: the A-type sphere partition function of $U(K)$ gauge theories with $N_f$ fundamental and $N_a\leq N_f$ antifundamental chiral multiplets is equal to a correlator in the $A_{N_f-1}$ Toda CFT.  Toda CFT charge conjugation reproduces precisely $K'=N_f-K$ and the map of FI parameters, R-charges and twisted masses.  This instance of the AGT correspondence goes further: one can include adjoint matter and obtain two other dualities.
\begin{itemize}
\item A generalization of $\cN{=}(2,2)^*$ dualities studied in \cite{BLBenini:2014mia}. The duality relates $U(K)$ and $U(K')$ gauge theories with $N$ fundamental, $N$ antifundamental and one adjoint chiral multiplet $X$ with a superpotential $\supo=\sum_{F=1}^{N} \anti{\phi}_F X^{l_F} \phi_F$ for arbitrary integers $l_F\geq 0$.  The magnetic theory has $K'=\sum_F l_F - K$ colors (for $K'<0$ supersymmetry is broken and there is no duality).  When all $l_F=1$, the theories are $\cN{=}(2,2)^*$ theories \big(mass deformations of $\cN{=}(4,4)$ SQCD\big).  A-type sphere partition functions of the two theories were proven to be equal in \cite{BLGomis:2014eya}. Chiral rings are generated by $\anti{\phi}_A X^k \phi_F$ with $0\leq k<l_A,l_F$ and by $\Tr X^k$ for $0\leq k<K$ or $0\leq k<K'$ depending on the theory.  This mismatch has not been investigated but might be cured by the superpotential. Just like Seiberg duality, (part of the) flavor symmetries can be gauged to produce dualities between quivers.
\item An $\cN{=}(2,2)$ Kutasov-Schwimmer duality. The electric theory has $U(K)$ gauge group with $N_f$ fundamentals, $N_a$ antifundamentals and one adjoint $X$ with superpotential $\supo=\Tr X^{l+1}$ for some $l\geq 1$.  The magnetic theory is identical with $(K,N_f,N_a,l)\to (\max(N_f,N_a)l-K,N_a,N_f,l)$ and $l N_f N_a$ gauge singlets $M_{jFA}$ for $0\leq j<l$, with a superpotential $\supo=M_{jAF}\anti{\phi}'_F X'^j \phi'_A+\Tr X'^{(l+1)}$. Chiral rings match under $X\to X'$ and $\anti{\phi}_A X^j \phi_F \to M_{jAF}$.  A-type sphere partition functions were proven to be equal in \cite{BLGomis:2014eya}. It would be interesting to investigate the existence of analogues of Brodie dualities, where the electric and magnetic theories have two adjoint chiral multiplets $X$ and $Y$ subject to a superpotential $\supo=\Tr(X^{k+1} + XY^2)$.
\end{itemize}

Orthogonal and symplectic gauge groups were considered by Hori \cite{BLHori:2011pd}.  For instance, the $USp(2k)$ gauge theory with $2p+1$ fundamentals $\phi_I$ and $m$ singlets $M^a$ subject to a cubic superpotential $\supo=A_a^{[IJ]}M^a\vev{\phi_I,\phi_J}$ with generic coefficients $A_a^{[IJ]}$ is dual to the theory with $k\to p-k$ and $m\to\binom{2p+1}{2}-m$ provided these numbers are positive. Sphere partition functions of these dual theories have not been compared: the usual method of comparing Higgs branch expressions fails due to the absence of FI parameter. Gauging a $U(1)$ flavor symmetry gives further dualities \cite{BLHori:2013gga, BLGerhardus:2015sla}, whose gauge group $U(1)\times USp(2k)$ (and its $\mathbb{Z}_2$ quotient) allow FI-theta parameters. Sphere partition functions have been compared for theories whose low-energy limit has a Calabi-Yau threefold target space.  Presumably, integrating over the FI-theta parameters as in \eqref{BLseibergZSU} should help prove that Hori duals have equal A-type sphere partition function.

\subsection[\texorpdfstring{$\cN{=}(0,2)$}{N=(0,2)} trialities]{$\boldsymbol{\cN{=}(0,2)}$ trialities}
\label{BLssec:dual-02}

We describe dualities \cite{BLGadde:2013lxa} between $\cN{=}(0,2)$ theories with a gauge group $U(K)$ and (anti)fundamental matter.  The theories are expected to flow at intermediate energy scales to NLSMs on bundles over Grassmannians \cite{BLGadde:2014ppa} and each duality is due to an isomorphism between bundles over $Gr(K,N)$ and $Gr(N-K,N)$ \cite{BLJia:2014ffa}.

We let $N_P$, $N_\Phi$, $N_\Psi$ denote the number of antifundamental chirals $P$, fundamental chirals $\Phi$, and antifundamental Fermi multiplets $\Psi$ (which could be mapped to fundamental ones by exchanging $E$ and $J$ interactions).  The $SU(K)$ gauge anomaly $N_P/2+N_\Phi/2-N_\Psi/2-K$ due to fermions in the matter and vector multiplets must vanish, thus $K=(N_P+N_\Phi-N_\Psi)/2$.  The $U(1)$ gauge anomaly $N_P K + N_\Phi K - N_\Psi K = 2 K^2$ of these multiplets is cancelled%
\footnote{One could consider $SU(K)$ gauge theories and omit $\Omega_{1,2}$.}
by adding two Fermi multiplets $\Omega_{1,2}$ with $U(1)$ charge $K$, \ie, in the determinant representation of $U(K)$.  For convenience, we also include $N_P N_\Phi$ neutral Fermi multiplets $\Gamma$ with a J-term interaction $\Gamma P \Phi$.

Theories with $(N_P,N_\Phi,N_\Psi)$ equal to the same $(N_1,N_2,N_3)$ up to cyclic permutations, depicted by the quivers in \autoref{BLfig:triality}, are expected to flow to the same infrared fixed point.  As evidence, we show that the elliptic genus%
\footnote{While \cite{BLGadde:2013lxa} work in the NSNS sector we work in the RR sector; results are related by spectral flow.}
is invariant under cyclic permutations of $(N_P,N_\Phi,N_\Psi)$.  The classical flavor symmetry is $S[U(N_P)\times U(N_\Phi)\times U(N_\Psi)\times U(2)]$ with holonomies $(\xi^P_i,\xi^\Phi_i,\xi^\Psi_i,\xi^\Omega_i)$ modulo gauge transformations, but mixed flavor-gauge anomalies reduce Abelian symmetries to a two-dimensional subgroup: this can be used for instance to fix the $U(1)_\Omega$ holonomy $\sum_i \xi^\Omega_i = - \sum_i \xi^P_i - \sum_i \xi^\Phi_i + \sum_i \xi^\Psi_i$ \big(we took $\Phi$ in the $\overline{\square}$ of $SU(N_\Phi)$\big).  Up to a constant, the elliptic genus is a sum of residues of
\begin{equation*}\label{BLtriality1loopA}
  \frac
    {\prod_{\ell=1}^{2}\theta(\xi^\Omega_\ell + \sum_a u_a) \prod_{a=1}^K \prod_{i=1}^{N_\Psi} \theta(\xi^\Psi_i-u_a) \prod_{j=1}^{N_P} \prod_{k=1}^{N_\Phi} \theta(\xi^P_j - \xi^\Phi_k)}
    {\prod_{a=1}^K \prod_{i=1}^{N_P} \theta(\xi^P_i-u_a) \prod_{j=1}^{N_\Phi} \theta(u_a - \xi^\Phi_j)}
  \prod_{a\neq b}^K \theta(u_a-u_b) \; 
  \theta'(0)^K\!\dd[^K]{u}
\end{equation*}
where $\theta(u) = \theta_1(\tau|u)/\I\eta(q)=-\theta(-u)$ has zeros at $\mathbb{Z}+\tau\mathbb{Z}$, no poles, and $\theta'(0)=2\pi\I\eta(q)^2$.  Several sets of poles give the same sum of residues: poles due to $P$ (at $\{u_a\}=\{\xi^P_j|j\in J\}$ for each set of $K$ distinct flavors $J\subset\{1,\ldots,N_P\}$) with residue
\begin{equation}\label{BLtrialityantires}
  \frac
    {\prod_{\ell=1}^{2}\theta(\xi^\Omega_\ell +\sum_{j\in J} \xi^P_j)
      \prod_{j\in J} \prod_{i=1}^{N_\Psi} \theta(\xi^\Psi_i-\xi^P_j)
      \prod_{j\not\in J} \prod_{k=1}^{N_\Phi} \theta(\xi^P_j-\xi^\Phi_k)}
    {\prod_{j\in J} \prod_{i\not\in J} \theta(\xi^P_i-\xi^P_j)} \;;
\end{equation}
or poles due to~$\Phi$ (at $\{u_a\}=\{\xi^\Phi_j|j\in J\}$ for $J\subset\{1,\ldots,N_\Phi\}$) with residue
\begin{equation}\label{BLtrialityfundres}
  \frac
    {\prod_{\ell=1}^{2}\theta(\xi^\Omega_\ell + \sum_{j\in J} \xi^\Phi_j)
      \prod_{j\not\in J} \prod_{i=1}^{N_P} \theta(\xi^P_i-\xi^\Phi_j)
      \prod_{j\in J} \prod_{k=1}^{N_\Psi} \theta(\xi^\Psi_k - \xi^\Phi_j)}
    {\prod_{j\in J} \prod_{i\not\in J} \theta(\xi^\Phi_j-\xi^\Phi_i)} \;.
\end{equation}
Up to a sign, \eqref{BLtrialityantires} is mapped to \eqref{BLtrialityfundres} under $(N_P,\xi^P)\to(N_\Phi,\xi^\Phi)\to(N_\Psi,\xi^\Psi)\to(N_P,\xi^P)$ and $J\to J^\complement$.  One also has a shift $\xi^\Omega_i \to \xi^\Omega_i - \sum_j \xi^\Omega_j - \sum_j \xi^P_j$ fixed by the above constraint on $\sum\xi^\Omega$.  Elliptic genera of theories in \autoref{BLfig:triality} are thus equal, with $S[ U(N_1)\times U(N_2)\times U(N_3)\times SU(2) ]$ flavor symmetries identified.  Another outcome of the calculation is that the elliptic genus vanishes if $K>N_P$ or $K>N_\Phi$ in any frame, which is equivalent to $K<0$ in a dual frame.  This suggests that supersymmetry is broken unless $(N_1,N_2,N_3)$ obey the triangle inequality.

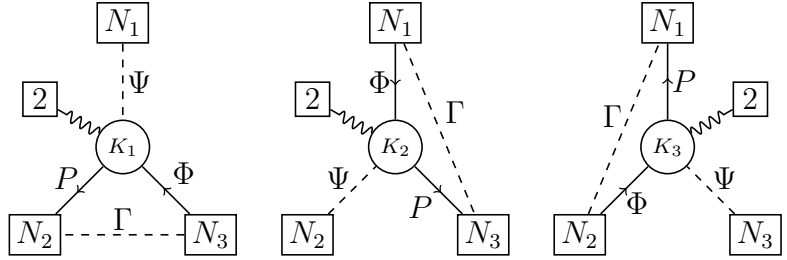
\begin{figure}
\newcommand{\FUN}{\square}\newcommand{\ANT}{\overline{\square}}\newcommand{\1}{\mathbf{1}}
\begin{tabular}[b]{*{6}{>$c<$}}
\toprule
          & P & \Phi & \Psi & \Gamma & \Omega
\\
U(K)      & \ANT & \FUN & \ANT & \1  & \det \\
U(N_P)    & \FUN & \1   & \1   & \FUN & \1  \\
U(N_\Phi) & \1   & \ANT & \1   & \ANT & \1  \\
U(N_\Psi) & \1   & \1   & \FUN & \1   & \1  \\
SU(2)     & \1   & \1   & \1   & \1   & \FUN
\\\bottomrule
\end{tabular}\hfill
\begin{tikzpicture}[semithick,node distance=4em]
  \node (K) [color-group] {$\scriptstyle K_1$};
  \node (1) [flavor-group, above of=K] {$N_1$};
  \node (2) [flavor-group, below left of=K] {$N_2$};
  \node (3) [flavor-group, below right of=K] {$N_3$};
  \node (Om)[flavor-group] at ([shift={(K)}]150:7ex) {$2$};
  \draw[->-=.55] (K) -- (2) node [midway, above left=-1ex] {$P$}; 
  \draw[->-=.55] (3) -- (K) node [midway, above right=-.5ex] {$\Phi$};
  \draw[dashed]  (1) -- (K) node [midway, right=-.5ex] {$\Psi$};
  \draw[dashed]  (3) -- (2) node [midway, above=-.5ex] {$\Gamma$};
  \draw[snake=coil,mirror snake,segment aspect=0,segment amplitude=2pt,segment length=4pt](K) -- (Om);
\end{tikzpicture}\hfill
\begin{tikzpicture}[semithick,node distance=4em]
  \node (K) [color-group] {$\scriptstyle K_2$};
  \node (1) [flavor-group, above of=K] {$N_1$};
  \node (2) [flavor-group, below left of=K] {$N_2$};
  \node (3) [flavor-group, below right of=K] {$N_3$};
  \node (Om)[flavor-group] at ([shift={(K)}]150:7ex) {$2$};
  \draw[->-=.55] (K) -- (3) node [midway, below left=-.5ex] {$P$}; 
  \draw[->-=.55] (1) -- (K) node [midway, left=-.5ex] {$\Phi$};
  \draw[dashed]  (2) -- (K) node [midway, above left=-1ex] {$\Psi$};
  \draw[dashed]  (1) -- (3) node [midway, above right=-.5ex] {$\Gamma$};
  \draw[snake=coil,mirror snake,segment aspect=0,segment amplitude=2pt,segment length=4pt](K) -- (Om);
\end{tikzpicture}\hfill
\begin{tikzpicture}[semithick,node distance=4em]
  \node (K) [color-group] {$\scriptstyle K_3$};
  \node (1) [flavor-group, above of=K] {$N_1$};
  \node (2) [flavor-group, below left of=K] {$N_2$};
  \node (3) [flavor-group, below right of=K] {$N_3$};
  \node (Om)[flavor-group] at ([shift={(K)}]30:7ex) {$2$};
  \draw[->-=.55] (K) -- (1) node [midway, right=-.5ex] {$P$}; 
  \draw[->-=.55] (2) -- (K) node [midway, below right=-1ex] {$\Phi$};
  \draw[dashed]  (3) -- (K) node [midway, above right=-1ex] {$\Psi$};
  \draw[dashed]  (2) -- (1) node [midway, above left=-1ex] {$\Gamma$};
  \draw[snake=coil,segment aspect=0,segment amplitude=2pt,segment length=4pt](K) -- (Om);
\end{tikzpicture}
\caption{\label{BLfig:triality}Quivers for three $\cN{=}(0,2)$ theories related by triality: $K_i=(N_1+N_2+N_3)/2-N_i$ and $(N_P,N_\Phi,N_\Psi)$ is one of $(N_2,N_3,N_1)$, $(N_3,N_1,N_2)$, $(N_1,N_2,N_3)$.  The table lists how the chirals $P,\Phi$ and Fermis $\Psi,\Gamma,\Omega$ transform under the gauge group $U(K)=U\bigl(\frac{1}{2}(N_P+N_\Phi-N_\Psi)\bigr)$ and the classical flavor symmetry $S[U(N_P)\times U(N_\Phi)\times U(N_\Psi)\times U(2)]$ which loses one Abelian factor due to a flavor-gauge anomaly.}
\end{figure}

Lack of space forces us to only mention variants of the Gadde-Gukov-Putrov triality.  Gauging flavor symmetries leads to dualities between $\cN{=}(0,2)$ quiver gauge theories with bifundamental chiral/Fermi multiplets and Fermi multiplets in determinant representations; however, most quivers have either gauge anomalies or spontaneous supersymmetry breaking.  Such quivers were obtained from brane brick models in \cite{BLFranco:2016nwv}.  As shown in \cite{BLPutrov:2015jpa}, a twisted dimensional reduction of the 6d (2,0) theory on $S^2\times\Sigma$ yields $\cN{=}(0,4)$ Lagrangians labelled by pants decompositions of $\Sigma$, and changes in pants decomposition give $\cN{=}(0,4)$ dualities.  Similarly, twisted dimensional reductions on $S^2$ of 4d $\cN{=}1$ and 4d $\cN{=}2$ dualities yield two-dimensional $(0,2)$ or $(0,4)$ or $(2,2)$ dualities \cite{BLGadde:2015wta}, in particular an $SU(K)$ variant of the $(0,2)$ triality above.  All of these dualities are checked by comparing elliptic genera.  The dimensional reduction is based on \cite{BLClosset:2013sxa, BLBenini:2015noa} (see also \cite{BLHonda:2015yha, BLHonda:2015ava, BLCecotti:2015lab}).


\section{Conclusion}
\label{BLsec:conclusion}

We have reviewed the main localization calculations in two dimensions on the sphere (\autoref{BLsec:sphere}), other curved backgrounds (\autoref{BLsec:curved}) and the torus (\autoref{BLsec:torus}), and discussed applications to mirror symmetry and gauge theory dualities (\autoref{BLsec:dual}).  We now conclude this review by mentioning other developments.

Two-dimensional gauge theory descriptions of several non-critical strings in 6d SCFTs have been tested by comparing topological vertex results to the elliptic genus of a string.  As discussed in \volcite{KL}, this was done for M-strings \cite{BLHaghighat:2013gba, BLHaghighat:2013tka, BLHosomichi:2014rqa, BLSugimoto:2015nha} (namely M2-branes suspended between M5-branes), E-strings \cite{BLKim:2014dza, BLKim:2015fxa} (M2-branes suspended between M5- and M9-branes), little strings \cite{BLKim:2015gha} (strings on type IIA or IIB NS5-branes), and for a class of 6d $\cN{=}(1,0)$ SCFTs engineered from F-theory \cite{BLHaghighat:2014vxa, BLGadde:2015tra}.

Nekrasov's instanton partition function of 4d $\cN{=}2$ theories such as $SU(N)$ super-Yang-Mills can be reproduced by an appropriate (R-preserving) $S^2$ partition function \cite{BLBonelli:2013rja}.%
\footnote{In a different approach \cite{BLPark:2012nn}, for 4d $\cN{=}2$ theories engineered by string theories on a Calabi-Yau three-fold $X$, the Seiberg-Witten K\"ahler potential can be obtained as that of $X$, itself derived from the $S^2$ partition function of a GLSM flowing to an NLSM on $X$.}
The 4d theory is engineered by $N$ fractional D3-branes on $\mathbb{C}^2/\mathbb{Z}_2$ and its instantons by $k$ D($-1$)-branes.  Blowing up the singular point yields a D1--D5 brane system, described in the gauge theory limit by a 2d $\cN{=}(2,2)$ $U(k)$ GLSM on the blown-up sphere $\mathbb{CP}^1$. In the zero-radius limit, its sphere partition function reproduces the equivariant volume of the ADHM moduli space of $k$ instantons, while for non-zero radius it captures genus zero Gromov-Witten invariants of the ADHM moduli space \cite{BLBonelli:2013mma, BLManabe:2014rma}.  This construction was later used to extract spectra of hydrodynamic quantum integrable systems \cite{BLBonelli:2014iza, BLBonelli:2015kpa}.

Another appearance of integrable models in relation to 2d localization is that elliptic genera give solutions to Yang-Baxter equations \cite{BLYamazaki:2015voa, BLYagi:2015lha}.  In this context, the Yang-Baxter equation amounts to the invariance of the genus under $\cN{=}(2,2)$ Seiberg-like dualities.

We have already mentioned that twisted dimensional reductions of 4d $\cN{=}1$ theories on a sphere yield 2d $\cN{=}(0,2)$ theories and that 4d dualities become 2d dualities. The 2d $\cN{=}(0,2)$ elliptic genus is thus a limit of a $T^2 \times S^2$ partition function \cite{BLBenini:2015noa}. Similarly, the partition function of 3d $\cN{=}2$ theories on Lens spaces, described in \volcite{WI}, reduces to an $S^2$ partition function when the circle fiber of the Lens space shrinks to zero size (see also \cite{BLBenini:2011nc, BLYamazaki:2013fva}). The $S^2$ partition function also appears when localizing 4d $\cN{=}2$ theories on $S^4$ on their Higgs branch \cite{BLChen:2015fta, BLPan:2015hza}.

In four-dimensional supersymmetric gauge theories, a class of surface operators can be constructed by coupling a two-dimensional theory to the bulk fields supersymmetrically (see also \volcite{HO}).  Superconformal indices of coupled 2d/4d systems were computed in \cite{BLGadde:2013ftv} (see \cite{BLGaiotto:2014ina, BLBullimore:2014awa} for related 5d calculations) and led to discovering the node-hopping duality (see also \cite{BLChen:2014rca}): coupling the same 2d $\cN{=}(2,2)$ theory to different fields in a 4d $\cN{=}2$ quiver theory gives the same surface operator at low energies. The AGT correspondence relates the node-hopping duality to crossing symmetry in a 2d CFT \cite{BLGomis:2014eya}: the $S^4$ partition function of the 4d theory is identified with a Toda CFT correlator, adding a surface operator on $S^2\subset S^4$ corresponds to inserting a (degenerate) vertex operator in the correlator, and different choices of couplings correspond to different OPEs of that vertex operator with others.  K\"ahler parameters (the 4d gauge coupling and 2d FI/theta parameters) correspond to positions of vertex operators in the correlator and are expected to transform non-trivially under node-hopping.  The map could be found by comparing $S^2\subset S^4$ partition functions of these 2d/4d systems, but instanton-vortex partition functions which appear when localizing \cite{BLLamy-Poirier:2014sea} are unknown.  It would be interesting to derive them.

\section*{Acknowledgements}

We are grateful to Nikolay Bobev, Cyril Closset, Stefano Cremonesi, Nima Doroud, Richard Eager, Jaume Gomis, Kentaro Hori, Sungjay Lee, Daniel S. Park, Mauricio Romo, Yuji Tachikawa, Ran Yacoby, Alberto Zaffaroni and Peng Zhao for collaborations and insightful discussions over the years on some of the material presented here. F.B. is supported in part by the Royal Society as a Royal Society University Research Fellowship holder, and by the MIUR-SIR grant RBSI1471GJ ``Quantum Field Theories at Strong Coupling: Exact Computations and Applications''.

\documentfinish

%% file: header.tex
\newcommand{\chapterauthor}[1]{
\begin{center}
{\bf \normalsize  #1}
\end{center}
}

\newcommand{\chapteraddress}[1]{
\begin{center}
{ \small \it \addressvalue}
\end{center}
}

\newcommand{\chapterabstract}[1]{
\vspace{\baselineskip}
\begin{center}
\textbf{\small Abstract}
\end{center}
#1}

\newcommand{\chapterheader}{

\chapter[\titlevalue{}  (by \shortauthorvalue)]{\titlevalue}
\label{Chapter\IDvalue}
\chapterauthor{\authorvalue}
\chapteraddress{\addressvalue}
\chapterabstract{\abstractvalue}
\tightmtctrue
\minitoc
}

\newcommand{\documentheader}{
\begin{flushright} \small
  \preprintvalue
 \end{flushright}

\begin{center}
{\bf \Large \titlevalue}
\end{center}

\chapterauthor{\authorvalue}
\chapteraddress{\addressvalue}
\chapterabstract{\abstractvalue}

\medskip

This is a contribution to the review volume ``Localization techniques
in quantum field theories'' (eds. V.~Pestun and M.~Zabzine) which
contains 17 Chapters available at \cite{ContributionSummary}

\tableofcontents
}

\newcommand{\ifvolume}[2]{\ifx\ifLONG\undefined#2\else#1\fi}

\newcommand{\documentfinish}{
\ifx\ifLONG\undefined
\bibliographystyle{bibreview} 
\bibliography{\IDvalue,review}  
\end{document}
\else
\addcontentsline{toc}{section}{References}
\input{\DIRvalue/\IDvalue.bbl}
\fi
}

\newcommand{\documentfinishBBL}{
\addcontentsline{toc}{section}{References}
\ifx\ifLONG\undefined
\input{\IDvalue.separate.bbl}
\end{document}
\else
\input{\DIRvalue/\IDvalue.volume.bbl}
\fi
}

\ifx\ifLONG\undefined
\def\volcite#1{Contribution \cite{Contribution#1}}
\else 
\def\volcite#1{Chapter \ref{Chapter#1}}
\fi

%% file: BL.bbl
\providecommand{\href}[2]{#2}\begingroup\raggedright\endgroup

%% file: BLdef.tex
\usepackage{microtype}                             
\usepackage{amsmath,amssymb,amsfonts,mathtools}    
\usepackage{array,booktabs,multirow}               
\usepackage{lmodern}                               
\usepackage{etoolbox}                              
\usepackage{xcolor}                                
\usepackage{slashed}                               
\usepackage{bbm}                                   
\usepackage{hyperref,bookmark}
\usepackage{tikz}
\usepackage{fullpage}

\newcommand{\unit}{\mathbbm{1}}

\newcommand{\eg}{\textit{e.g.\@}}

\newcommand{\ie}{\textit{i.e.\@}}

\newcommand{\bC}{\mathbb{C}}

\newcommand{\bP}{\mathbb{P}}

\newcommand{\bR}{\mathbb{R}}
\newcommand{\bZ}{\mathbb{Z}}
\newcommand{\bfQ}{\mathbf{Q}}

\newcommand{\fg}{\mathfrak{g}}
\newcommand{\fh}{\mathfrak{h}}
\newcommand{\fk}{\mathfrak{k}}
\newcommand{\fm}{\mathfrak{m}}
\newcommand{\fM}{\mathfrak{M}}
\newcommand{\fQ}{\mathfrak{Q}}
\newcommand{\fR}{\mathfrak{R}}

\newcommand{\ft}{\mathfrak{t}}

\newcommand{\se}{{\sf e}}

\newcommand{\sQ}{{\sf Q}}		
\newcommand{\sT}{{\sf T}}

\newcommand{\lie}[1]{\mathfrak{\lowercase{#1}}}                         
\newcommand{\parfrac}[2]{\frac{\partial #1}{\partial #2}}               
\newcommand{\du}[2]{_{#1}^{\phantom{#1}#2}}                             
\newcommand{\wt}{\widetilde}
\newcommand{\anti}[1]{\texorpdfstring{{\protect\widetilde{#1}}}{#1'}}   
\newcommand{\I}{i}                                                      
\newcommand{\dd}[2][]{\mathop{d#1#2}\nolimits}

\newcommand{\dvol}{d\mathrm{vol}}                                       
\newcommand{\repr}{{\mathfrak{R}}}                                      
\newcommand{\twsupo}{\widetilde{W}}                                     
\newcommand{\supo}{W}                                                   
\newcommand{\flux}{\mathfrak{m}}                                        
\newcommand{\flavorflux}{\mathfrak{n}}                                  
\newcommand{\fluxshift}{\mathfrak{u}}                                   
\newcommand{\twmass}{\tau}                                              
\newcommand{\Rcharge}{q}                                                
\newcommand{\supercharge}{\mathcal{Q}}                                  
\newcommand{\bosonic}{\text{b}}                                         
\newcommand{\fermionic}{\text{f}}                                       
\newcommand{\mat}[1]{\begin{pmatrix}#1\end{pmatrix}}                    
\newcommand{\smat}[1]{\big(\begin{smallmatrix}#1\end{smallmatrix}\big)} 
\newcommand{\BLF}[1]{\ifhmode\unskip\fi}                                
\newcommand{\FB}[1]{\ifhmode\unskip\fi}                                 

\numberwithin{equation}{section}

%% file: BLdefcommon.tex
\newcommand{\cA}{\mathcal{A}}
\newcommand{\cB}{\mathcal{B}}
\newcommand{\cC}{\mathcal{C}}
\newcommand{\cD}{\mathcal{D}}
\newcommand{\cE}{\mathcal{E}}

\newcommand{\cG}{\mathcal{G}}
\newcommand{\cH}{\mathcal{H}}

\newcommand{\cK}{\mathcal{K}}
\newcommand{\cL}{\mathcal{L}}
\newcommand{\cM}{\mathcal{M}}
\newcommand{\cN}{\mathcal{N}}
\newcommand{\cO}{\mathcal{O}}

\newcommand{\cQ}{\mathcal{Q}}
\newcommand{\cR}{\mathcal{R}}

\newcommand{\cV}{\mathcal{V}}
\newcommand{\cW}{\mathcal{W}}

\newcommand{\cZ}{\mathcal{Z}}

\DeclareMathOperator{\Tr} {Tr}
\DeclareMathOperator{\re}{Re}         
\DeclareMathOperator{\im}{Im}         
\DeclareMathOperator{\sign}{sign}     
\DeclareMathOperator{\rank}{rank}     
\DeclareMathOperator{\coker}{coker}   
\DeclareMathOperator{\Cone}{Cone}     
\DeclareMathOperator*{\JKres}{JK-Res} 
\DeclareMathOperator*{\Res}{Res}      
\DeclareMathOperator*{\ind}{ind}      
\DeclareMathOperator{\Weyl}{Weyl}     
\DeclarePairedDelimiter{\abs}{\lvert}{\rvert}   
\DeclarePairedDelimiter{\vev}{\langle}{\rangle} 
